\documentclass[%
preprint,
 amsmath,amssymb,
 aps,
prb,
floatfix,
]{revtex4-1}

\usepackage{graphicx}
\usepackage{dcolumn}
\usepackage{bm}
\usepackage{subfig}
\DeclareMathOperator{\Tr}{Tr}

\begin{document}
\title{Thermal Transport Across Metal Silicide-Silicon Interfaces: First-Principles Calculations and Green's Function Transport Simulations}

\author{Sridhar Sadasivam}
\affiliation{Department of Mechanical Engineering and Birck Nanotechnology Center, Purdue University, West Lafayette, IN 47907, USA}
\author{Ning Ye}
\affiliation{Department of Mechanical Engineering, University of Delaware, Newark, DE, 19716, USA}
\author{James Charles}
\affiliation{Network for Computational Nanotechnology and Department of Electrical and Computer Engineering, Purdue University, West Lafayette, Indiana 47907, USA}
\author{Kai Miao}
\affiliation{Network for Computational Nanotechnology and Department of Electrical and Computer Engineering, Purdue University, West Lafayette, Indiana 47907, USA}
\author{Joseph P. Feser}
\affiliation{Department of Mechanical Engineering, University of Delaware, Newark, DE, 19716, USA}
\author{Tillmann Kubis}
\affiliation{Network for Computational Nanotechnology (NCN), Purdue University, West Lafayette, Indiana 47907, USA}
\author{Timothy S. Fisher}
\email{tsfisher@purdue.edu}
\affiliation{Department of Mechanical Engineering and Birck Nanotechnology Center, Purdue University, West Lafayette, IN 47907, USA}
\begin{abstract}
Heat transfer across metal-semiconductor interfaces involves multiple fundamental transport mechanisms such as elastic and inelastic phonon scattering, and electron-phonon coupling within the metal and across the interface. The relative contributions of these different transport mechanisms to interface conductance remains unclear in the current literature. In this work, we use a combination of first-principles calculations under the density functional theory framework and heat transport simulations using the atomistic Green's function (AGF) method to quantitatively predict the contribution of the different scattering mechanisms to the thermal interface conductance of epitaxial CoSi$_2$-Si interfaces. An important development in the present work is the direct computation of interfacial bonding from density functional perturbation theory (DFPT) and hence the avoidance of commonly used `mixing rules' to obtain the cross-interface force constants from bulk material force constants. Another important algorithmic development is the integration of the recursive Green's function (RGF) method with B\"{u}ttiker probe scattering that enables computationally efficient simulations of inelastic phonon scattering and its contribution to the thermal interface conductance. First-principles calculations of electron-phonon coupling reveal that cross-interface energy transfer between metal electrons and atomic vibrations in the semiconductor is mediated by delocalized acoustic phonon modes that extend on both sides of the interface, and phonon modes that are localized inside the semiconductor region of the interface exhibit negligible coupling with electrons in the metal. We also provide a direct comparison between simulation predictions and experimental measurements of thermal interface conductance of epitaxial CoSi$_2$-Si interfaces using the time-domain thermoreflectance technique. Importantly, the experimental results, performed across a wide temperature range, only agree well with predictions that include all transport processes: elastic and inelastic phonon scattering, electron-phonon coupling in the metal, and electron-phonon coupling across the interface.
\end{abstract}
\maketitle
\section{Introduction}
Interfaces between heterogeneous materials provide a plethora of possibilities for the design of devices with engineered electronic and optical properties. This work concerns the study of heat transport across metal-semiconductor heterojunctions that form a technologically important class of interfaces used in electronic devices. The understanding of charge and heat transport through metal contacts to semiconductor channels is critical to ensure reliable operation of field effect transistors that form the basic building block of high-power electronic devices. Understanding of thermal transport through metal-semiconductor interfaces is also important in the design of modern memory storage devices such as heat assisted magnetic recording \cite{kryder2008heat} and phase change memory \cite{reifenberg2008impact}. Apart from their technological relevance, metal-semiconductor interfaces also provide a material system in which various physical mechanisms of heat transport such as elastic interfacial phonon scattering, inelastic phonon scattering, and electron-phonon coupling co-exist. In this work, we provide a rigorous modeling framework to understand the contribution of various interfacial scattering mechanisms to thermal transport across cobalt silicide (CoSi$_2$) - silicon interfaces that are extensively used in microelectronic devices \cite{murarka1995silicide}. 
 
Elastic scattering of phonons at an interface is the most widely studied framework to understand and predict thermal interface conductance at heterojunctions. Under the elastic transport framework, a phonon of energy $\hbar\omega$ incident from one side of an interface is either transmitted across the interface or reflected back into the same material. For elastic interfacial transport, the primary quantity of interest is the phonon transmission function that represents the probability that a phonon incident from one side of the interface transmits to the other side. Anharmonic or three-phonon scattering processes typically become important at room temperature and above, in which a phonon of energy $\hbar\omega$ incident on the interface could transmit or reflect multiple phonons with appropriate energies to ensure energy conservation. This mechanism has been postulated to be important in acoustically mismatched interfaces such as Pb-diamond \cite{kosevich1995fluctuation,hohensee2015thermal}. 

\begin{figure}[htb]
\begin{center}
\includegraphics[height=60mm]{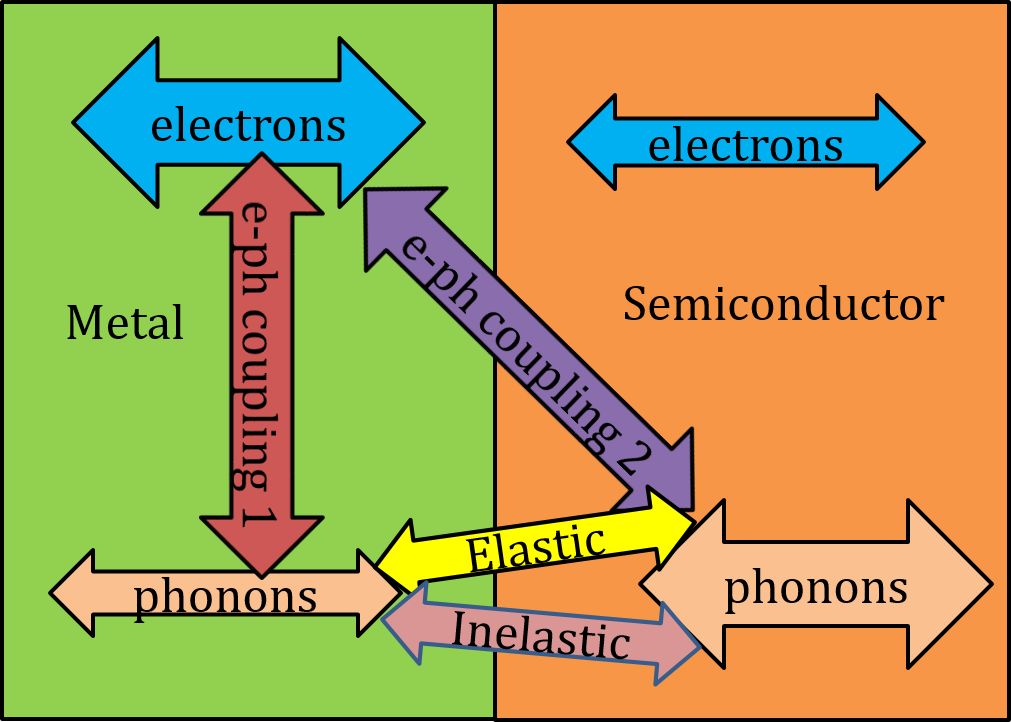}
\caption{Schematic of various mechanisms involved in heat transfer between the dominant energy carriers, i.e., electrons in the metal and phonons in the semiconductor. Phonon-phonon energy transfer across the interface could involve elastic and inelastic interfacial scattering processes. Electron-phonon coupling could involve coupling between electrons in the metal with phonons in the metal and with phonons in the semiconductor.}\label{mechanism_schematic1}
\end{center}
\end{figure}

Electrons are the primary energy carriers in metals while phonons are dominant in intrinsic semiconductors. Hence, electron-phonon coupling can be another important energy transfer mechanism that affects thermal interface conductance in metal-semiconductor interfaces. Electron-phonon coupling within the metal provides an additional resistance to heat transfer \cite{majumdar2004role}. However, electron-phonon coupling across an interface, i.e., coupling between metal electrons and semiconductor phonons, provides a parallel heat flow path in addition to phonon-phonon heat transfer across the interface. Time domain thermoreflectance (TDTR) experiments in the literature \cite{hopkins2009effects,guo2012heat} suggest that direct electron-phonon coupling can contribute significantly to heat transport across metal-semiconductor interfaces, and models \cite{huberman1994electronic,sergeev1999inelastic,mahan2009kapitza,ren2013heat,lombard2014influence,lu2016interfacial} have been developed to quantify its contribution. The different mechanisms of heat transport at a metal-semiconductor interface are summarized in Figure \ref{mechanism_schematic1}. 

Simplified empirical models are commonly used to interpret experimental thermal conductance data for metal-semiconductor interfaces. Elastic interfacial phonon scattering is commonly modeled using the acoustic \cite{little1959transport} and diffuse \cite{swartz1989thermal} mismatch models (AMM, DMM) which are heuristic approaches applicable for smooth and rough interfaces respectively. Also, simplifying assumptions such as the Debye approximation to phonon dispersion can compromise the quantitative accuracy of such models. Even atomistic simulation approaches for elastic interfacial thermal transport such as the atomistic Green's function (AGF) method often involve empirical force constant models that can produce significant discrepancies when compared to calculations that employ harmonic force constants obtained from \textit{ab initio} approaches \cite{tian2012enhancing}. The contribution of inelastic phonon scattering to thermal interface conductance has also been modeled in a simplified manner using heuristic extensions to the elastic diffuse mismatch model \cite{hopkins2009multiple}. The strength of electron-phonon coupling is typically modeled using idealized approximations such as bulk metal deformation potentials \cite{huberman1994electronic,sergeev1998electronic}, and such approximations are expected to be inaccurate for the direct coupling of metal electrons with joint or interface phonon modes. Little work exists on rigorous first-principles determination of the strength of coupling between electrons and joint interface phonon modes at a metal-semiconductor interface. 

Apart from the complexity of various thermal transport mechanisms described above, the uncertainty in interfacial atomic structure has historically made direct comparisons between simulations and experiments difficult. Much of the existing experimental data \cite{stoner1993kapitza,stevens2005measurement,guo2012heat} on thermal conductance of metal-semiconductor interfaces involves materials with mismatched lattice constants, for which the interface atomic structure is likely to be at least partially amorphous. Experimental studies that simultaneously characterize interfacial atomic structure along with interface conductance are scarce \cite{costescu2003thermal,wilson2015thermal}. However, predictive atomistic transport simulations that involve first-principles approaches are typically limited to crystalline epitaxial interfaces because of the associated computational tractability. This disparity between simulations and experimental studies often makes quantitative comparisons challenging. To overcome this difficulty, we choose to work with CoSi$_2$ (metal) - Si (semiconductor) interfaces in the present work. Both CoSi$_2$ and Si have FCC lattice structures with similar lattice constants of 5.36 $\text{\AA}$ and 5.43 $\text{\AA}$ respectively. Measurements of thermal interface conductance on CoSi$_2$ (111)/ Si (111) interface using the TDTR technique have been reported in our recent work \cite{ning_prep}, and the interface has been verified to be epitaxial and smooth using TEM imaging (see ref.~\onlinecite{ning_prep} for a TEM image of the interface). We use the same experimental data to compare with the present simulation predictions on a lattice-matched CoSi$_2$ (111)/Si (111) interface; the interfacial atomic configuration was also chosen to match with the atomic configuration of samples used in the experiment (see Section \ref{methods_agf} for details of the various interfacial atomic configurations). The close correspondence between the atomic structures used in the present work and the experimental data reported in ref.~\onlinecite{ning_prep} enables a direct comparison between simulations and experiments.  

Although the primary focus of the present work is the study of thermal transport across metal-semiconductor interfaces, the methods developed and reported here are also expected to be useful for a broad class of problems that use the non-equilibrium Green's function (NEGF) method for atomistic transport simulations. From a methodology standpoint, we report a framework that combines first-principles calculations of interatomic force constants with the atomistic Green's function method and evaluate the validity of the `mixing rule' that is commonly used to approximate interfacial bonding at a heterojunction. The conventional AGF method that is suitable for elastic phonon transport is extended to include anharmonic phonon scattering using a B\"{u}ttiker probe approach \cite{Miao_2016_buttiker}. Since the B\"{u}ttiker probe approach is not directly compatible with the conventional recursive Green's function (RGF) method (see Section \ref{rgf_bp_algo}), we develop a modification that enables the use of the RGF method in simulations that involve B\"{u}ttiker probe scattering. The new RGF algorithm enables computationally efficient simulations of phonon-phonon scattering using the B\"{u}ttiker probe approach and is expected to be applicable for a wide range of problems that require efficient representation of dephasing processes under the NEGF framework. \textit{Ab initio} calculations of electron-phonon coupling are also integrated into the AGF transport simulations.

Apart from the development of new methods, the present work also provides useful insights into the physics of thermal transport across metal-semiconductor junctions. Rigorous first-principles calculations indicate that elastic phonon transport under-predicts the experimental data over a wide temperature range. Analysis of the cross-interface heat flux accumulation function provides useful insights on the microscopic mechanisms responsible for increased interface conductance due to anharmonic phonon scattering in the bulk materials forming the interface. First-principles calculations of electron-phonon coupling on an interface supercell along with a detailed analysis of the contribution from different kinds of phonon modes to the Eliashberg function reveal that delocalized phonon modes mediate cross-interface energy transfer between metal electrons and the semiconductor lattice. We also obtain an effective length scale of electron-phonon interaction in the semiconductor by comparing simulation predictions with experimental data and evaluate the accuracy of prior approximations to the length scale of joint or interface phonon modes. 

\section{Methods}
\subsection{First-principles Calculations}
\label{methods_first_principles}
All first-principles calculations in this paper were performed under the framework of density functional theory (DFT) using a planewave basis set as implemented in the Quantum Espresso suite of codes \cite{Giannozzi_JPhysics_2009}. Rappe-Rabe-Kaxiras-Joannopoulos (RRKJ) ultrasoft pseudopotentials were used for both Co and Si atoms, and the exchange correlation energy was approximated under the generalized gradient approximation (GGA) using the Perdew-Burke-Ernzerhof (PBE) functional form. Three sets of first-principles calculations are performed for the results reported in this paper; these involve calculations on bulk Si (6 atom non-primitive unit cell along [111] direction), bulk strained CoSi$_2$ (9 atom unit cell along [111] direction) where a tensile strain is applied along the (111) plane, and a Si (111)-CoSi$_2$ (111) interface supercell. The relaxed lattice constants of bulk Si and bulk CoSi$_2$ are 5.44 $\text{\AA}$ and 5.36 $\text{\AA}$ respectively. For all simulations considered in this paper, a tensile strain of 1.5\% is applied on CoSi$_2$ along the (111) plane to match the lattice constants of Si and CoSi$_2$. Table \ref{calculation_params} shows the cutoff energies and k-point grids used for DFT calculations on all three systems. Structural relaxation is carried out to reduce the Hellmann-Feynman forces on every atom below 10$^{-3}$ eV/$\text{\AA}$. A full stress relaxation is carried out for bulk Si while the stresses on bulk strained CoSi$_2$ and the interface supercell are relaxed only along the transport direction. CoSi$_2$ is stretched along the in-plane direction to match its lattice with Si, and hence the in-plane stresses are not relaxed. 

Phonons are analyzed using density functional perturbation theory (DFPT) where the dynamical matrices are obtained on a $4 \times 4 \times 3$ q-point grid for bulk Si (6 atom unit cell along the [111] direction), bulk strained CoSi$_2$ (9 atom unit cell along the [111] direction) and a $4 \times 4 \times 1$ q-point grid for the interface supercell. The real-space inter-atomic force constants (IFCs) needed for Green's function transport calculations are obtained from a Fourier transform of the dynamical matrices. 
\begin{table}[ht]
\centering
\caption{Parameters used for DFT, DFPT calculations on bulk Si, bulk strained CoSi$_2$, and the Si-CoSi$_2$ interface supercell.}
\label{calculation_params}
\begin{tabular}{|l|l|l|l|}
\hline
Parameter                  & Bulk Si                 & Bulk strained CoSi$_2$   & Interface supercell      \\ \hline
Kinetic energy cutoff (eV) & 680                     & 820                      & 820                      \\ \hline
Charge density cutoff (eV) & 6800                    & 8200                     & 8200                     \\ \hline
Electron k-point grid      & $12 \times 12 \times 9$ & $16 \times 16 \times 12$ & $16 \times 16 \times 1$ \\ \hline
Phonon q-point grid        & $4 \times 4 \times 3$   & $4 \times 4 \times 3$    & $4 \times 4 \times 1$    \\ \hline
\end{tabular}
\end{table}
\subsection{Coherent Phonon Transport using the Atomistic Green's Function Method}
\label{methods_agf}
The terms `coherent' and `ballistic' phonon transport are used interchangeably in the present manuscript and refer to simulations performed under the conventional AGF framework. These simulations do not model phonon dephasing (cf., coherent) and inelastic phonon scattering (cf., ballistic). However, elastic interfacial scattering, i.e., reflection and transmission of phonon waves at the interface is included in this framework. The next section describes modifications to the conventional AGF approach to model phonon dephasing and inelastic phonon scattering. The details of the AGF method are available in prior reports \cite{zhang2007atomistic,sadasivam2014atomistic}, and a brief description is provided in this section. The basic conceptual framework for the AGF method involves a `device' region that is connected to semi-infinite `leads'. The device Green's function $G$ is given by:
\begin{equation}
G(\omega) = (\omega^2I-H_d-\Sigma_1-\Sigma_2)^{-1}
\end{equation}
where $H_d$ is the force constant matrix corresponding to the device region, and $\Sigma_1$, $\Sigma_2$ are the contact self-energies. The contact self-energies are obtained from the surface Green's functions $g_1$, $g_2$ as follows:
\begin{equation}
\Sigma_1 = \tau_1g_1\tau_1^\dagger \qquad \Sigma_2 = \tau_2g_2\tau_2^\dagger
\end{equation}
where $\tau_1$, $\tau_2$ represent the force constant matrices for interaction between atoms in the device region and the semi-infinite contacts. $g_1$ and $g_2$ are the surface Green's functions of the contacts that are obtained using the Sancho-Rubio method \cite{sancho1985highly,guinea1983effective}. The phonon transmission function across the device is obtained from the Caroli formula:
\begin{equation}
\mathcal{T}(\omega) = \Tr(\Gamma_1G\Gamma_2G^\dagger);
\end{equation}
where $\Gamma_{1,2} = i[\Sigma_{1,2}-\Sigma_{1,2}^\dagger]$ denotes the imaginary part of the contact self-energies and physically represents the phonon `escape rate' \cite{sadasivam2014atomistic} from the device into the respective contacts. In all the above expressions, the dependence of the Green's function, contact self-energy, and the transmission function on the transverse phonon wavevector $q_{||}$ (for structures with periodicity in the transverse or in-plane direction) and frequency $\omega$ is implicitly assumed. After obtaining the transmission function, the thermal interface conductance $G_Q$ can be obtained using the Landauer formula:
\begin{equation}
G_{Q} = \sum\limits_{q_{||}}{\frac{1}{2\pi}\int\limits_{0}^{\infty}{\hbar\omega}\mathcal{T}(\omega,q_{||})\frac{\partial f_{BE}^o}{\partial T}d\omega}
\label{cond}
\end{equation}
\subsection{Inelastic Scattering Using B{\"u}ttiker Probe Approach}
\label{methods_agf_bp}
The AGF formulation presented in the previous section is applicable only for elastic phonon transport, i.e., anharmonic scattering mechanisms such as Umklapp scattering are not considered. The formulation for extension of AGF to include anharmonic phonon scattering has been developed in ref.~\onlinecite{mingo2006anharmonic}; however, the approach is computationally expensive, and we are not aware of its application to study phonon transport through realistic three-dimensional crystals. The authors recently proposed a phenomenological B{\"u}ttiker probe approach to model anharmonic phonon scattering within the AGF method \cite{Miao_2016_buttiker}. The approach is an extension to phonons of the widely used B{\"u}ttiker probe method to model inelastic electron scattering processes in the NEGF framework \cite{venugopal2003simple,maassen2009effects,afzalian2011computationally}. Although the method is heuristic, it provides a computationally efficient alternative to the self-consistent Born approximation (SCBA) \cite{luisier2012atomistic} that is not phenomenological but is computationally intensive in both memory and time. The essence of the B{\"u}ttiker probe approach involves attaching fictitious contact probes to every atom in the device, and the temperatures of these fictitious contacts are then iteratively solved to ensure energy conservation in the device region. The B{\"u}ttiker probes contribute an additional self-energy to the device Green's function (in addition to the self-energies due to the real contacts):
\begin{equation}
G = (\omega^2I-H_d-\Sigma_1-\Sigma_2-\Sigma_{BP})^{-1}
\end{equation}

In the present formulation, the B{\"u}ttiker probe self-energy is assumed to be a diagonal matrix whose diagonal elements are of the form:
\begin{equation}
\Sigma_{BP(j,p)}(\omega) = -i\frac{2\omega}{\tau(\omega)}
\label{bp_self_energy}
\end{equation}
where $\Sigma_{BP(j,p)}$ denotes the B{\"u}ttiker probe self-energy at atom $j$ and vibrational direction $p$ ($x$, $y$, $z$). Similar to the matrices $\Gamma_1$, $\Gamma_2$, we also define $\Gamma_{BP} = i(\Sigma_{BP}-\Sigma_{BP}^\dagger)$ that represents the imaginary part of the B\"{u}ttiker probe self-energy. $\tau(\omega)$ denotes the frequency dependent scattering time due to Umklapp scattering and is assumed to be of the form $\tau^{-1}(\omega) = A\omega^2$ for both Si and CoSi$_2$. Quadratic frequency dependence of the Umklapp scattering rate has been used in prior studies involving the BTE \cite{singh2011effect} and Landauer approach \cite{mingo2003calculation,jeong2012thermal}. The parameter $A$ is chosen by fitting (see Supplemental Material) to the experimental thermal conductivity of bulk Si (at different temperatures) and bulk CoSi$_2$ (at room temperature). Due to lack of experimental data on the temperature dependence of the lattice thermal conductivity, the scattering parameter $A$ is assumed to be independent of temperature for CoSi$_2$. This assumption is acceptable because the thermal conductivity of CoSi$_2$ is dominated by electrons, and the interface conductance is found to exhibit a weak dependence on the lattice thermal conductivity of CoSi$_2$ (see Supplemental Material).
\subsubsection{Recursive Green's Function Method for Efficient Solution of B{\"u}ttiker Probe Temperatures}
\label{rgf_bp_algo}
B\"{u}ttiker probes offer a heuristic but efficient method to implement scattering in NEGF simulations. However, the popular recursive Green's function (RGF) method \cite{anantram2008modeling} that avoids full inversion in the calculation of the retarded Green's function $G$ and the lesser Green's function $G^{n} = G(\Gamma_1+\Gamma_2+\Gamma_{BP})G^\dagger$ is not compatible with B\"{u}ttiker probes. This incompatibility can be understood from the following equation used to enforce heat current conservation in each B\"{u}ttiker probe $i$. 
\begin{equation}
Q_i = \sum\limits_{j}\sum\limits_{q_{||}}{ \int\limits_{0}^{\infty}{\frac{\hbar\omega}{2\pi}\Tr({\Gamma_iG\Gamma_jG^{\dagger}})\left[f_{BE}^o(\omega,T_i)-f_{BE}^o(\omega,T_j)\right]d\omega}}=0
\label{energy_cons_bp}
\end{equation}
where the summation in the variable $j$ runs over all other B\"{u}ttiker probes ($i\neq j$) and the contacts. The computation of the transmission function matrix $\Tr({\Gamma_iG\Gamma_jG^{\dagger}})$ between every pair of B\"{u}ttiker probes requires calculation of the full Green's function matrix $G$. The equation for charge current conservation in electronic transport is similar to the foregoing equation for heat current conservation \cite{venugopal2003simple,afzalian2011computationally}. Hence, prior implementations \cite{venugopal2003simple,afzalian2011computationally,Miao_2016_buttiker} of the B{\"u}ttiker probe formalism have employed direct matrix inversion instead of the RGF method to calculate the full device Green's function matrix. 

Eq.~(\ref{energy_cons_bp}) enforces the condition that the total integrated energy flux in each B\"{u}ttiker probe is zero, i.e., inelastic scattering between different energy levels is allowed. Alternative implementations of the B\"{u}ttiker probe approach invoke energy flux conservation at each phonon frequency instead of the total integrated energy flux over all phonon frequencies \cite{venugopal2003simple,maassen2009effects}. Such an approach is suitable only for elastic dephasing and is hence not adopted in this work. 

Eq.~(\ref{energy_cons_bp}) can be cast in a slightly different form as:
\begin{equation}
Q_i = \sum\limits_{q_{||}}\int\limits_{0}^{\infty}{\frac{\hbar\omega}{2\pi}\Tr(\Sigma_i^{in}A-\Gamma_iG^n)d\omega}=0
\label{curr_eqn_new}
\end{equation}
where $\Sigma_i^{in} = f_{BE}^o(\omega,T_i)\Gamma_i$ and $A = i(G-G^\dagger)$ denotes the spectral function. In the above equation, the matrices $\Sigma_{i}^{in}$ and $\Gamma_i$ are block-diagonal. Hence, only the block-diagonals of $A$ and $G^n$ need to be computed, and this can be done using the RGF algorithm. However, the computation of $\Sigma_{i}^{in}$ and $G^n$ require knowledge of the B\"{u}ttiker probe temperatures. Hence the use of RGF with B\"{u}ttiker probes requires that $\Sigma_{i}^{in}$ and $G^n$ are recalculated during every Newton iteration of the solution for B\"{u}ttiker probe temperatures. Although this step is not needed in a conventional B\"{u}ttiker probe implementation with full matrix inversion, the computational advantage of RGF over full inversion far outweighs the computational expense of repeated RGF calculations in every Newton iteration. Also, the memory required to store and invert the full Green's function matrix can become prohibitively large with increasing device length. 

Anantram et al. \cite{anantram2008modeling} provide a detailed discussion of the RGF methodology for computation of the block diagonals of $G$ and $G^n$. Here, we provide only the modifications needed to combine RGF with the B\"{u}ttiker probe formalism. The RGF algorithm for computation of the block-diagonal elements of the retarded Green's function $G$ remains unchanged. However, the RGF algorithm for computation of $G^n$ requires modification to also calculate the derivative of the diagonal elements of $G^n$ with respect to the B\"{u}ttiker probe temperatures. We also assume that all B\"{u}ttiker probes within a RGF `block' have the same temperature, i.e., the number of B\"{u}ttiker probe temperatures that need to be solved is equal to the number of blocks in the device region. 

The equation for left-connected $g^{nL}$ is given by \cite{anantram2008modeling}:
\begin{equation}
\label{gn_left}
g^{nLi+1}_{i+1,i+1} = g^{Li+1}_{i+1,i+1}\left(\Sigma^{in}_{i+1,i+1}+\sigma^{in}_{i+1,i+1}\right)g^{Li+1\dagger}_{i+1,i+1}
\end{equation}
where $\sigma^{in}_{i+1,i+1} = B_{i+1,i}g^{nLi}_{i,i}B_{i,i+1}^{\dagger}$, $B = (\omega^2I-H_d-\Sigma_1-\Sigma_2-\Sigma_{BP})$. Our terminology follows that of ref.~\onlinecite{anantram2008modeling} where $g^L$ denotes the left-connected retarded Green's function, and $\Sigma^{in}$ denotes the lesser self-energy. The derivative of $g^{nLi+1}_{i+1,i+1}$ with respect to the B\"{u}ttiker probe temperature $T_j$ is given by:
\begin{equation}
\label{gn_left_derv}
\frac{\partial g^{nLi+1}_{i+1,i+1}}{\partial T_j} = 
\begin{cases}
g^{Li+1}_{i+1,i+1}B_{i+1,i}\frac{\partial g^{nLi}_{i,i}}{\partial T_j}B_{i,i+1}^{\dagger}g^{Li+1\dagger}_{i+1,i+1}\qquad\text{, if }j<i+1 \\
g^{Li+1}_{i+1,i+1}\Gamma_{BP,i+1}g^{Li+1\dagger}_{i+1,i+1}\frac{\partial f_{BE}^o(\omega,T)}{\partial T}\bigg|_{T_j}\qquad\text{, if }j=i+1 \\
0 \qquad \text{, if } j>i+1 
\end{cases}
\end{equation}
The equation for $G^n_{i,i}$ is given by:
\begin{equation}
\label{Gnqq}
G^n_{i,i} = g^{nLi}_{i,i}+g^{Li}_{i,i}\left(B_{i,i+1}G^n_{i+1,i+1}B^{\dagger}_{i+1,i}\right)g^{\dagger Li}_{i,i}-\left(g^{nLi}_{i,i}B^{\dagger}_{i,i+1}G^{\dagger}_{i+1,i}+G_{i,i+1}B_{i+1,i}g^{nLi}_{i,i}\right)
\end{equation}
The derivative of $G^n_{i,i}$ with respect to B\"{u}ttiker probe temperatures can be computed using the derivatives of the left connected Green's function $g^{nL}$ computed in Eq.~(\ref{gn_left_derv}).
\begin{eqnarray}
\label{Gnqq_derv}
\frac{\partial G^n_{i,i}}{\partial T_j} &=& \frac{\partial g^{nLi}_{i,i}}{\partial T_j}+g^{Li}_{i,i}\left(B_{i,i+1}\frac{\partial G^n_{i+1,i+1}}{\partial T_j}B^{\dagger}_{i+1,i}\right)g^{\dagger Li}_{i,i}- \nonumber \\
& & \left(\frac{\partial g^{nLi}_{i,i}}{\partial T_j}B^{\dagger}_{i,i+1}G^{\dagger}_{i+1,i}+G_{i,i+1}B_{i+1,i}\frac{\partial g^{nLi}_{i,i}}{\partial T_j}\right)
\end{eqnarray}
Overall, the RGF algorithm for $G^n$ needs to be modified to compute the derivatives of $G^n$ with respect to the B\"{u}ttiker probe temperatures. The new RGF algorithm's commutation involves the following steps:
\begin{enumerate}
\item $g^{nL1}_{11} = g^{L1}_{11}\Sigma^{in}_{11}g^{L1\dagger}_{11}$, $\frac{\partial g^{nL1}_{11}}{\partial T_1} = g^{L1}_{11}\Gamma_{BP,1}g^{L1\dagger}_{11}\frac{\partial f_{BE}^o(\omega,T)}{\partial T}\bigg|_{T_1}$, $\frac{\partial g^{nL1}_{11}}{\partial T_j} = 0 \quad (j>1)$
\item For $i=1,2,\dots ,N-1$ and $j=1,2,\dots,N$, compute Eqs.~(\ref{gn_left}) \& (\ref{gn_left_derv})
\item $G^n_{NN} = g^{nLN}_{NN}$, $\frac{\partial G^n_{NN}}{\partial T_j} = \frac{\partial g^{nLN}_{NN}}{\partial T_j}$ for $j=1,2,\dots,N$
\item For $q=N-1,N-2,\dots,1$ and $j=1,2,\dots,N$, compute Eqs.~(\ref{Gnqq}) \& (\ref{Gnqq_derv}).
\end{enumerate}
The algorithm proceeds as follows. 
\begin{enumerate}
\item Start with an initial guess for the B\"{u}ttiker probe temperatures.
\item For each phonon frequency, compute $G^R_{ii}$, $G^n_{ii}$, and $\frac{\partial G^n_{ii}}{\partial T_j}$ using the RGF algorithm described above.
\item Compute energy current densities in each B\"{u}ttiker probe using Eq.~(\ref{curr_eqn_new}).
\item Compute the Jacobian matrix whose $(i,j)^{th}$ element is given by the following equation:
\begin{equation}
J_{i,j} = \sum\limits_{q_{||}}\int\limits_{0}^{\infty}{\frac{\hbar\omega}{2\pi}\Tr\left(\Gamma_iA_{ii}\frac{\partial f_{BE}^o}{\partial T}\bigg|_{T_j}\delta_{ij}-\Gamma_i\frac{\partial G^n_{ii}}{\partial T_j}\right)d\omega}
\end{equation}
where $\delta_{ij}$ is the Kronecker Delta function.
\item Update the temperature of B\"{u}ttiker probes using the Newton equation:
\begin{equation}
T_{new} = T_{old}-J^{-1}f
\end{equation}
\item If $\lVert T_{new}-T_{old} \rVert > \epsilon$, go back to Step 1 with the new guess for B\"{u}ttiker probe temperatures as $T_{new}$. 
\end{enumerate}
An alternative to the Newton-Raphson method is the secant method in which the exact Jacobian needs to be computed only in the first iteration. For further iterations, the Jacobian could be updated using the Broyden's update formula \cite{broyden1965class}, and this method was also found to give satisfactory convergence. With the secant method, the derivative of the lesser Green's function with respect to B\"{u}ttiker probe temperatures need only be computed in the first iteration. For the remaining iterations, the traditional RGF function is sufficient. For large device lengths, the computation and storage of the full Jacobian matrix can become prohibitively expensive; an alternative approach involves approximation of the Jacobian by a sparse block diagonal matrix. Such an approximate Jacobian was also found to lead to convergence, however with an increased number of iterations compared to the exact Jacobian. The approximate Jacobian provides a memory-time tradeoff as storage of the sparse block diagonal matrix involves lesser memory but the computational time increases relative to the Newton Raphson scheme with exact Jacobian. All the results presented in this paper involve the secant method in which the exact Jacobian is computed in the first iteration and the Broyden's update formula is used for further iterations.

\begin{figure}[ht]
\begin{center}
\includegraphics[height=60mm]{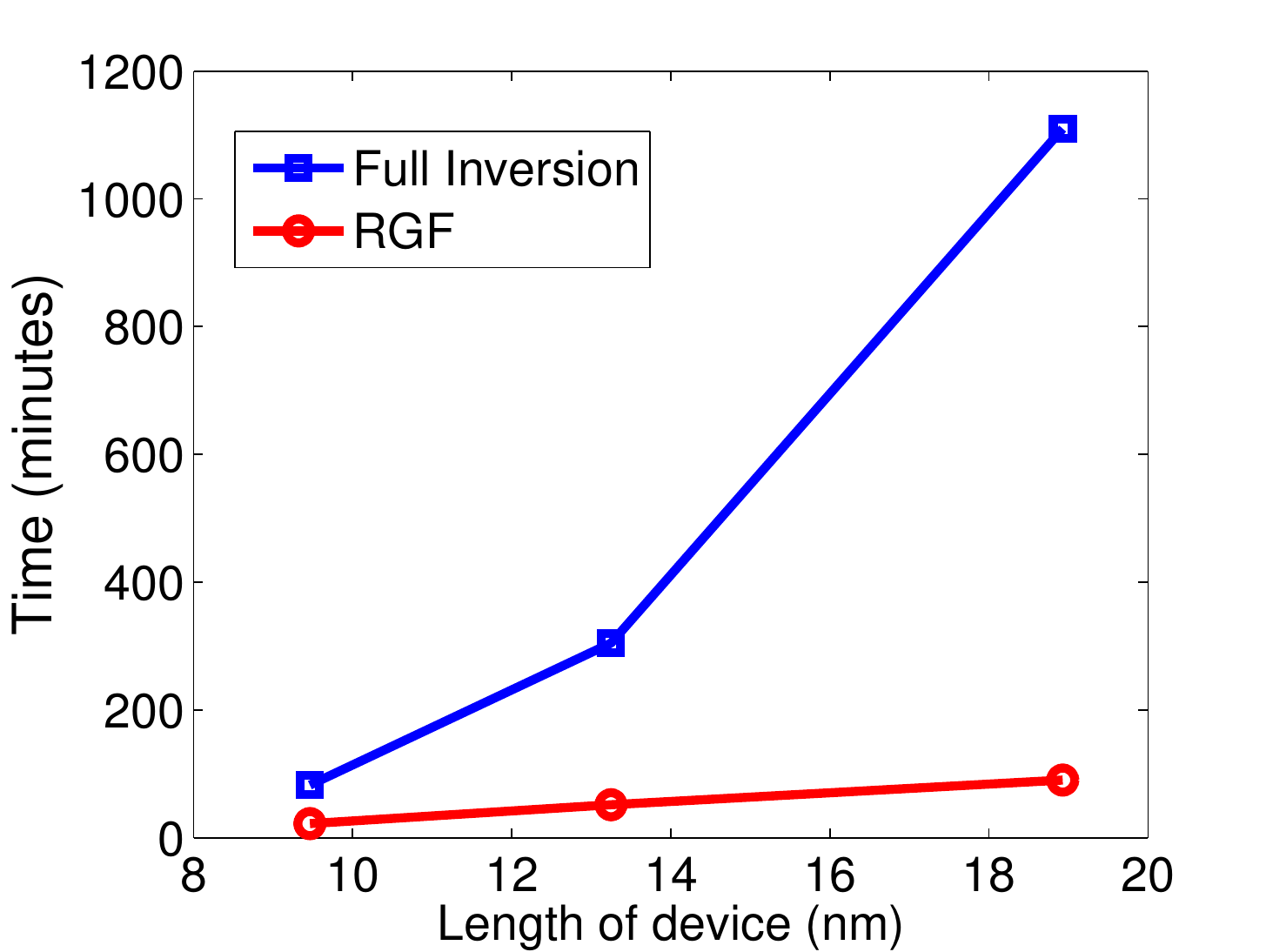}
\caption{Comparison of computational times for different device lengths obtained using direct inversion and the RGF algorithm.}
\label{speed_comp}
\end{center}
\end{figure}

Figure \ref{speed_comp} shows a comparison of the computational times for AGF simulations of bulk silicon with B\"{u}ttiker probe scattering using direct inversion and the RGF algorithm described above. The computational times were obtained using MATLAB, and both direct inversion and the RGF methods were parallelized over transverse wavevectors. The computational time for full matrix inversion increases rapidly with device length (matrix inversion scales as $\mathcal{O}(n_{||}^3n_z^3)$ where $n_{||}$, $n_z$ denote the number of atoms per slab and the number of slabs in the transport direction respectively) while that for the RGF algorithm proposed above shows a more gradual scaling with device length (RGF scales as $\mathcal{O}(n_{||}^3n_z)$). 

Apart from the computational time improvement, another important advantage of the RGF method over full inversion is the reduced memory needed to store and invert full Green's function matrices. The device sizes considered in the present work (see Section \ref{anharmonic_scat_sec}) are computationally intractable with direct matrix inversion. Hence, the extension of the RGF algorithm to B\"{u}ttiker probes is expected to be critical for application of the B\"{u}ttiker probe method to realistic device sizes. Also, the results in Figure \ref{speed_comp} confirm that the computational expense of repeated AGF calculations of $G^n (\omega;q_{||})$ for each RGF iteration is far less than the computational expense for a single calculation of the full retarded Green's function of the device $G(\omega;q_{||})$ through direct inversion. 
\subsection{Fourier Diffusion of Electrons Coupled with Phonons}
\label{methods_ttm_agf_bp}
\label{ttm_agf_algo}
Electrons are the primary heat carriers in metals, and they transfer energy to phonons near the interface between metal and semiconductor. Intrinsic Si is the semiconductor of interest in this paper, and hence the contribution of electrons in Si to thermal transport is neglected. We neglect any cross-interface electron tranport through the CoSi$_2$-Si interface and consider diffusive transport of electrons in CoSi$_2$. Electrons are included in the AGF simulation within the framework of a two-temperature model that is commonly used to interpret ultra-fast laser experimental data and is also used to model energy transfer between electron and phonon subsystems within the Eliashberg function framework. The primary assumption involved in the definition of a local electron and phonon temperature is the existence of electron-electron and phonon-phonon collisions that enable local equilibrium separately within electron and phonon subsystems. The electron and phonon subsystems exchange energy through electron-phonon coupling that is expressed in terms of the Eliashberg function.The steady state Fourier diffusion equation for electrons in the metal $(x>0)$ is given by:
\begin{equation}
k_e\frac{d^2T_e(x)}{dx^2}+Q_{ep}(x)=0
\label{fourier_diff}
\end{equation}
where $Q_{ep}(x)$ denotes the volumetric heat source term due to coupling between electrons and phonons and is expressed in terms of the Eliashberg function $\alpha^2F(\omega)$ as derived by Allen in ref.~\onlinecite{allen1987theory}:
\begin{equation}
Q_{ep}(x) = 2\pi D(E_f)\int\limits_{0}^{\infty}{(\hbar\omega)^2\alpha^2F(\omega)\left[f_{BE}^o(T_p(x))-f_{BE}^o(T_e(x))\right]d\omega}
\label{Qep}
\end{equation}
In the foregoing equation $T_e(x)$, $T_p(x)$ denote the local electron and lattice temperatures respectively at location $x$, and $D(E_f)$ is the electronic density of states at the Fermi energy. The above term can be included as a source term in the B\"{u}ttiker probe at location $x$, i.e., the energy current conservation equation for the $i^{\text{th}}$ B\"{u}ttiker probe in the metal is given by:
\begin{equation}
\sum\limits_{q_{||}}\int\limits_{0}^{\infty}{\frac{\hbar\omega}{2\pi}\Tr(\Sigma_i^{in}A-\Gamma_iG^n)d\omega}+Q_{ep,i}\Delta x_i = 0
\label{energy_curr_eph}
\end{equation}
where $\Delta x_i$ is the length that the B\"{u}ttiker probe occupies along the transport direction. In a phonon-only simulation, the total energy current in each B\"{u}ttiker probe is set to zero to ensure energy flux conservation (see Eq.~(\ref{curr_eqn_new})), i.e., B\"{u}ttiker probes redistribute the energy, but with no net transfer of energy between electrons and phonons via the fictitious B\"{u}ttiker contacts. When electrons are included in the transport calculation, the total energy current in each B\"{u}ttiker probe is set to the electron-phonon energy exchange given by Eq.~(\ref{Qep}).

The local lattice temperature $T_p$ in Eq.~(\ref{Qep}) is obtained by equating the local phonon energy density to the product of a local Bose-Einstein distribution at temperature $T_p$ and the local phonon density of states:
\begin{equation}
\sum\limits_{q_{||}}\int\limits_{0}^{\infty}{\omega^2 G^n(\omega;q_{||})d\omega} = \sum\limits_{q_{||}}\int\limits_{0}^{\infty}{\omega^2 A(\omega;q_{||})f_{BE}^o(\omega,T_p)d\omega}
\label{local_dev_temp}
\end{equation} 
The above equation makes use of the following expressions for the local phonon number density $\rho(\omega)$ and the local phonon DOS $D(\omega)$ in terms of the lesser Green's function $G^n(\omega)$ and the spectral function $A(\omega)$ respectively:
\begin{equation}
\rho(\omega) = \sum\limits_{q_{||}}\frac{\omega G^n(\omega;q_{||})}{\pi} \qquad D(\omega) = \sum\limits_{q_{||}}\frac{\omega A(\omega;q_{||})}{\pi}
\end{equation}
Eqs.~(\ref{fourier_diff}), (\ref{energy_curr_eph}), (\ref{local_dev_temp}) constitute a set of coupled non-linear equations that are solved iteratively to obtain the electron temperature, the B\"{u}ttiker probe temperature, and the local device temperatures. Similar to the methodology for B\"{u}ttiker probe temperatures in Section \ref{rgf_bp_algo}, the Newton-Raphson method is used for the solution of the above equation and details of the algorithm are provided in the Supplemental Material.
\section{Coherent Phonon Transport}
\label{agf_results}
This section contains results for the phonon transmission function and thermal interface conductance of Si-CoSi$_2$ interface from ballistic phonon transport calculations (i.e., B\"{u}ttiker probe scattering turned off). Different interfacial atomic configurations are possible for the Si-CoSi$_2$ interface depending on the coordination number of the Co atom closest to the interface (possible values of 5, 7, 8) and the relative crystal orientation between the (111) surfaces of Si and CoSi$_2$. The `A' type orientation occurs when the Si-CoSi$_2$ stacking is continuous while the `B' orientation occurs when the CoSi$_2$ crystal is rotated by 180$^{\circ}$ about the [111] direction. Previous first-principles calculations in the literature \cite{stadler1999ab,wardle2005structural} indicate that the 8A and 8B configurations have the lowest interfacial energies and are hence the most probable interfacial atomic structures. Both configurations are considered for the present phonon transport calculations using AGF. 

While bulk IFCs are quite commonly obtained from first-principles calculations, little work exists on the use of rigorous DFPT calculations to determine the force constants between atoms belonging to different materials across a heterogeneous interface. Force constants (in AGF) and inter-atomic potentials (in molecular dynamics) between atoms belonging to different bulk materials are commonly represented using simplifying approximations without rigorous calculations of the actual strength of interfacial bonding \cite{huang2011atomistic,ong2010molecular}. In the present work, both bulk and cross-interface IFCs needed for AGF transport simulations are determined entirely from DFPT calculations. 

\begin{figure}
\centering
\subfloat[]{\includegraphics[height=25mm]{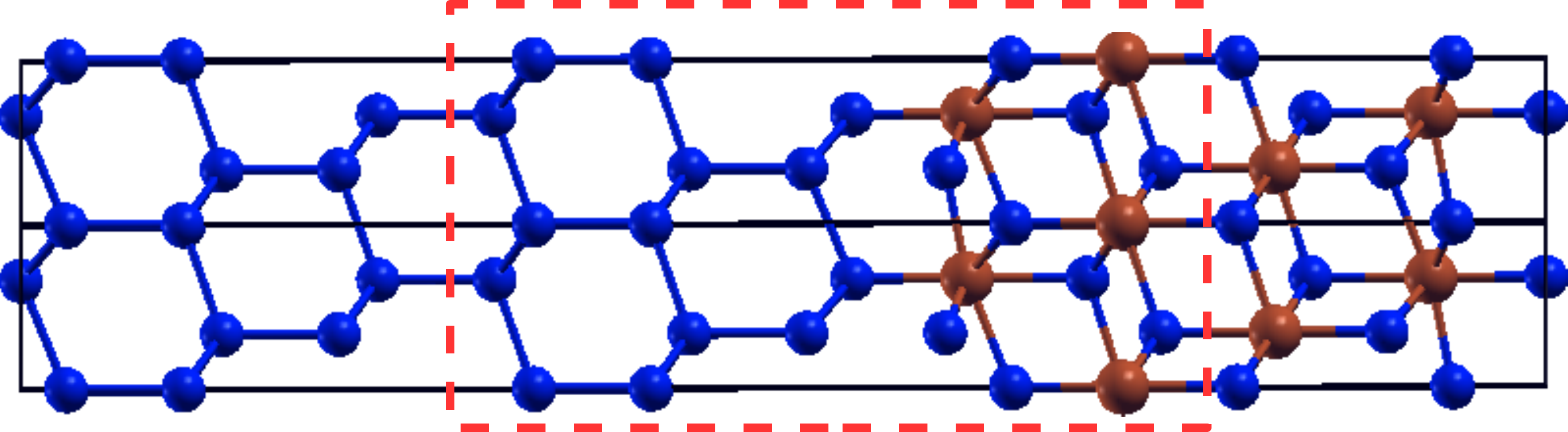}}\\
\subfloat[]{\includegraphics[height=25mm]{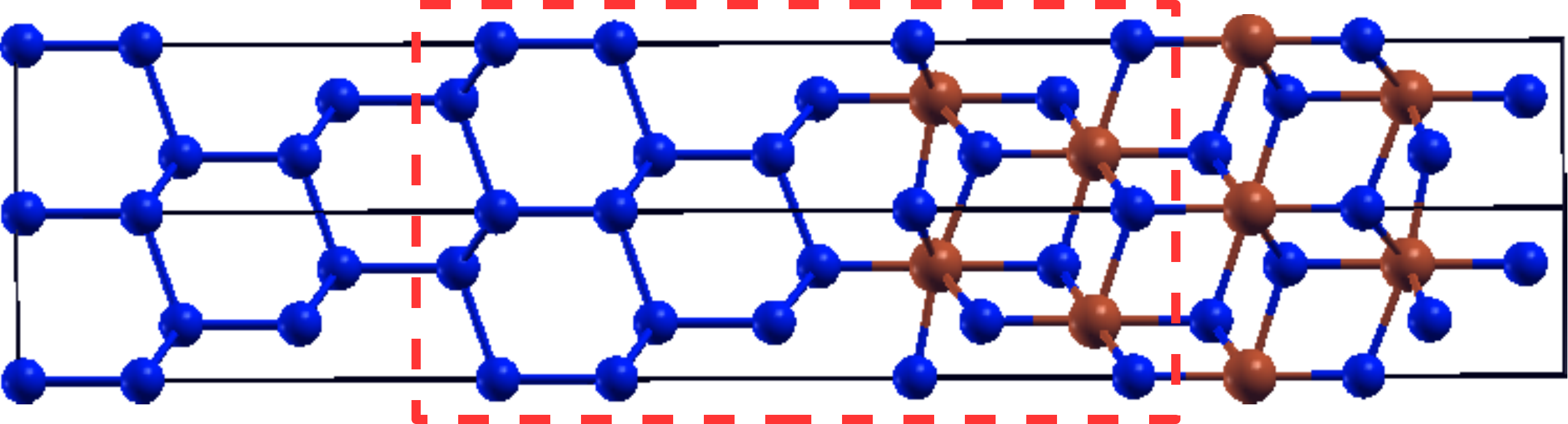}}
\caption{Atomic structures of Si-CoSi$_2$ interface supercells used in DFPT calculations (two unit cells shown along the in-plane direction for clarity). The red dotted boxes indicate the region around the interface for which IFCs are extracted from the interface supercell calculation. a) 8A configuration. b) 8B configuration.}\label{interface_structure_CoSi2}
\end{figure}

Figure \ref{interface_structure_CoSi2}a,b shows supercells of the 8A and 8B interfacial atomic configurations respectively. Each supercell contains two interfaces because of periodic boundary conditions. Although DFPT calculations are performed on a finite interface supercell, transport simulations are performed on a single Si-CoSi$_2$ interface formed by semi-infinite Si and CoSi$_2$ crystals. The red dotted boxes enclose atoms around the interface for which the force constants are obtained from the interface supercell DFPT calculation. In the atomic structure considered for transport calculations, the IFCs for atoms outside the red dotted box are assumed to equal the bulk IFCs of Si (left of the box) and CoSi$_2$ (right of the box). Results that illustrate the convergence of cross-interface force constants with respect to the size of interface supercell are provided in the Supplemental Material. 

Enforcement of the acoustic sum rules is an important consideration when IFCs obtained from DFPT calculations are used in thermal transport simulations. Acoustic sum rules constitute a set of translational invariance conditions on the IFCs to ensure that long wavelength acoustic modes of a crystal have zero vibrational frequency:
\begin{equation}
\sum\limits_{j}H_{i\alpha,j\beta} = 0
\label{asr}
\end{equation}
where $i$, $j$ denote atom indices while $\alpha$, $\beta$ denote the directions. The spatial range of inter-atomic interactions is artificially truncated in DFPT by the finite q-point grid used in the calculations. Although the neglected long-range interactions are typically small, this procedure results in small violations of the translational invariance conditions. Hence, the raw force constants obtained from DFPT do not satisfy the acoustic sum rules exactly and need to be enforced as a post-processing step on the IFCs \cite{mingo2008phonon}. Common DFT codes such as Quantum Espresso automatically enforce translational invariance for the IFCs of the crystal on which DFPT calculations are performed. However in the present calculations, the IFCs obtained for bulk Si and bulk CoSi$_2$ are combined with that obtained for the interface supercell. Hence the IFCs for Si and CoSi$_2$ unit cells nearest to the interface will require modifications to ensure that the acoustic sum rules are satisfied. In the present work, the diagonal blocks of the force constant matrix are modified to ensure that Eq.~(\ref{asr}) is satisfied. 

Cross-interface force constants between heterogeneous materials are commonly approximated using simplifying assumptions due to the computational complexity of performing direct DFPT calculations on an interface supercell with a large number of atoms. If the materials on both sides of the interface have the same lattice structure such as Si-Ge interfaces, a common approximation is to assume the same force constants for both materials with the assumption that interfacial scattering is primarily affected by the change in atomic mass across the interface \cite{tian2012enhancing}. Other approximations include the use of empirical corrections to obtain the cross-interface force constants from the bulk force constants \cite{huang2011atomistic}. Another common approximation involves the use of mixing rules to obtain the strength of cross-interface interactions from an average of the bulk material parameters \cite{luckyanova2012coherent}. 

\begin{figure}
\centering
\subfloat[]{\includegraphics[height=60mm]{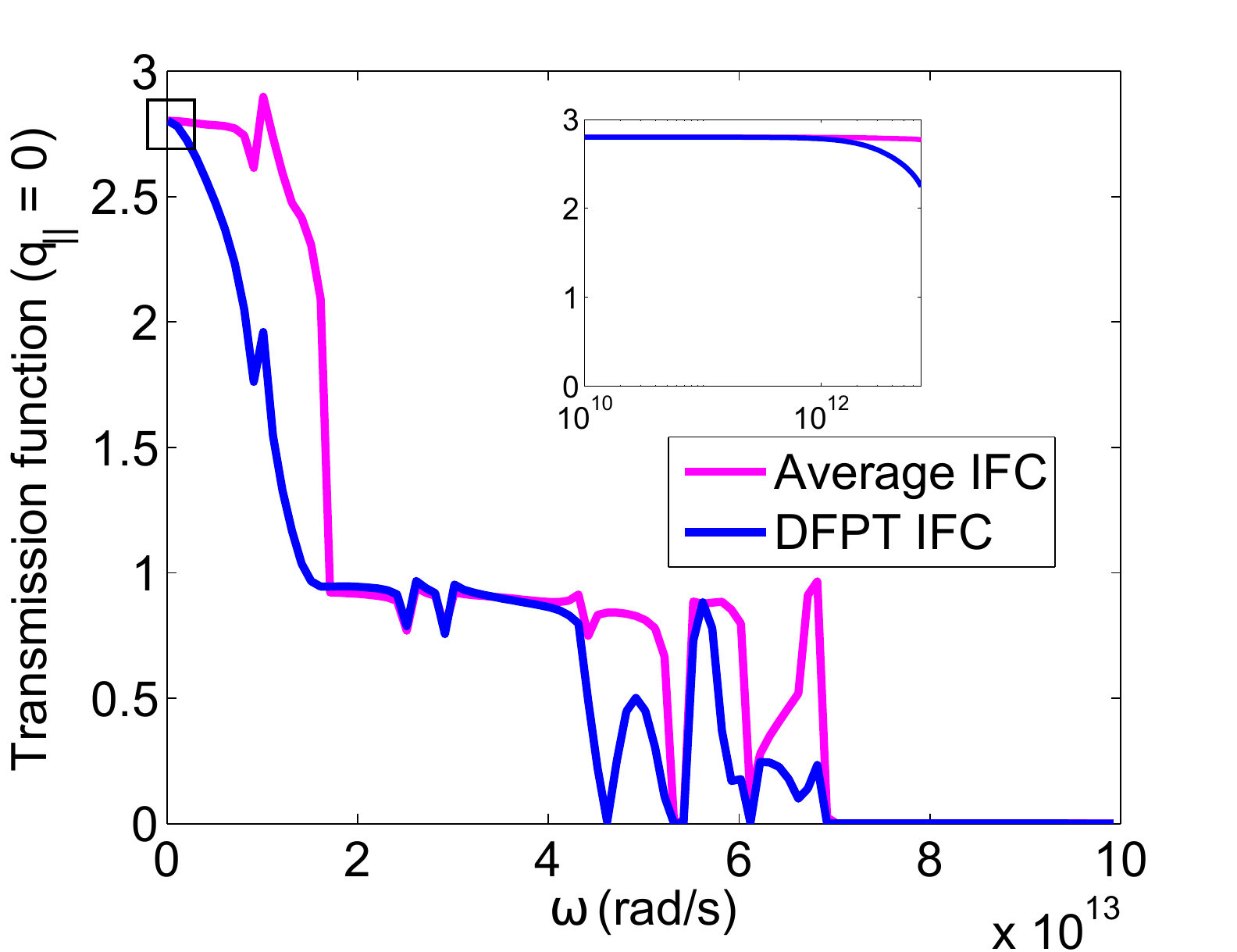}}\quad
\subfloat[]{\includegraphics[height=70mm]{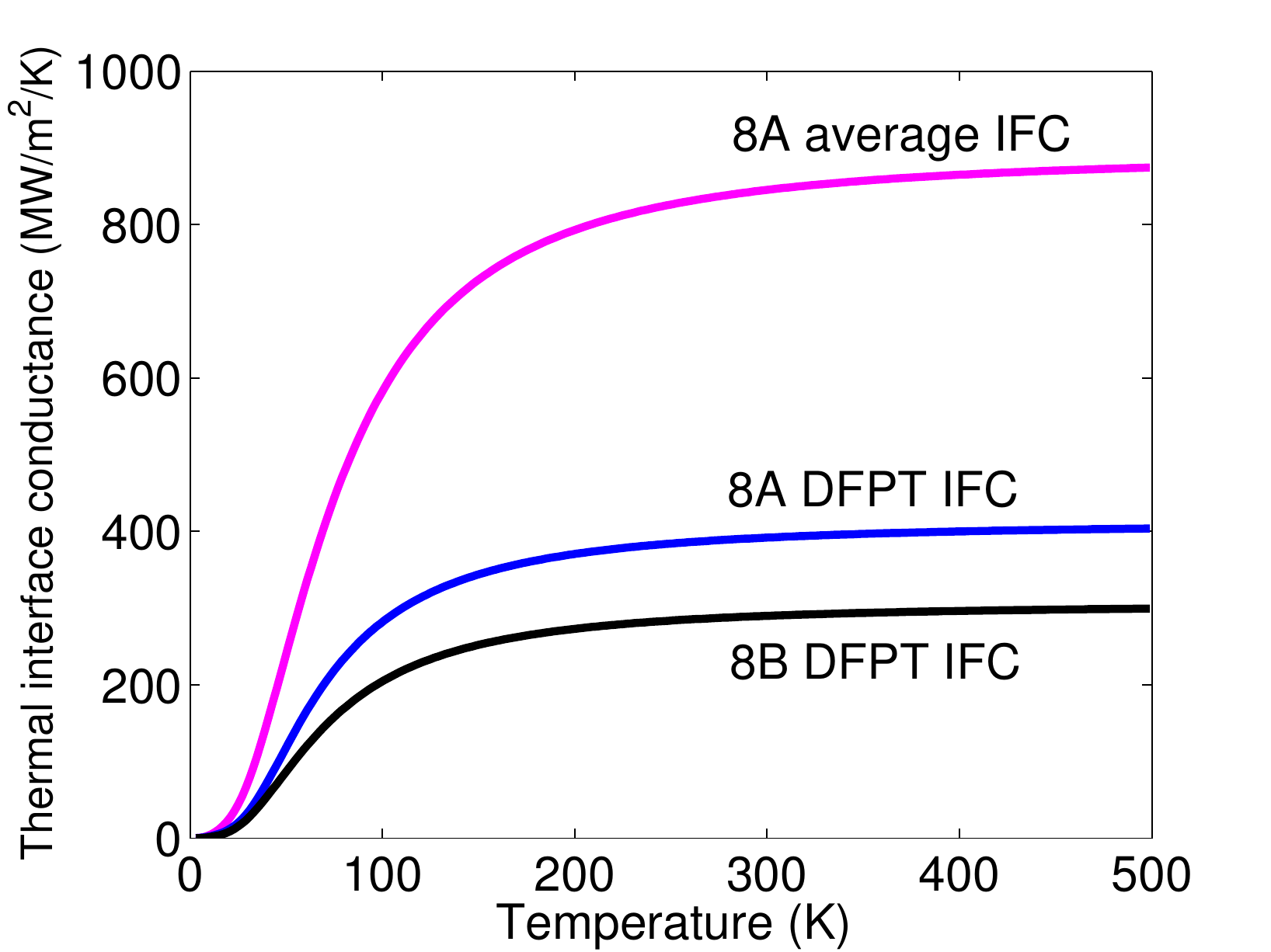}}
\caption{Results from ballistic phonon transport calculations for a Si-CoSi$_2$ interface. a) Phonon transmission function at normal incidence computed using average and DFPT force constants for the 8A interface. The inset shows the same graph for small phonon frequencies or long wavelengths. b) Thermal interface conductance for 8A and 8B Si-CoSi$_2$ interfaces.}\label{ballistic_results}
\end{figure}

In order to evaluate the validity of a simple averaging approximation for the cross-interface force constants, Figure \ref{ballistic_results}a compares the phonon transmission function at normal incidence ($q_{||}=0$) when the cross interface IFCs are assumed to be a simple arithmetic average of the bulk IFCs and when the cross-interface IFCs are obtained from DFPT on the interface supercell shown in Figure \ref{interface_structure_CoSi2}a. In the averaging approach, the cross-interface Co-Si and Si-Si IFCs are obtained by averaging the interactions in bulk Si and bulk CoSi$_2$. The averaging approximation is found to over-estimate the transmission function for most of the frequency range except at very low frequencies or long wavelengths (see inset in Figure \ref{ballistic_results}a) where the predictions from both the average and DFPT IFCs converge to the acoustic mismatch limit. Long wavelength phonons are insensitive to the local details of interfacial bonding, and hence the transmission function at low phonon frequencies is expected to be insensitive to the exact interfacial force constants. However, rigorous predictions of cross-interface force constants are necessary for accurate prediction of transmission at higher frequencies that are expected to dominate phonon transport at room temperature and beyond. The thermal interface conductance computed directly from Eq.~(\ref{cond}) includes contributions from the ballistic contact conductances at the contact-device interfaces in addition to the conductance of the Si-CoSi$_2$ interface in the middle of the device region. To obtain the conductance of the Si-CoSi$_2$ interface alone, we use the following expression to subtract the ballistic contact resistances from the total resistance obtained from Eq.~(\ref{cond}) \cite{tian2012enhancing}:
\begin{equation}
G_Q'(T) = \frac{G_Q(T)}{1-\frac{1}{2}\left(\frac{G_Q(T)}{G_{Q,\text{Si}}(T)}+\frac{G_Q(T)}{G_{Q,\text{CoSi}_2}(T)}\right)}
\end{equation}
where $G_Q(T)$ is the interface conductance computed from Eq.~(\ref{cond}), and $G_Q'(T)$ is the thermal interface conductance of a single Si-CoSi$_2$ interface and plotted in Figure \ref{ballistic_results}b. $G_{Q,\text{Si}}(T)$ and $G_{Q,\text{CoSi}_2}(T)$ are the ballistic conductances of homogeneous Si and CoSi$_2$ slabs respectively. 

The use of a simple arithmetic average for cross-interface IFCs over-estimates the thermal interface conductance (see Figure \ref{ballistic_results}b) by almost 70\% at room temperature, and the errors increase at higher temperatures. Hence, the prediction of phonon thermal interface conductance for temperatures beyond a few tens of K requires the rigorous prediction of interfacial bonding strength, and simple averaging approximations are not expected to be quantitatively accurate. Similar conclusions on the over-estimation of interface conductance due to simple approximations that neglect local changes in the force field near a heterogeneous interface were found in ref.~\onlinecite{gu2015phonon}. 

\section{Effect of Anharmonic Scattering on Thermal Interface Conductance}
\label{anharmonic_scat_sec}
In this section, the effect of anharmonic phonon scattering in Si and CoSi$_2$ on the thermal interface conductance is presented. This section contrasts with results in the previous section for which phonon transport in Si and CoSi$_2$ were assumed to be ballistic. The B\"{u}ttiker probe scattering rates for both Si and CoSi$_2$ were assumed to be of the form $\tau^{-1}(\omega) = A\omega^2$, and the parameter $A$ was fitted to obtain the bulk thermal conductivity of Si and CoSi$_2$ (see Supplemental Material). Since Si has a relatively high phonon thermal conductivity compared to CoSi$_2$, this circumstance corresponds to a low scattering rate on the Si side of the interface and a high scattering rate in CoSi$_2$. To understand the effect of bulk scattering rates on the interface conductance, we also performed simulations in which the scattering rate in CoSi$_2$ is reduced by a factor of 100 while the Si scattering rate is maintained the same (Case A) and the scattering rate in Si is increased by a factor of 100 while that in CoSi$_2$ is maintained the same (Case C). Case B corresponds to the nominal scattering rate in both Si and CoSi$_2$, i.e., the scattering rate that reproduces the bulk thermal conductivity of Si and CoSi$_2$. The present simulations assume that all the B\"{u}ttiker probes on the Si and CoSi$_2$ sides of the interface have scattering rates of bulk Si and bulk CoSi$_2$ respectively. However, the local anharmonicity near the Si-CoSi$_2$ interface is likely to differ from the bulk anharmonicities of Si and CoSi$_2$. Future work on determining the change in Umklapp scattering rates near a heterogeneous interface is needed to improve the present model. 

\begin{figure}
\centering
\subfloat[]{\includegraphics[height=40mm]{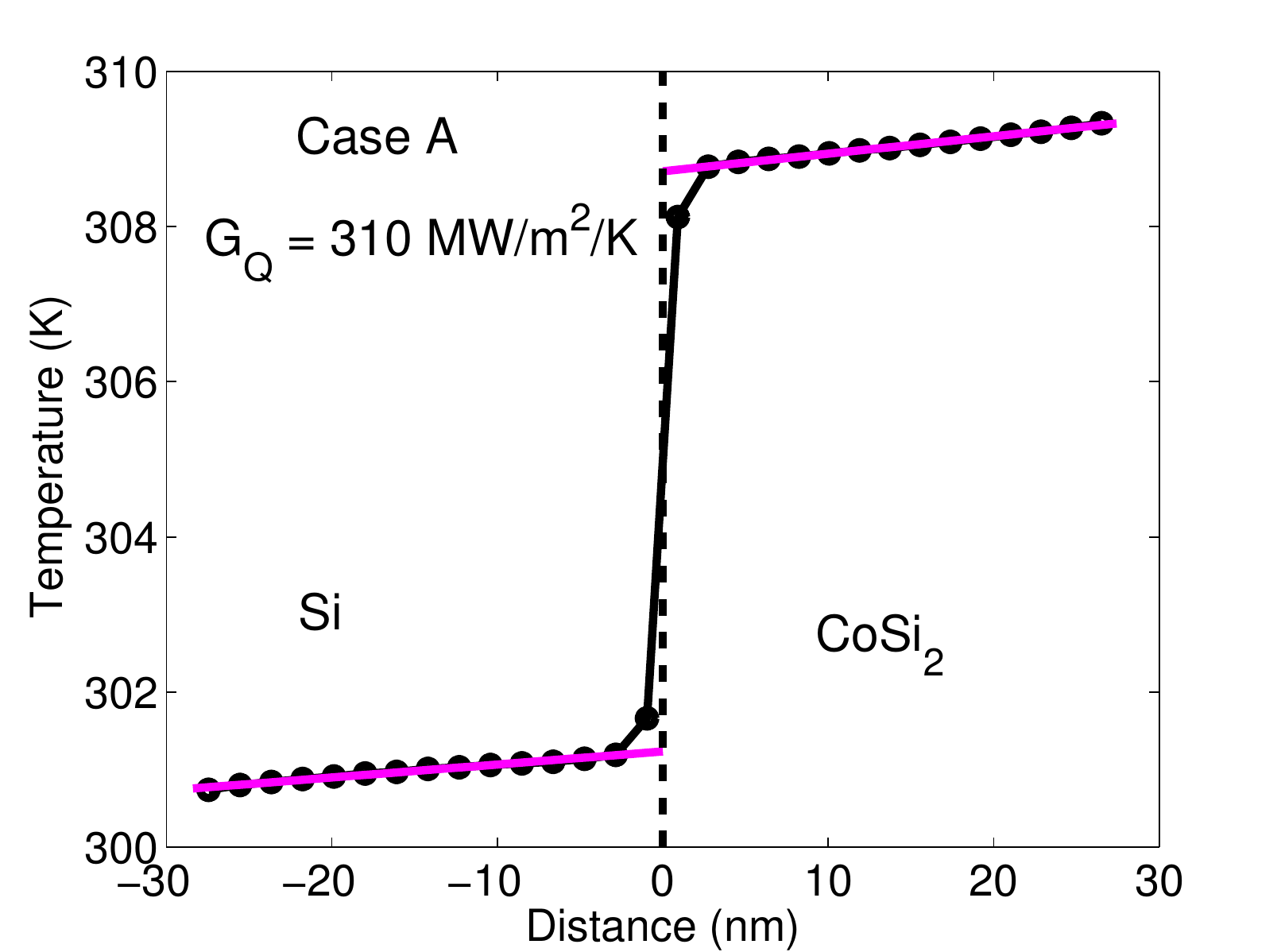}}
\subfloat[]{\includegraphics[height=40mm]{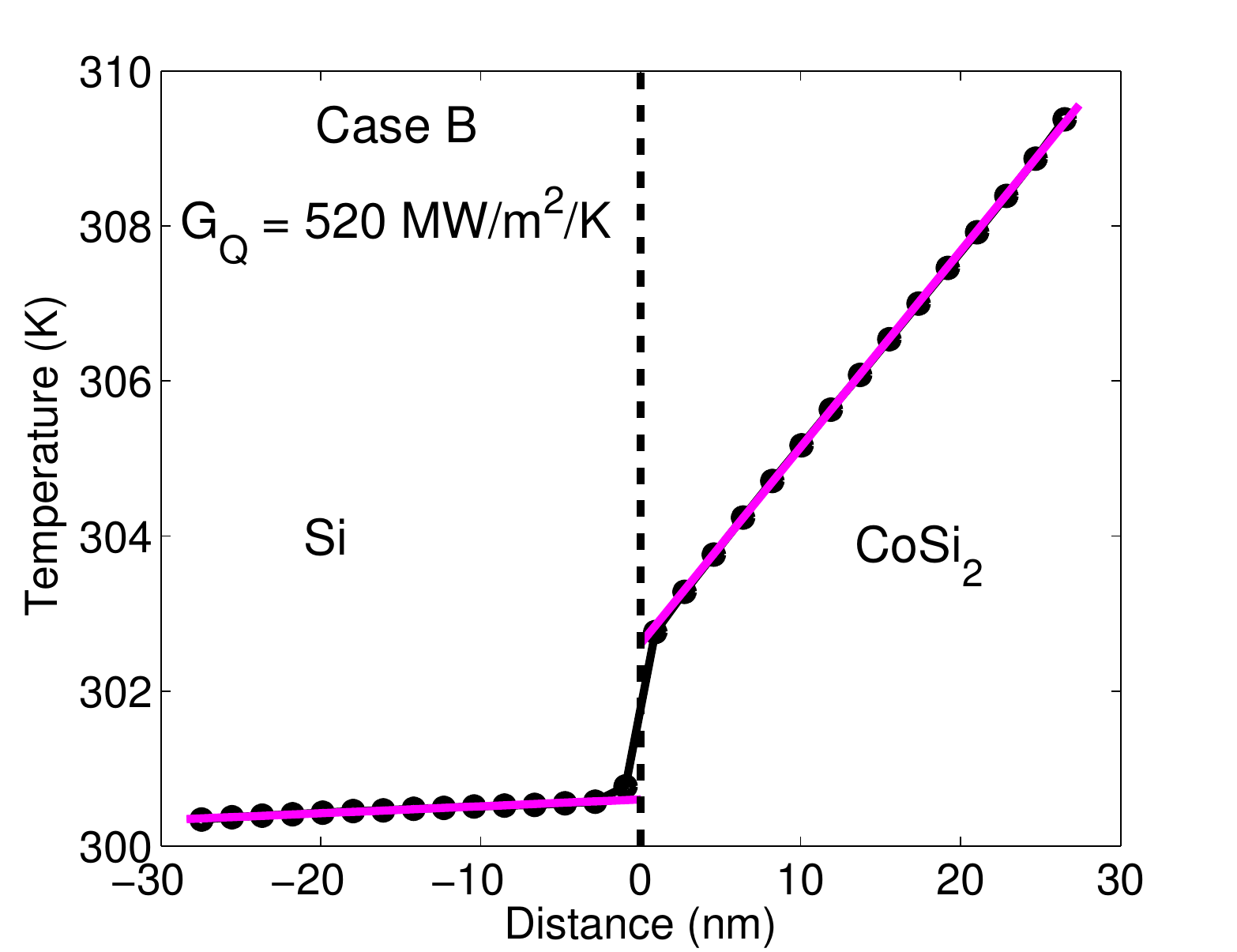}}
\subfloat[]{\includegraphics[height=40mm]{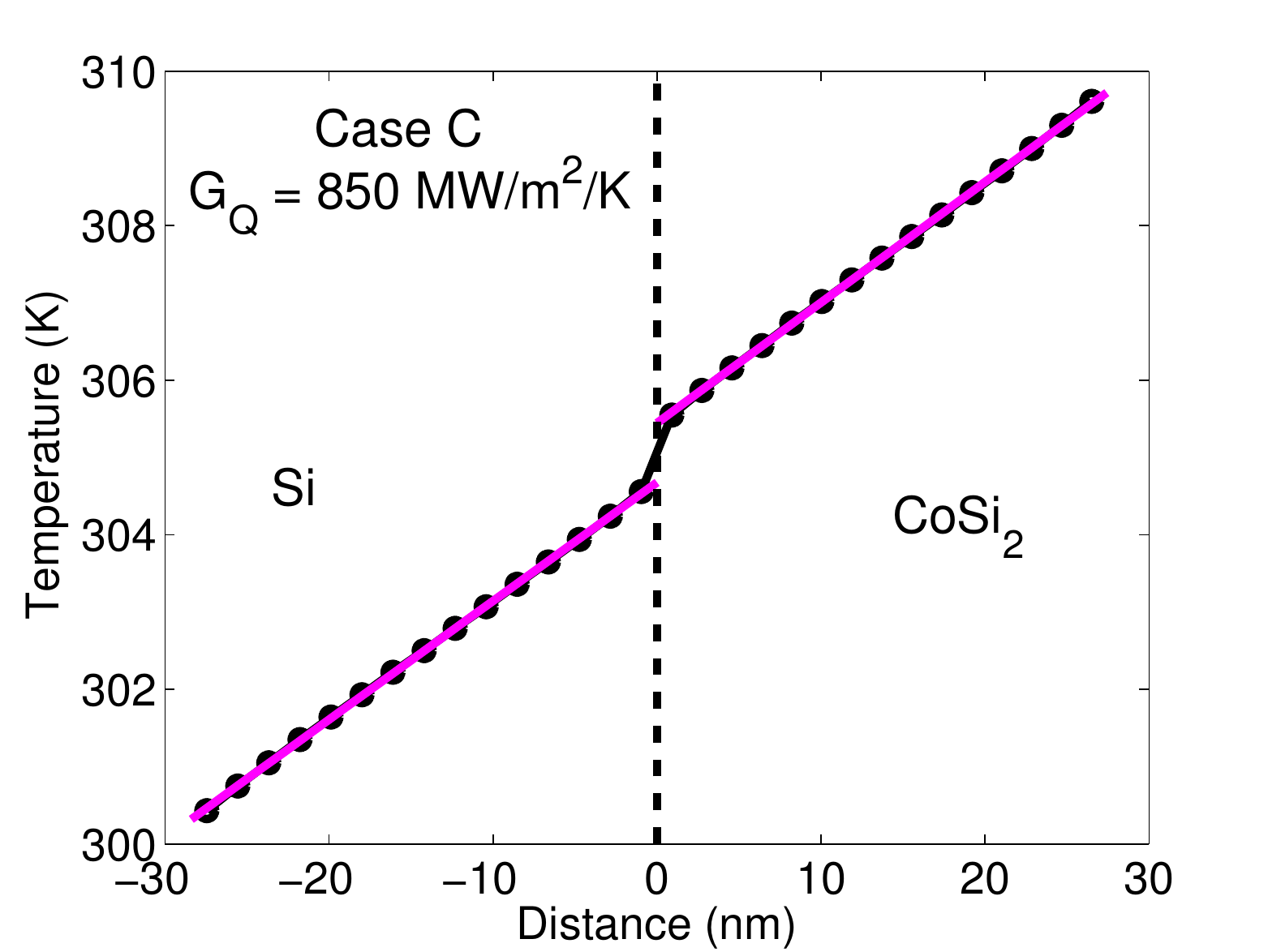}}\\
\subfloat[]{\includegraphics[height=40mm]{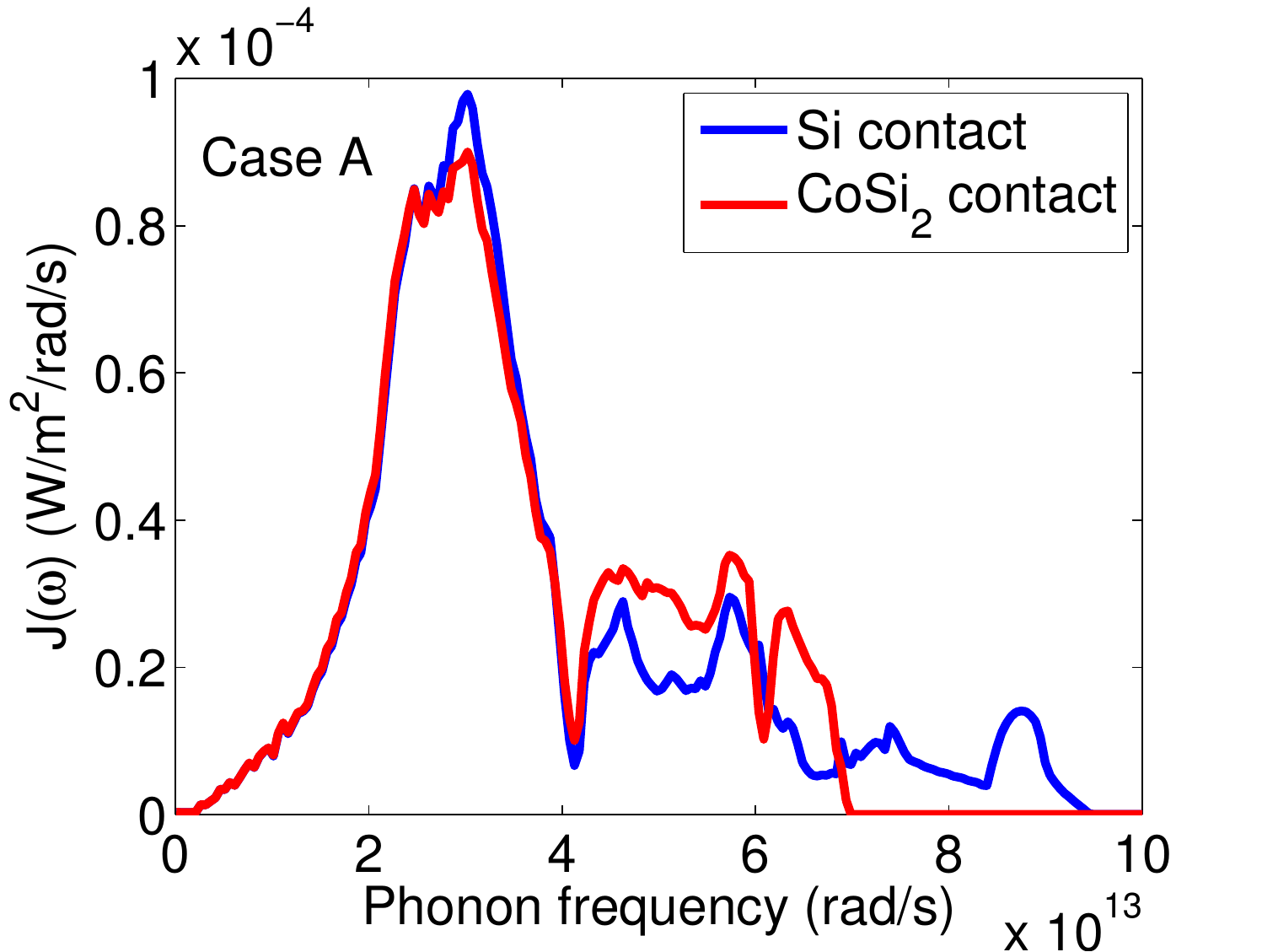}}
\subfloat[]{\includegraphics[height=40mm]{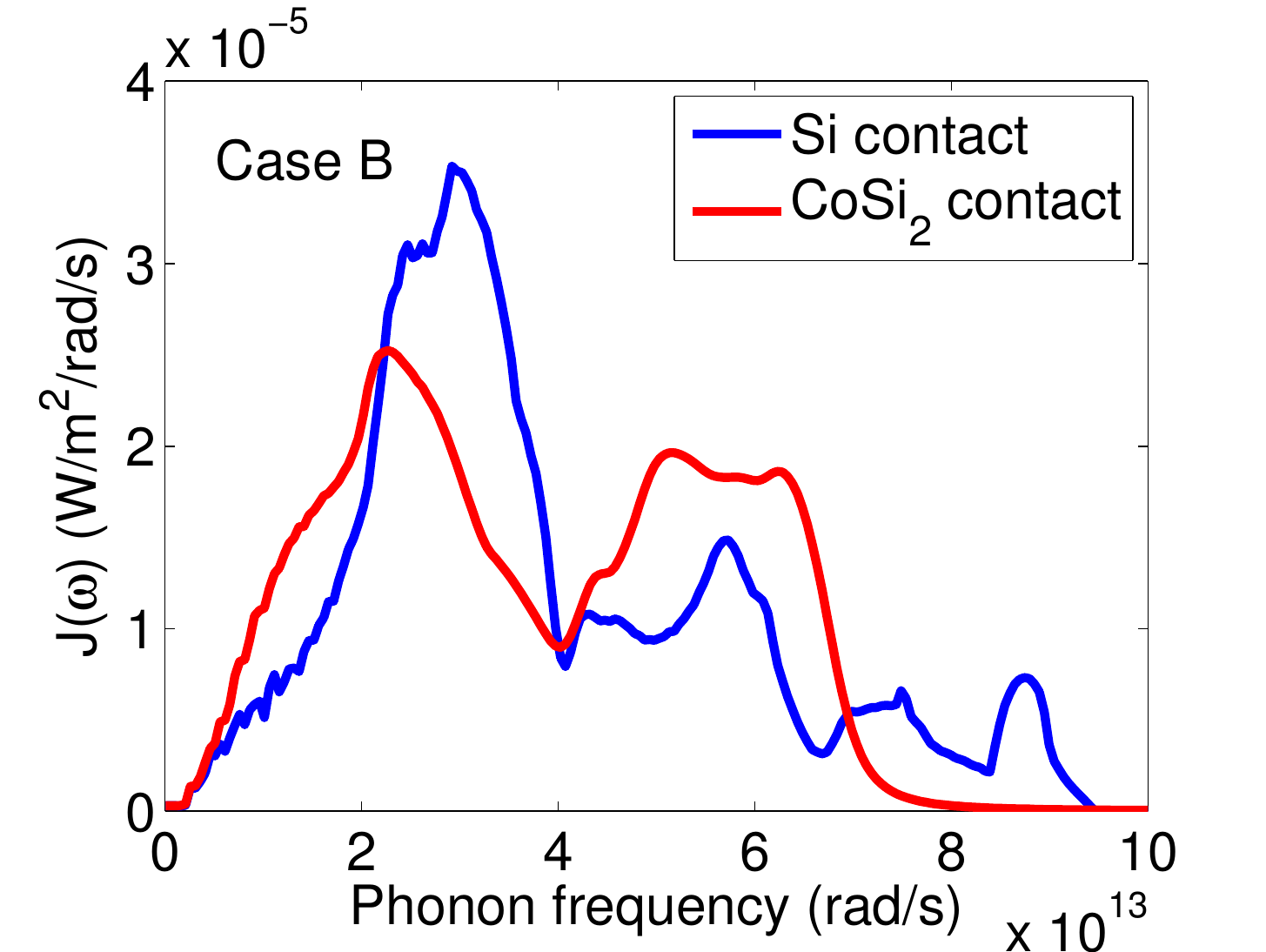}}
\subfloat[]{\includegraphics[height=40mm]{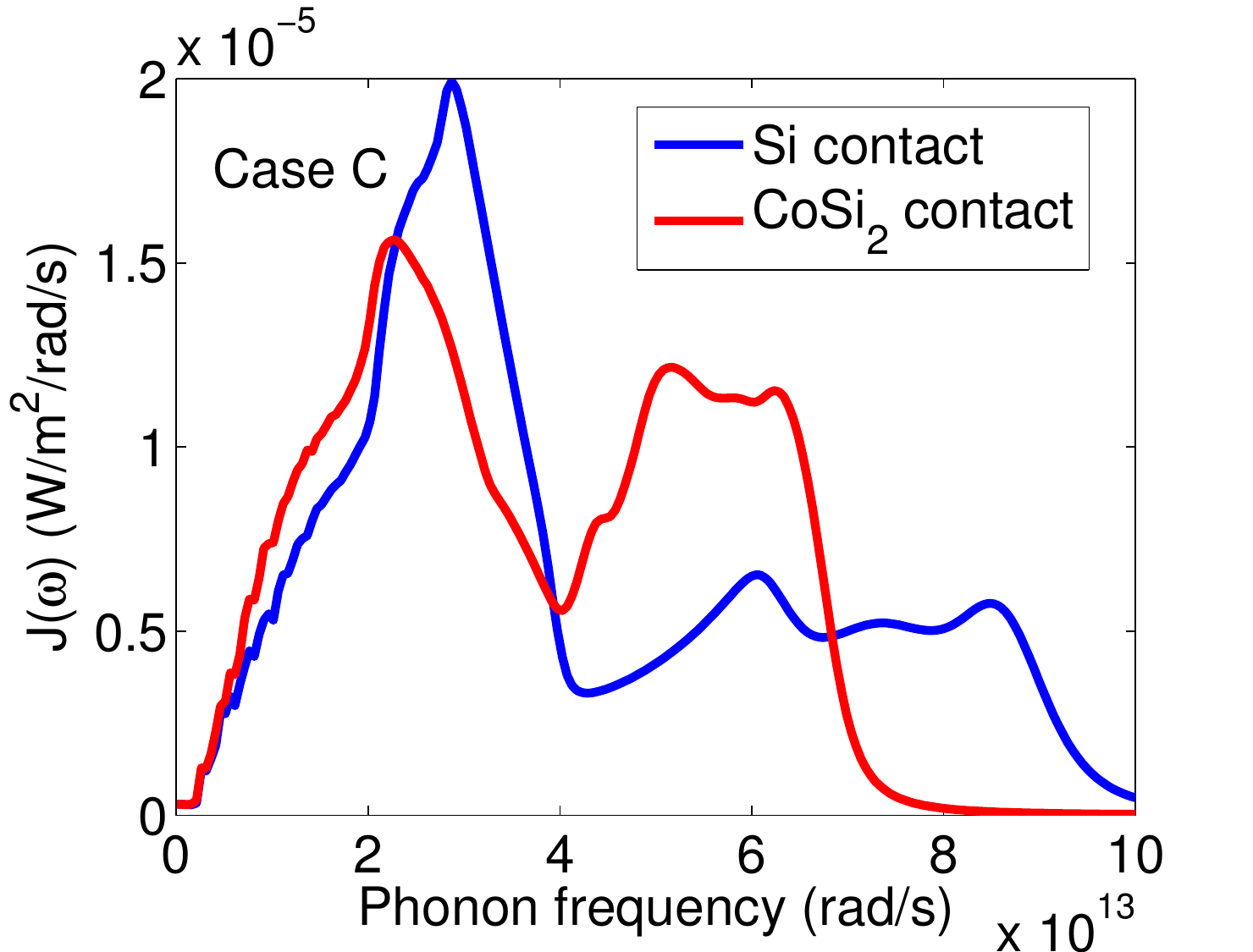}}\\
\subfloat[]{\includegraphics[height=40mm]{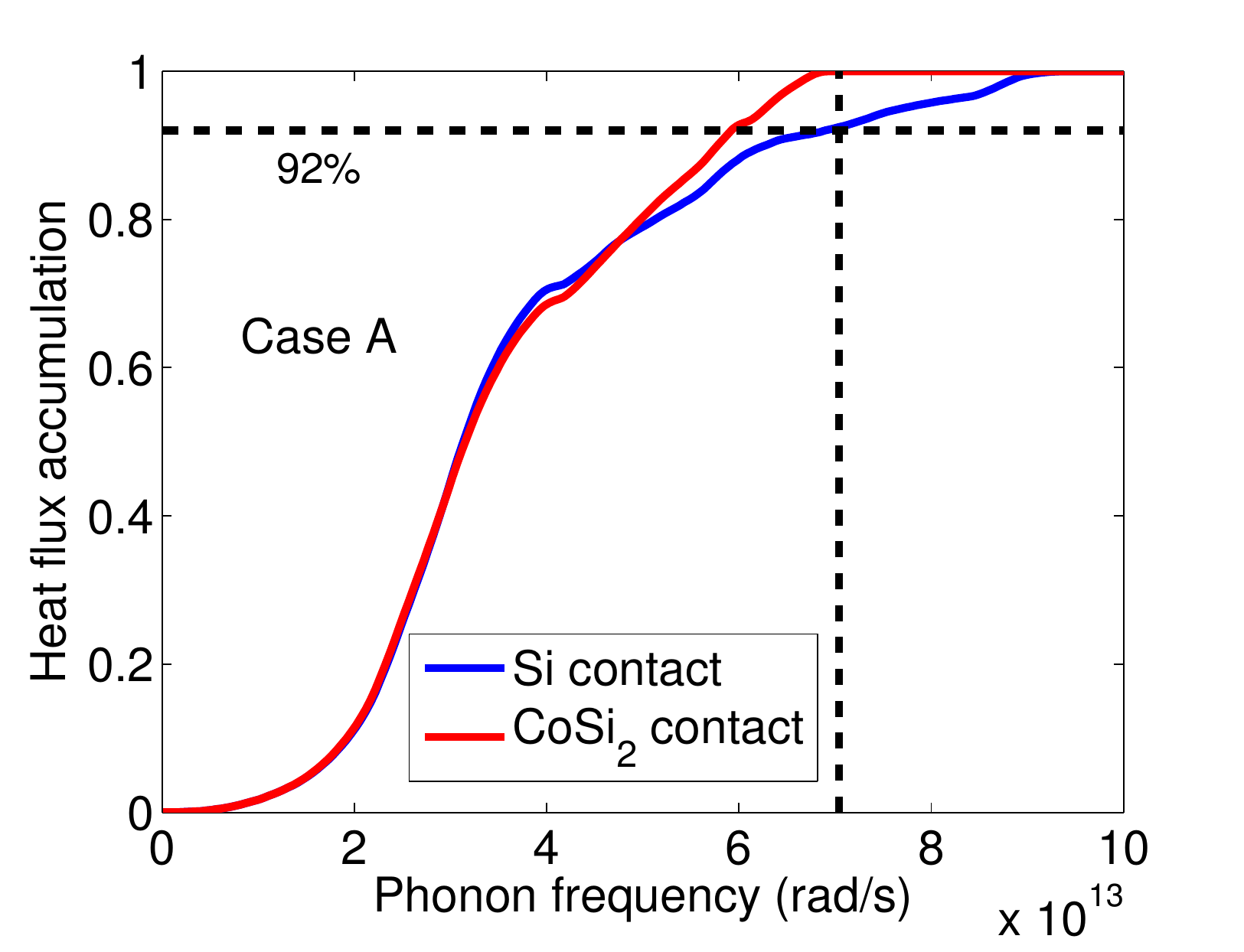}}
\subfloat[]{\includegraphics[height=40mm]{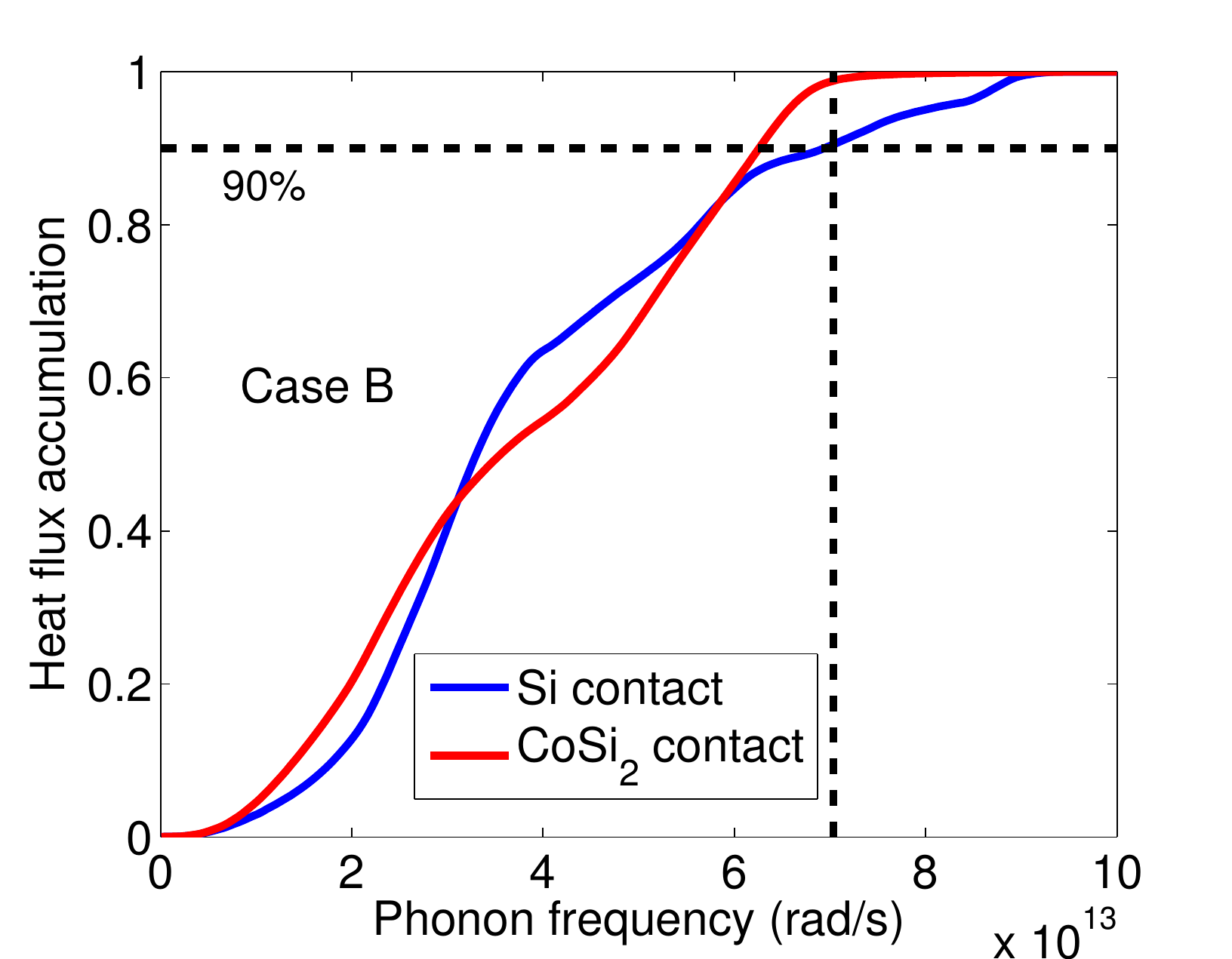}}
\subfloat[]{\includegraphics[height=40mm]{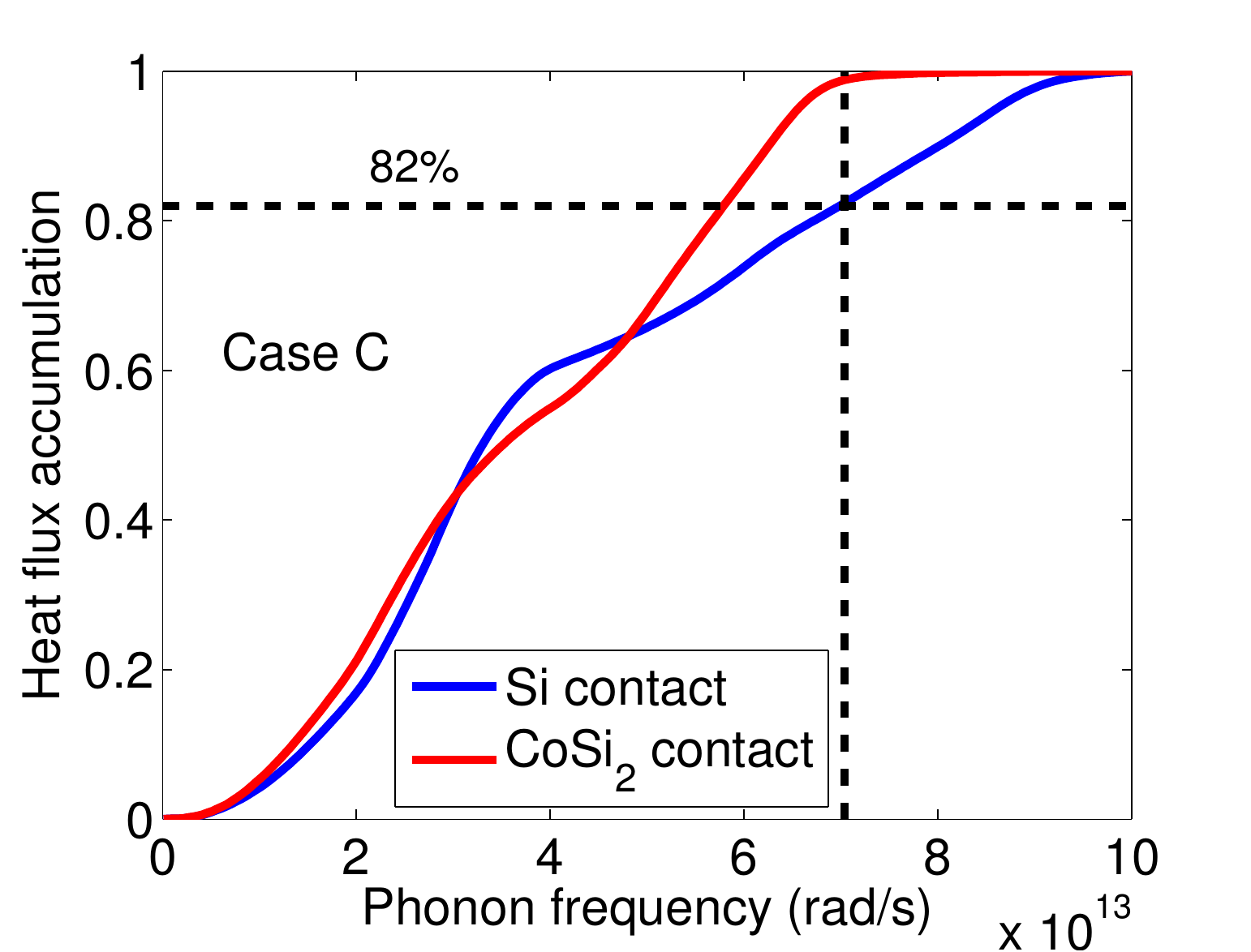}}
\caption{a,b,c) Device temperature profile for Cases A, B, C where Case B corresponds to nominal scattering rates in Si and CoSi$_2$ while Cases A and C correspond to artificially decreased and increased scattering rates respectively. The magenta lines correspond to linear fits of the temperature profiles on either side of the interface. d,e,f) Spectral variation of the energy flux from Si and CoSi$_2$ contacts for Cases A, B, C respectively. g,h,i) Accumulation of energy flux in the Si and CoSi$_2$ contacts with respect to phonon frequency for Cases A, B, C respectively.}\label{anharmonic_scat}
\end{figure}

Figures \ref{anharmonic_scat}a,b,c show the local device temperature profile corresponding to all three cases and the associated thermal interface conductance. A temperature difference of 10 K is applied across the leads in all cases. The conductance is enhanced with increase in the bulk scattering rates of the materials comprising the interface. As expected from conventional scattering theory, the bulk material conductances however decrease with increased scattering rates (observe the progressive rise in temperature drops within Si and CoSi$_2$ in Figures \ref{anharmonic_scat}a,b,c). The foregoing results suggest that the inclusion of inelastic phonon scattering in the AGF simulations produces contrasting effects on the interface and bulk material conductances. 

To elucidate the microscopic mechanisms responsible for the enhancement in interface conductance with inelastic scattering, Figures \ref{anharmonic_scat}d,e,f show the spectral variation of heat flux from the Si and CoSi$_2$ contacts. For Case A with low scattering on both sides of the interface, the spectral heat fluxes from the two contacts follow each other, suggesting that `vertical transport', i.e., mixing between different energy levels is insignificant. However, higher scattering rates result in a shift of the frequencies at which the spectral heat flux is a maximum. Also, the maximum allowed phonon frequency in strained CoSi$_2$ is about $7\times 10^{13}$ rad/s while that in Si is close to $10^{14}$ rad/s. Hence the phonons in Si between $7\times 10^{13}$ rad/s and $10^{14}$ rad/s do not contribute to cross-interface heat transport in a ballistic simulation (see Figure \ref{ballistic_results}a). Inelastic scattering enables phonon scattering into the high energy optical modes of Si whose contribution is enhanced with a rise in the scattering rates of Si and CoSi$_2$. The elevated participation of high-energy optical phonons in Si can also be observed in Figures \ref{anharmonic_scat}g,h,i where phonons with frequencies larger than $7\times 10^{13}$ rad/s contribute 8\%, 10\%, and 18\% of the total energy flux in Si for cases A, B, and C respectively. Although the spectral heat flux shows spatial variation, the total energy flux integrated over all phonon frequencies is independent of position. Similar conclusions on the enhancement of thermal interface conductance due to inelastic scattering have been reported in prior work \cite{kosevich1995fluctuation,landry2009thermal,hohensee2015thermal}. 

\section{Effect of Electron-Phonon Coupling on Thermal Interface Conductance}
\label{eph_results}
\subsection{First-Principles Calculations}
The results from first-principles calculations of electron-phonon coupling, both in bulk strained CoSi$_2$ and Si-CoSi$_2$ interface supercells, are reported in this section. The phonon linewidth $\gamma_{\boldsymbol{q}p}$ due to electron-phonon scattering is given by \cite{allen1987theory,bauer1998electron}:
\begin{equation}
\label{linewidth}
\gamma_{\boldsymbol{q}p} = 2\pi\omega_{\boldsymbol{q}p}\sum\limits_{\nu\nu'}\int\frac{d^3\boldsymbol{k}}{\Omega_{BZ}}|g_{\boldsymbol{k}\nu,\boldsymbol{k}+\boldsymbol{q}\nu'}^{\boldsymbol{q}p}|^2\delta(E_{\boldsymbol{k}\nu}-E_f)\delta(E_{\boldsymbol{k}+\boldsymbol{q}\nu'}-E_f)
\end{equation}
where $g_{\boldsymbol{k}\nu,\boldsymbol{k}+\boldsymbol{q}\nu'}^{\boldsymbol{q}p}$ is the electron-phonon coupling matrix element for scattering of an electron with energy $E_{\boldsymbol{k}}$ in band $\nu$ by a phonon of energy $\hbar\omega_{\boldsymbol{q}p}$ into a state with energy $E_{\boldsymbol{k}+\boldsymbol{q}}$ in band $\nu'$. The above expression is valid at low temperatures when electron-phonon scattering is restricted to a narrow energy window around the Fermi surface. The phonon linewidth can be used to compute the spectral Eliashberg function $\alpha^2F(\omega)$ which quantifies the strength of electron-phonon coupling:
\begin{equation}
\label{eliashberg}
\alpha^2F(\omega) = \frac{1}{2\pi D(E_f)}\sum\limits_{\boldsymbol{q},p}\frac{\gamma_{\boldsymbol{q}p}}{\hbar\omega_{\boldsymbol{q}p}}\delta(\omega-\omega_{\boldsymbol{q}p})
\end{equation}
The spectral Eliashberg function can be used to obtain an effective volumetric electron-phonon coupling coefficient $G_{ep}$:
\begin{equation}
G_{ep} = 2\pi D(E_f)\int\limits_{0}^{\infty}{(\hbar\omega)^2\alpha^2F(\omega)\frac{\partial f_{BE}^o}{\partial T}d\omega}
\end{equation}
To understand the spatial variation of electron-phonon coupling across a semiconductor-metal interface, we also define a local Eliashberg function $\alpha^2F_l(\omega)$ as \cite{sadasivam2015electron}:
\begin{eqnarray}
\label{Eliashberg_local_decomp}
\alpha^2F(\omega) &=& \frac{1}{2\pi D(E_f)}\sum\limits_{\boldsymbol{q},p}\frac{\gamma_{\boldsymbol{q}p}}{\hbar\omega_{\boldsymbol{q}p}}\delta(\omega-\omega_{\boldsymbol{q}p})\nonumber \\
&=& \frac{1}{2\pi D(E_f)}\sum\limits_{\boldsymbol{q},p}\frac{\gamma_{\boldsymbol{q}p}}{\hbar\omega_{\boldsymbol{q}p}}\delta(\omega-\omega_{\boldsymbol{q}p})\sum\limits_{l}\sum\limits_{m={x,y,z}}\phi_{\boldsymbol{q}p,lm}\phi_{\boldsymbol{q}p,lm}^* \quad \left(\sum\limits_{l}\sum\limits_{m={x,y,z}}\phi_{\boldsymbol{q}p,lm}\phi_{\boldsymbol{q}p,lm}^*=1\right) \nonumber \\
&=& \frac{1}{2\pi D(E_f)}\sum\limits_{l}\sum\limits_{m={x,y,z}}\sum\limits_{\boldsymbol{q},p}\frac{\gamma_{\boldsymbol{q}p}}{\hbar\omega_{\boldsymbol{q}p}}\delta(\omega-\omega_{\boldsymbol{q}p})\phi_{\boldsymbol{q}p,l}\phi_{\boldsymbol{q}p,l}^* \nonumber \\
&=& \sum\limits_{l}\alpha^2F_l(\omega)
\end{eqnarray}
where $\phi_{\boldsymbol{q}p}$ denotes the phonon eigenvector, the index $l$ runs over all the atoms in the unitcell, and the index $m$ represents the vibrational degrees of freedom ($x$, $y$, $z$) for each atom. The local Eliashberg function is then used to define a local volumetric electron-phonon coupling coefficient $G_{ep,l}$:
\begin{equation}
G_{ep,l} = \left(2\pi D(E_f)\int\limits_{0}^{\infty}{(\hbar\omega)^2\alpha^2F_l(\omega)\frac{\partial f_{BE}^o}{\partial T}d\omega}\right)\frac{V_{unitcell}}{V_l}
\end{equation}
where the additional factor $V_{unitcell}/V_l$ ensures that $G_{ep,l}$ is the local volumetric coupling coefficient around atom $l$ that occupies a volume $V_l$. 

\begin{figure}[htb]
\centering
\subfloat[]{\includegraphics[height=61mm]{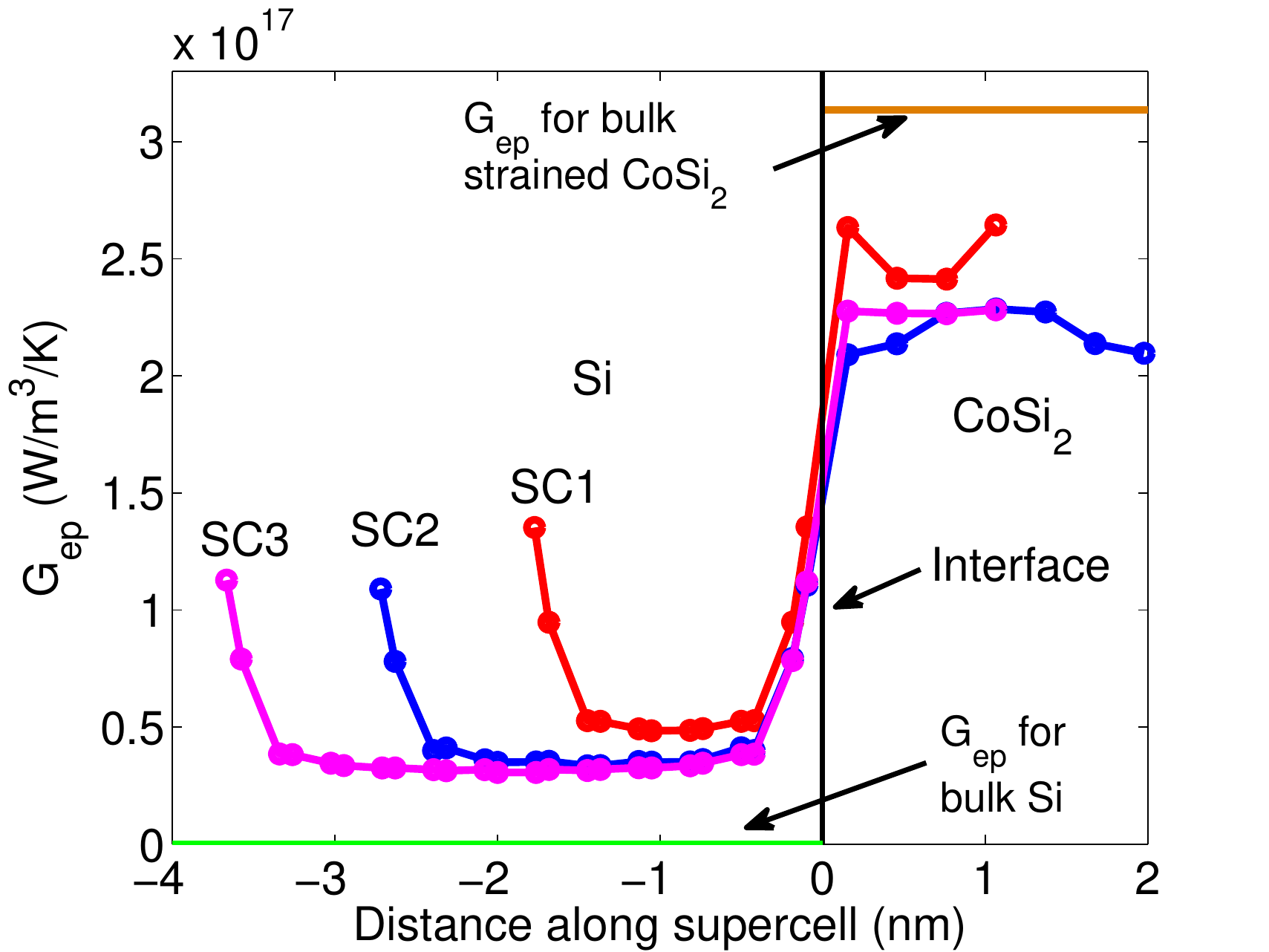}}
\subfloat[]{\includegraphics[height=61mm]{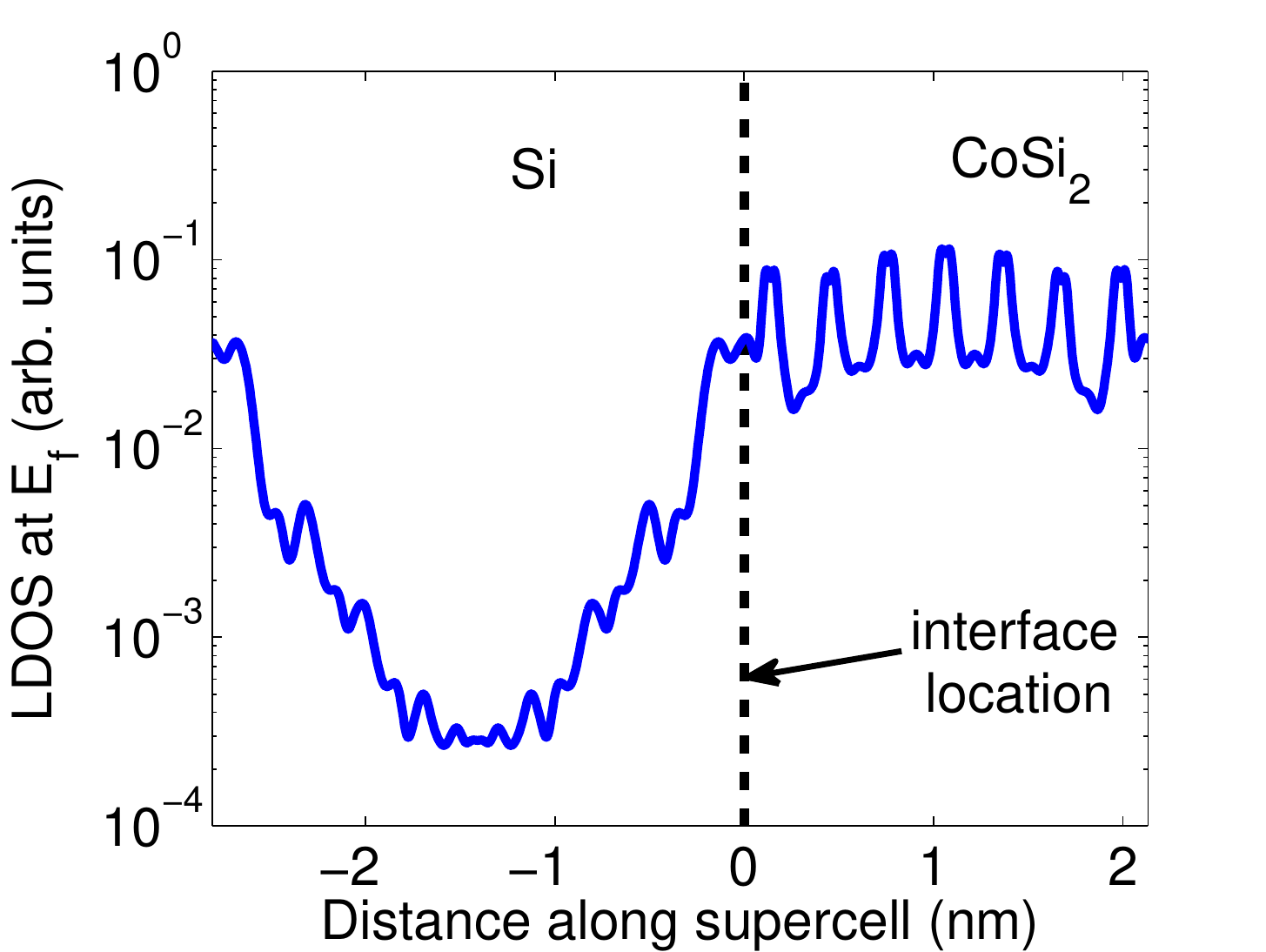}}
\caption{a) Spatial variation of the electron-phonon coupling coefficient $G_{ep}$ across a Si-CoSi$_2$ interface for different supercell lengths. The coupling coefficient in bulk strained CoSi$_2$ and bulk Si are also shown for comparison. b) Spatial variation of the local electron DOS at the Fermi energy across the Si-CoSi$_2$ structure shown in Figure \ref{interface_structure_CoSi2}c.}\label{Gep_local_fig}
\end{figure}

The Eliashberg functions of bulk strained CoSi$_2$ and the interface supercell calculated from DFPT are provided in the Supplemental Material along with a discussion on convergence with respect to k-point grid and smearing parameters. Eqs.~(\ref{linewidth}) and (\ref{eliashberg}) are appropriate for bulk materials in which translational periodicity is assumed in the scattering matrix elements $g_{\boldsymbol{k}\nu,\boldsymbol{k}+\boldsymbol{q}\nu'}^{\boldsymbol{q}p}$. These equations also apply for the Si-CoSi$_2$ interface supercells because these supercells represent Si-CoSi$_2$ superlattices with periodicities of the order of a few nm. The Eliashberg function for the supercells physically represents the coupling between electron and phonon modes of the Si-CoSi$_2$ superlattice. To ensure that the electron-phonon coupling coefficients obtained from DFT/DFPT calculations on superlattices are transferrable to transport simulations of a single Si-CoSi$_2$ interface, we performed calculations on a series of Si-CoSi$_2$ supercells with varying Si and CoSi$_2$ slab thicknesses. Figure \ref{Gep_local_fig}a shows the spatial variation of the electron-phonon coupling coefficient for three interface supercells (SC) of the 8B configuration with different lengths of the Si and CoSi$_2$ slabs forming the interface. The local coupling coefficient on the CoSi$_2$ side of the interface is averaged over one Co and two Si atoms to remove atomistic fluctuations in the local coupling coefficient. The electron-phonon coupling coefficients for bulk strained CoSi$_2$ and bulk intrinsic Si are also shown in Figure \ref{Gep_local_fig}a. The electron-phonon coupling in intrinsic bulk Si is zero since the Fermi level lies in the middle of the bandgap, and the delta functions around the Fermi surface in Eq.~(\ref{linewidth}) are zero. An important observation from Figure \ref{Gep_local_fig}a is the appearance of a non-zero coupling coefficient on the Si side of the interface. Also, the magnitude of this coupling is approximately constant in Si beyond two atomic layers from the interface for all three supercells considered in Figure \ref{Gep_local_fig}a. The convergence of the plateau on the Si side of the interface for supercells SC2 and SC3 in Figure \ref{Gep_local_fig}a indicates that the electronic wavefunctions of CoSi$_2$ are sufficiently localized within the CoSi$_2$ slab and do not tunnel across the Si slabs of the superlattice. The Si-CoSi$_2$ interface forms a Schottky barrier with a p-type Schottky barrier height of 0.2 eV. Hence, the local electronic DOS at the Fermi level decays rapidly in Si away from the Si-CoSi$_2$ interface (see Figure \ref{Gep_local_fig}b). However, such a two-order-of-magnitude decay in local electronic DOS does not result in a commensurate reduction in the local electron-phonon coupling coefficient shown in Figure \ref{Gep_local_fig}a. 

\begin{figure}
\centering
\subfloat[]{\includegraphics[height=40mm]{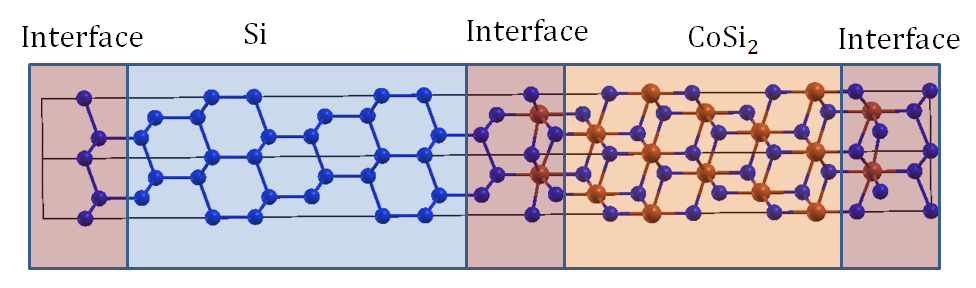}}\\
\subfloat[]{\includegraphics[height=60mm]{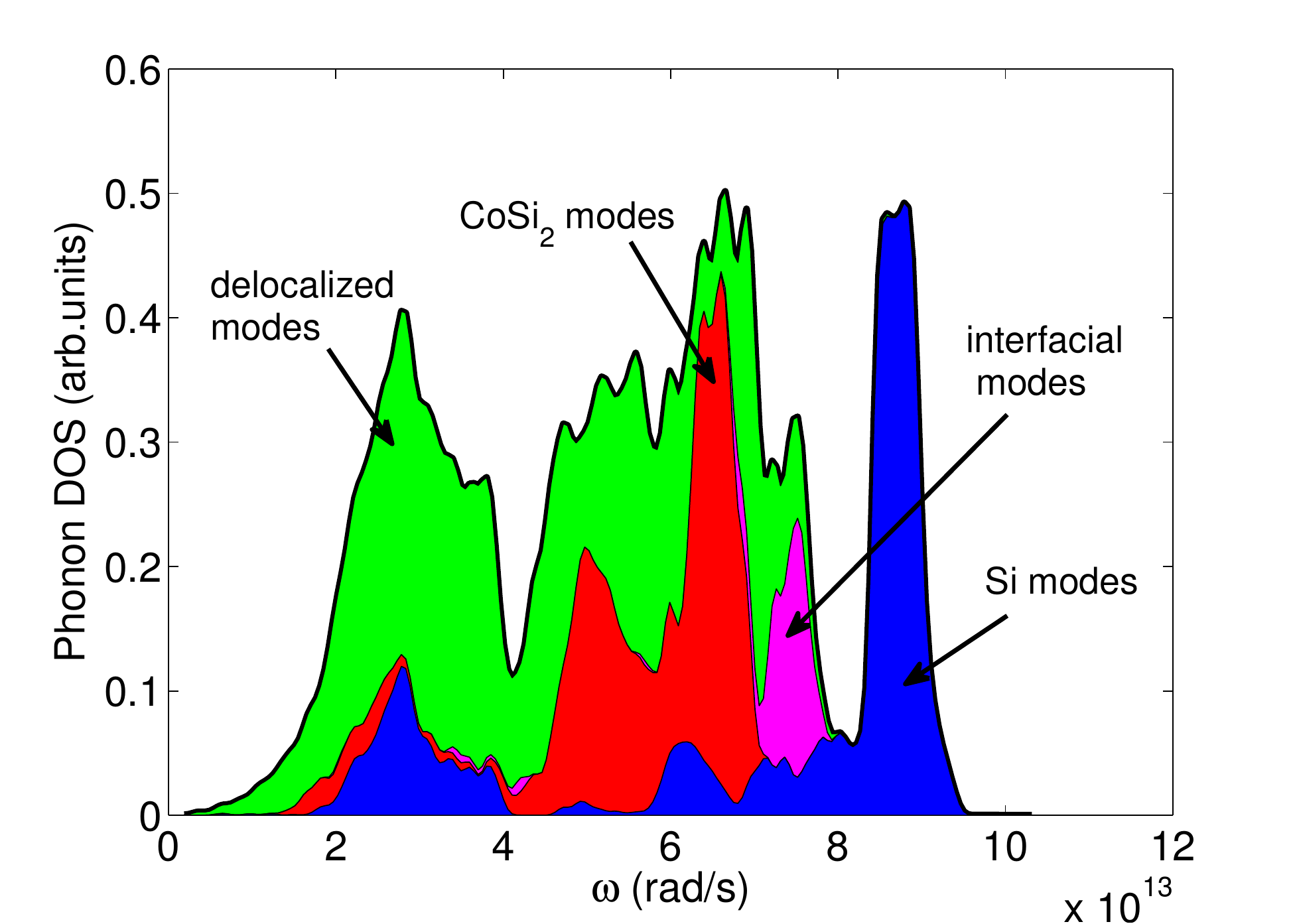}}
\subfloat[]{\includegraphics[height=60mm]{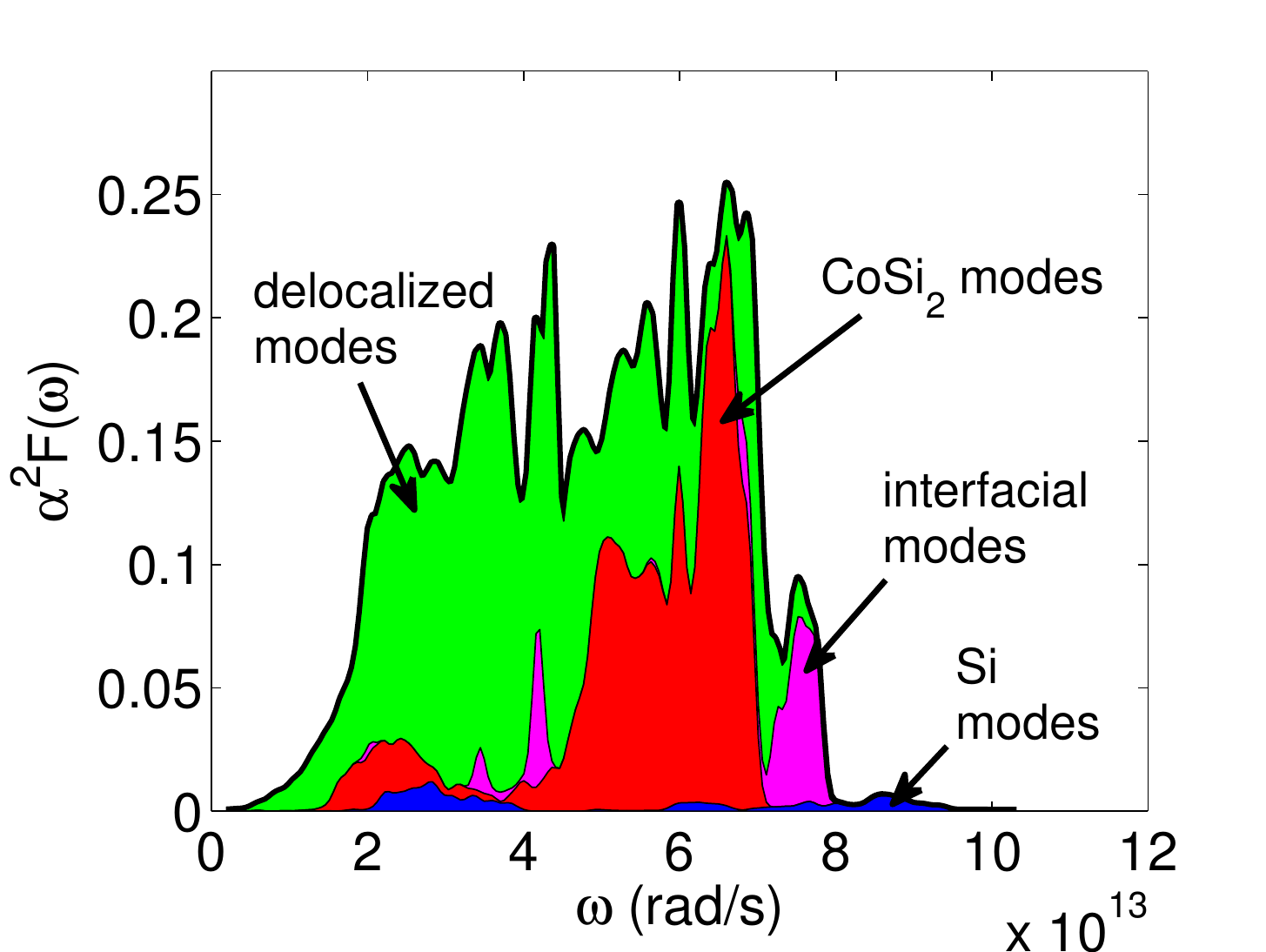}}
\caption{a) Partitioning of different regions in the Si-CoSi$_2$ interface supercell used for the classification of phonon modes. b,c) Contribution of Si, CoSi$_2$, interfacial, and delocalized phonon modes to the total DOS (b) and Eliashberg function (c) of the Si-CoSi$_2$ interface supercell.}\label{mode_contribution}
\end{figure}

This unusual result can be understood by considering the relative contributions of different types of phonon modes of the Si-CoSi$_2$ interface supercell to the overall Eliashberg function. The different phonon modes of the interface supercell are classified into four types based on the spatial localization of the phonon eigenvector corresponding to the mode. The present approach is analogous to the classification of interface modes in ref.~\onlinecite{gordiz2016phonon}. The interface supercell (SC 2 in Figure \ref{Gep_local_fig}a) shown in Figure \ref{mode_contribution}a is decomposed into three regions consisting of Si, CoSi$_2$, and interfacial atoms. The criteria for classification of a phonon mode $\phi_{\boldsymbol{q}p}$ is defined as follows:
\begin{equation}
\phi_{\boldsymbol{q}p} = 
\begin{cases}
\text{Si mode, if } \frac{\| \phi_{\boldsymbol{q}p,\text{Si}} \|}{\| \phi_{\boldsymbol{q}p,\text{tot}} \|}  >0.85 \\ 
\text{CoSi$_2$ mode, if } \frac{\| \phi_{\boldsymbol{q}p,\text{CoSi}_2} \|}{\| \phi_{\boldsymbol{q}p,\text{tot}} \|} >0.85 \\
\text{interfacial mode, if } \frac{\| \phi_{\boldsymbol{q}p,\text{int}} \|}{\| \phi_{\boldsymbol{q}p,\text{tot}} \|} >0.85 \\
\text{delocalized mode, if } \text{none of the above} \\ 
\end{cases}
\label{mode_classifier}
\end{equation}
where $\| \phi_{\boldsymbol{q}p,\text{Si}} \|$, $\| \phi_{\boldsymbol{q}p,\text{CoSi}_2} \|$, and $\| \phi_{\boldsymbol{q}p,\text{int}} \|$  denote the norm of the phonon eigenvector within the Si, CoSi$_2$ and interfacial regions respectively in Figure \ref{mode_contribution}a. $\| \phi_{\boldsymbol{q}p,\text{tot}} \|$ denotes the norm of the phonon eigenvector of the entire interface supercell and is normalized to unity. The choice of spatial extent of the interfacial region and the value 0.85 in Eq.~(\ref{mode_classifier}) are arbitrary and used only to provide a physical understanding of the mechanism of cross-interface coupling between electrons in metal and phonons in the semiconductor. 

The contribution of the different types of phonon modes to the total phonon DOS and the total Eliashberg function of the interface supercell are shown in Figures \ref{mode_contribution}b,c respectively. Figure \ref{mode_contribution}b indicates that phonon modes in the frequency range of $\omega = (8-10)\times 10^{13}$ rad/s are localized in the Si region of the interface. These high-frequency modes correspond to optical modes of Si and are above the maximum allowed phonon frequency of bulk CoSi$_2$. Although the optical modes of Si contribute to phonon DOS of the interface supercell, their contribution to the Eliashberg function shown in Figure \ref{mode_contribution}c is negligible. This result demonstrates that modes localized in Si do not couple with metal electrons. However, the significant volumetric coupling coefficient on the Si side of the interface in Figure \ref{Gep_local_fig}a can be attributed to delocalized modes whose vibrational energy is distributed across Si and CoSi$_2$ atoms of the interface supercell. Metal electrons transfer energy to delocalized phonon modes whose vibrational patterns dictate that a portion of the energy is transferred to silicon atoms across the interface. 

Hence, our results suggest that energy exchange between electrons in metal and atomic vibrations in the semiconductor is manifested primarily by the coupling between electrons and delocalized interface modes whose vibrational energy is distributed across Si and CoSi$_2$ atoms. An important implication of this result is that strength of direct electron-phonon coupling is intimately tied to the strength of interfacial bonding and the phonon-phonon conductance across the interface. For an interface with weak or van der Waals bonding, the contribution of such delocalized modes to phonon DOS is expected to be much smaller, and the phonon modes will be localized on either side of the interface. 

Figure \ref{mode_contribution}c also suggests that the mechanism of energy transfer from metal electrons to phonons in Si is primarily mediated by acoustic delocalized phonon modes and the contribution from coupling between electrons and optical modes of Si is negligible. This result contrasts with electron-phonon coupling in bulk Si where the contributions from acoustic and optical modes are similar in magnitude (see section VI in Supplemental Material). In the interface supercell considered here, optical modes of Si are localized to the Si side of the interface where the electron DOS at Fermi level is very small (see Figure \ref{Gep_local_fig}b). The acoustic modes in Si are delocalized with the acoustic modes of CoSi$_2$, leading to their stronger coupling with electrons in CoSi$_2$. Figure \ref{mode_contribution}c also indicates that coupling between metal electrons and CoSi$_2$ optical phonon modes contributes significantly to the Eliashberg function of the interface supercell. However, such coupling is localized within the metal and contributes little to energy transfer across the interface. Localized interfacial modes, i.e., modes with vibrational energy localized to a few atomic layers around the interface are observed to contribute to the Eliashberg function in a small frequency range $\omega = (7-8)\times 10^{13}$ rad/s. 

\subsection{Effect of Electron-Phonon Coupling on Thermal Interface Conductance}
In this section, results from first-principles calculations of electron-phonon coupling are incorporated into the AGF transport simulations. The details of the approach are described in Section \ref{ttm_agf_algo} and Supplemental Material. The primary difference between the results presented here and those in Section \ref{anharmonic_scat_sec} is the presence of non-zero energy fluxes in the B\"{u}ttiker probes to represent the energy exchanged between electrons and phonons. Hence, the simulation results presented in this section include contributions from both anharmonic phonon scattering and electron-phonon coupling. 

\begin{figure}
\centering
\subfloat[]{\includegraphics[height=55mm]{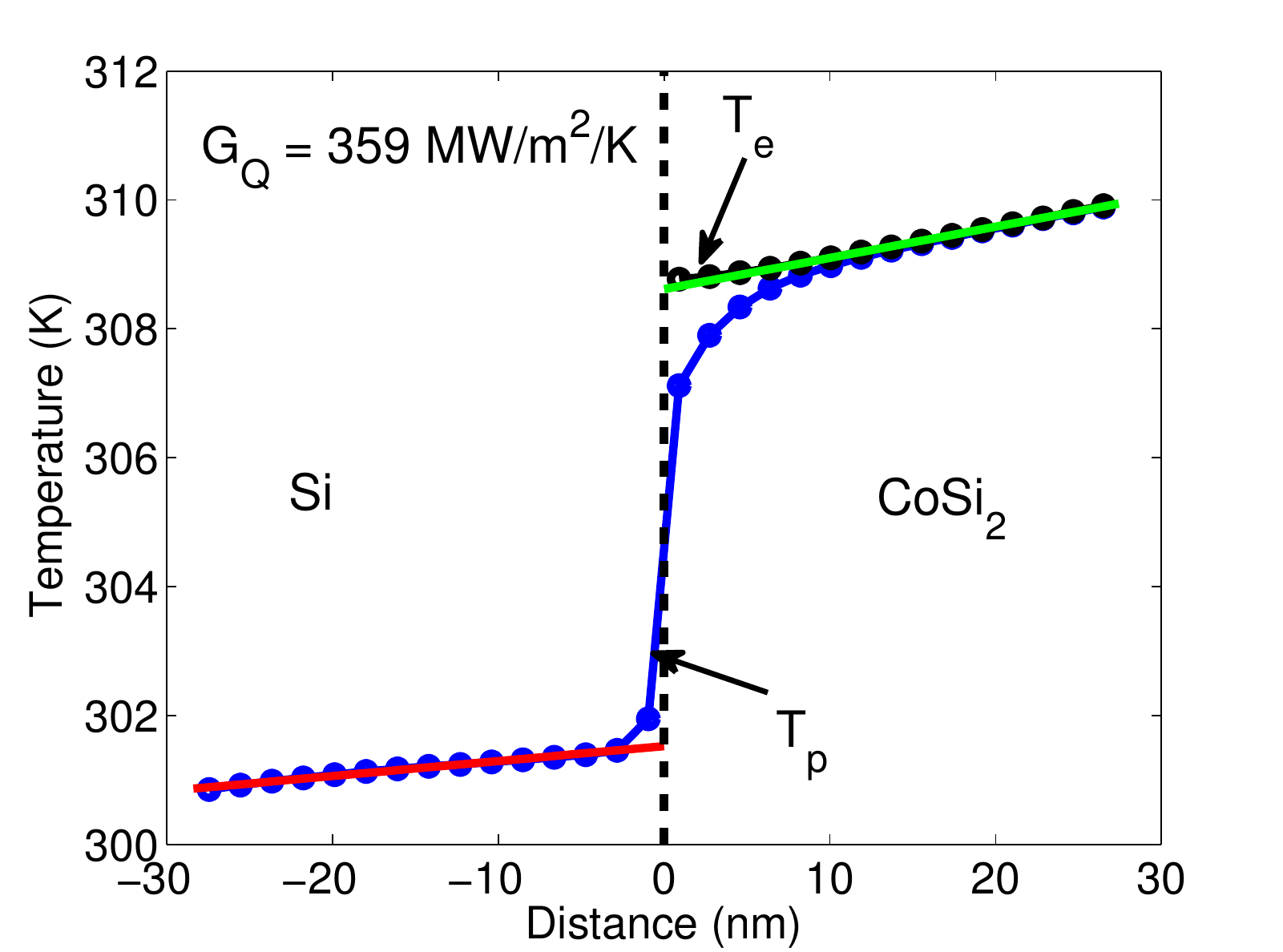}}
\subfloat[]{\includegraphics[height=55mm]{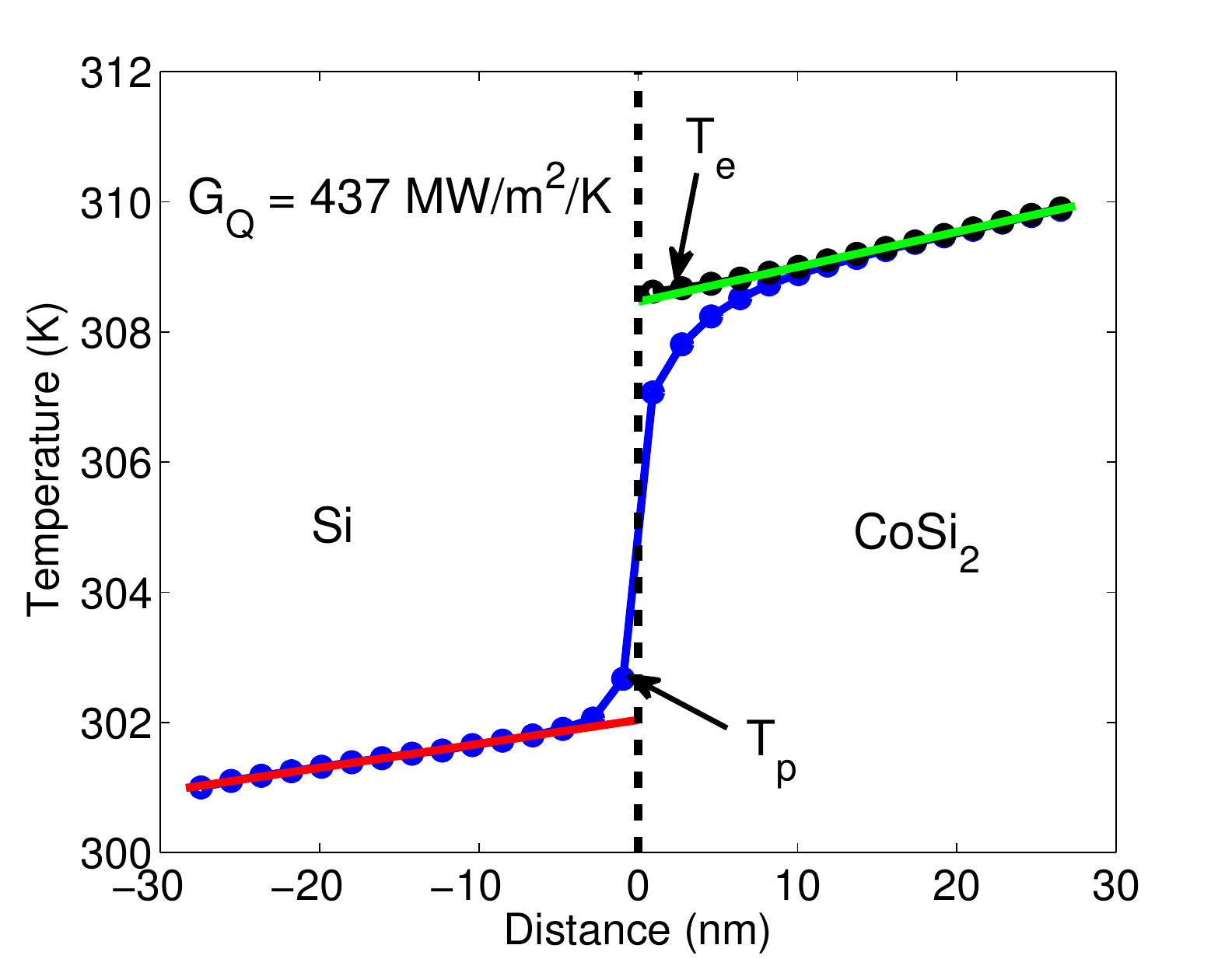}}\\
\subfloat[]{\includegraphics[height=55mm]{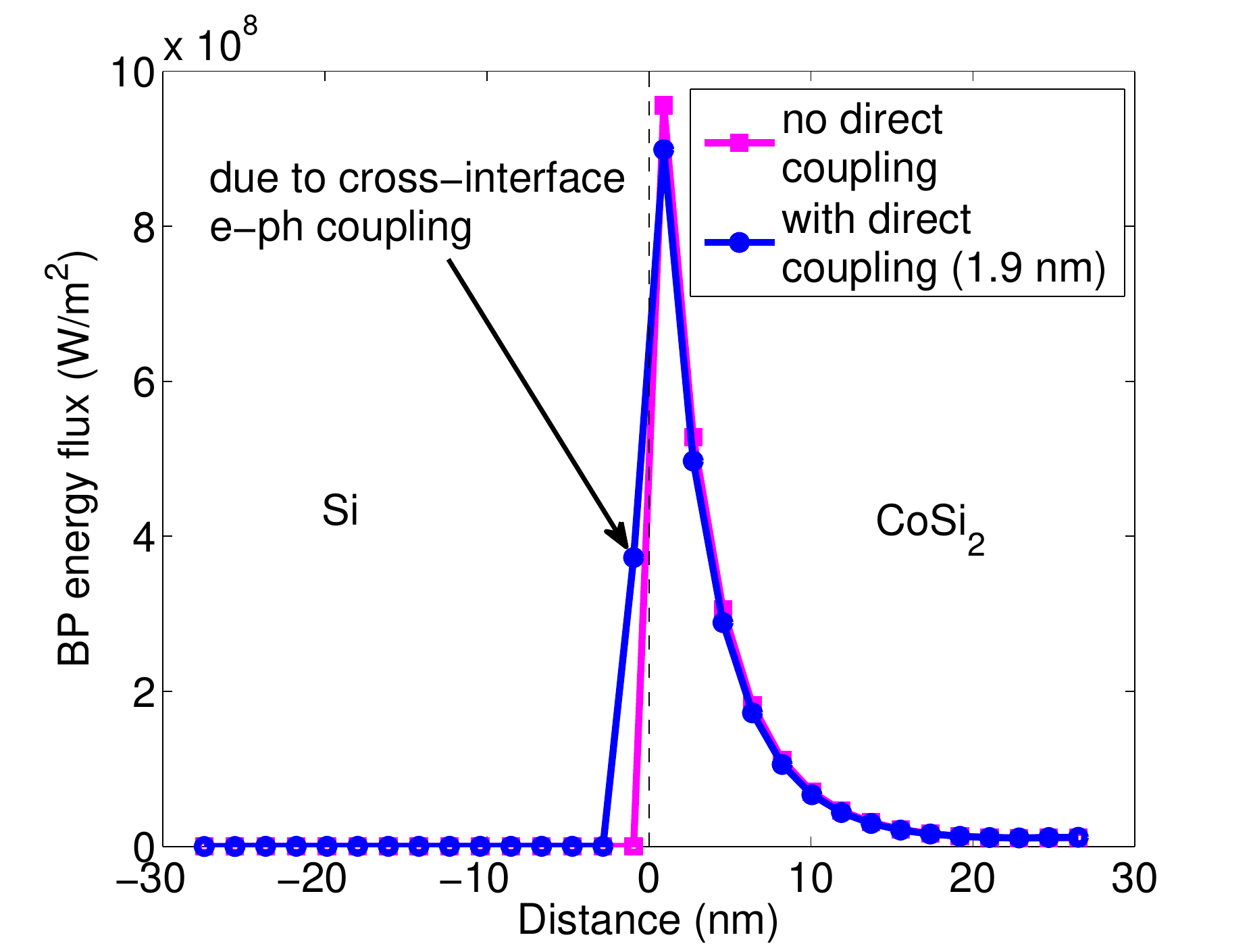}}
\caption{a) Electron and lattice temperature profile across Si-CoSi$_2$ interface with electron-phonon coupling inside the metal region only. b) Electron and lattice temperature profile across Si-CoSi$_2$ interface with electron-phonon coupling inside the metal region and in two unit cells of Si closest to the interface. In both a) and b), The red line corresponds to a linear fit of the lattice temperature profile in Si and the green line corresponds to a linear fit of the electron temperature profile in CoSi$_2$ away from the interface region. c) Heat flux distribution in the B\"{u}ttiker probes across the Si-CoSi$_2$ interface corresponding to the temperature profiles in a) and b). For the simulation with direct electron-phonon coupling, the first B\"{u}ttiker probe in Si closest to the interface has a non-zero energy flux.}\label{temp_profiles}
\end{figure}

We consider first the case where electrons exchange energy only through the B\"{u}ttiker probes in the metal, i.e., no direct coupling between metal electrons and semiconductor phonons. The Eliashberg function of bulk strained CoSi$_2$ is used to calculate the energy exchange term $Q_{ep}$ in Eq.~(\ref{Qep}). Figure \ref{temp_profiles}a shows a typical electron and lattice temperature profile obtained from such a simulation along with the heat fluxes from the various B\"{u}ttiker probes in Si and CoSi$_2$ (see Figure \ref{temp_profiles}c). The heat fluxes in all the B\"{u}ttiker probes on the Si side of the interface are zero while the heat fluxes in the B\"{u}ttiker probes of CoSi$_2$ decrease away from the interface. This decay in the electron-phonon energy transfer away from the interface is a consequence of the equilibrium between electrons and phonons away from the interfacial region (see temperature profile in Figure \ref{temp_profiles}a).

Also, comparison of the lattice temperature profiles in Figure \ref{anharmonic_scat}b with that in Figure \ref{temp_profiles}a shows that for the same scattering rates and applied temperature difference across the Si and CoSi$_2$ contacts, the lattice temperature drop in CoSi$_2$ is reduced when electrons are included in the simulation. The reduced lattice temperature drop in CoSi$_2$ is a consequence of electrons in metal providing a parallel heat flow path with lower resistance compared to phonons ($\kappa_{e,\text{CoSi}_2} = 46$ W/m/K, $\kappa_{p,\text{CoSi}_2} = 4.9$ W/m/K). Hence, a significant fraction of energy in CoSi$_2$ is carried by electrons that transfer energy to the lattice near the metal-semiconductor interface. 

The present simulation is conceptually similar to the analytical model developed by Majumdar and Reddy \cite{majumdar2004role} who suggested that electron-phonon coupling within the metal effectively provides a resistance in series with the phonon-phonon resistance across the interface. Hence the interface conductance in Figure \ref{temp_profiles}a is smaller than the phonon-only conductance in Figure \ref{anharmonic_scat}b. Majumdar and Reddy's model for the effective conductance with electron-phonon coupling is given by:
\begin{equation}
G_{Q} = \frac{\sqrt{G_{ep}\kappa_p}}{1+\frac{\sqrt{G_{ep}\kappa_p}}{G_{pp}}}
\label{majumdar_model}
\end{equation} 
where $G_{ep}$ is the effective electron-phonon coupling efficient in the metal, $\kappa_p$ is the lattice thermal conductivity of the metal, and $G_{pp}$ is the phonon interfacial conductance. The electron-phonon coupling coefficient in bulk CoSi$_2$ is $3.1 \times 10^{17}$ W/m$^3$/K (see Figure \ref{Gep_local_fig}a), $\kappa_p = 4.9$ W/m/K, and $G_{pp} = 5.2\times 10^8$ W/m$^2$/K (see Figure \ref{anharmonic_scat}b). Substituting these values in Eq.~(\ref{majumdar_model}), we obtain $G_{Q} = 365$ MW/m$^2$/K which is close to the value from the simulation in Figure \ref{temp_profiles}a. 

Although the temperature profiles presented so far in Figures \ref{anharmonic_scat} and \ref{temp_profiles} involve conditions near room temperature, similar simulations were also performed at temperatures of 100, 150, 200, and 250 K to obtain the temperature dependence of interface conductance. At each temperature, the B\"{u}ttiker probe scattering rate in Si was changed to match the bulk thermal conductivity corresponding to that temperature (see Supplemental Material). Figure \ref{expt_comp} shows a comparison of simulation predictions with experimental measurements using the time-domain thermoreflectance (TDTR) technique \cite{ning_prep}. 

The simulation predictions using the various models are presented to provide a quantitative understanding of the contributions from each heat transfer mechanism to the thermal interface conductance. Ballistic AGF simulations with only coherent interface scattering (black solid curve denoted by `A' in Figure \ref{expt_comp}) under-predict the thermal interface conductance for all temperatures with a 33\% difference at room temperature. Also, an elastic transport model does not capture the temperature dependence of the interface conductance. Experimental data suggests that the thermal interface conductance increases by 37\% from 150 K to room temperature; however the AGF simulation predicts a modest 15\% increase in interface conductance for the same change in temperature. The stronger dependence of the experimental data on temperature suggests the importance of inelastic scattering processes in cross-interface energy transport. The inclusion of inelastic phonon scattering (magenta curve with circles denoted by `B' in Figure \ref{expt_comp}) in the AGF simulations increases the interface conductance by about 80\% at room temperature, and the simulation predictions are closer to experimental data. However, if electrons in metal are also considered in the simulation with electron-phonon coupling limited to the metal region only (red curve with hexagrams denoted by `C' in Figure \ref{expt_comp}), the thermal interface conductance decreases by about 30\% at room temperature, and the simulation under-predicts the experimental data. We note that this simulation considers the contributions from both anharmonic phonon scattering and electron-phonon coupling within the metal. 

The DFPT calculations of electron-phonon coupling presented in the previous section do not consider anharmonicity of phonon modes in the interface supercell. In a single Si-CoSi$_2$ interface with semi-infinite Si and CoSi$_2$ slabs on either side, the interface phonon modes will be localized around the interface. The spatial extent of these modes will depend on the anharmonic interaction strength with bulk Si and bulk CoSi$_2$ modes. The local electron-phonon coupling coefficient $G_{ep}$ is expected to equal the bulk values for Si and CoSi$_2$ beyond the spatial extent of these interface modes. Different approximations for the extent of joint or interface phonon modes have been proposed in the literature. Huberman and Overhauser \cite{huberman1994electronic} proposed that the joint modes extend to a distance equal to the bulk mean free path of the materials forming the interface. For Si, the average phonon mean free path is of the order of 40 nm and the use of this length predicts a large contribution to thermal transport from cross-interface electron-phonon coupling \cite{sadasivam2015electron}. Results from application of the analytical model developed by Huberman \& Overhauser to the present Si-CoSi$_2$ interface is discussed in the Supplemental Material. More recently, Lu et al. \cite{lu2016metal} argued that the extent of interfacial phonon modes should equal the distance over which the temperature profile obtained in molecular dynamics simulations is non-linear. This length is typically of the order of 1-2 nm, and this model predicts a much smaller contribution of cross-interface electron-phonon coupling to interface conductance. In the present work, we obtain an approximate estimate of this length by fitting the simulation predictions to experimental data. 
\begin{figure}[!htb]
\centering
\includegraphics[height=80mm]{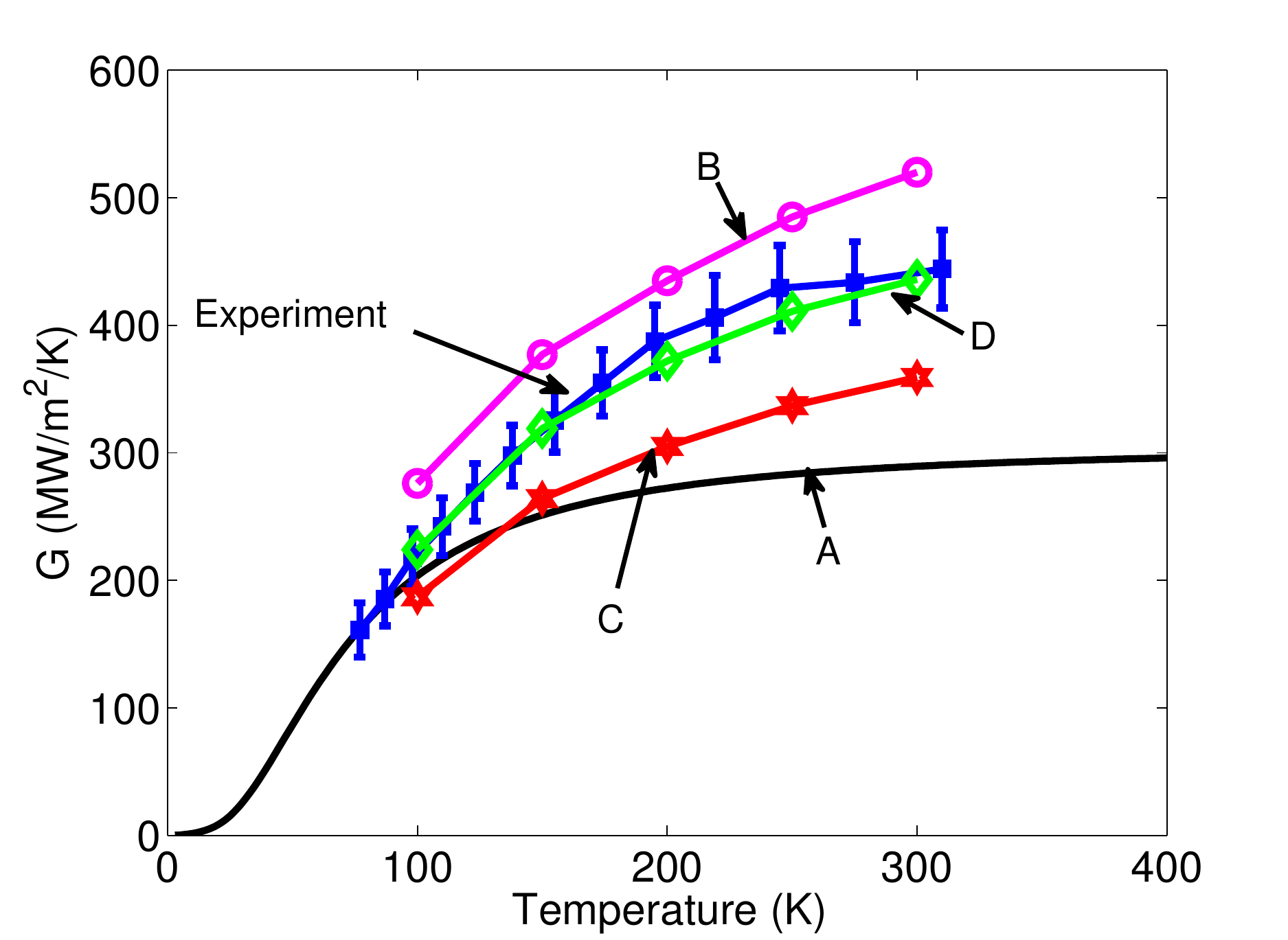}
\caption{a) Comparison of simulation predictions with experimental data (blue squares with error bars). A (black solid curve) - Phonon-only simulation with elastic interface scattering. B (magenta circles) - Phonon-only simulation with anharmonic phonon scattering in both Si and CoSi$_2$. C (red hexagrams) - Electrons and phonons considered in the simulation with electron-phonon energy transfer inside the metal region only. D (green diamonds) - Electrons and phonons considered in the simulation with electron-phonon energy transfer included in two (1.9 nm) unit cells of Si closest to the interface.}\label{expt_comp}
\end{figure}

With the assumption that cross-interface electron-phonon coupling is responsible for the difference between experimental data and the simulation results represented by the red curve in Figure \ref{expt_comp}, we use the coupling coefficient on the Si side of the interface (see Figure \ref{Gep_local_fig}a) to model energy transfer between electrons in metal and the semiconductor lattice. Curve `D'  in Figure \ref{expt_comp} represents the thermal interface conductance obtained by coupling electrons in metal with two unit cells of Si closest to the interface along the transport direction. Direct coupling with two unitcells of Si, which represents a length of approximately 1.9 nm, is found to be sufficient to obtain a close match with experimental data at various temperatures. The close match to experimental data suggests that the extent of joint interface modes in Si is much smaller than the bulk mean free path of Si. The small spatial extent of joint modes is likely due to the increased anharmonicity of interfacial phonon modes as compared to the bulk phonon modes. Similar conclusions regarding increased anharmonicity of the interfacial region are discussed in ref.~\onlinecite{gordiz2016phonon_sr} by computing the anharmonic contribution to the potential energy of interfacial atoms in Si/Ge interfaces. The temperature profile corresponding to the simulation with direct electron-phonon coupling (see Figure \ref{temp_profiles}b) is similar to that obtained from the simulation with electron-phonon coupling only in the metal region (see Figure \ref{temp_profiles}a). However, the non-zero energy flux in the B\"{u}ttiker probe closest to the interface in Si (see Figure \ref{temp_profiles}c) is indicative of direct electron-phonon energy transfer, and this effect contributes to the enhancement in thermal interface conductance. 
\section{Conclusions}
This work reports first-principles calculations of phonons and electron-phonon coupling at a Si-CoSi$_2$ interface and compares simulation predictions of thermal interface conductance to experimental measurements using the TDTR technique. TEM imaging of the Si-CoSi$_2$ interface confirms the epitaxial nature of the interface and thus enables a quantitative comparison between simulation and experiment. From a methodology standpoint, important contributions from the present work include the development of computationally efficient methods to include inelastic phonon scattering in a Green's function transport simulation and the incorporation of results from first-principles calculations of electron-phonon coupling into the AGF framework. We also evaluate the validity of the `mixing rule', a heuristic approximation to interfacial bonding at heterojunctions, using comparisons to results obtained from rigorous first-principles calculations of interfacial bonding, and find that simple averaging of interfacial force constants can result in errors of approximately 100\% in thermal interface conductance at room-temperature. 

Elastic scattering of phonons at an interface is the most widely used framework to understand and predict the thermal interface conductance of heterojunctions, but the need to include inelastic phonon and coupled electron-phonon processes has become apparent, largely due to the lack of agreement between models and experiments. The present work provides a rigorous evaluation of the contributions from various transport processes for a Si-CoSi$_2$ interface. Importantly, the experimental results, performed across a wide temperature range, only agree well with predictions that include all transport processes: elastic and inelastic phonon scattering, electron-phonon coupling only in the metal, and electron-phonon coupling across interface. The relative contributions of the various transport mechanisms would however be specific to the metal-semiconductor interface. For example, the extent of joint phonon modes is expected to be strongly sensitive to the strength of bonding at the interface (e.g., van der Waals vs. covalent bonding). Also, the polarity of interfacial bonds could have a significant impact on the strength of direct electron-phonon coupling. An interesting possibility for future work would involve a systematic study of the effect of interfacial bonding parameters on the relative contributions from the various cross-interface thermal transport mechanisms.
\section*{Acknowledgements}
SS acknowledges financial support from the Office of Naval Research (Award No: N000141211006) and Drs. Helen and Marvin Adelberg fellowship from the School of Mechanical Engineering at Purdue University. 
\bibliography{references}
\end{document}


\title{Thermal Transport Across Metal Silicide-Silicon Interfaces: First-Principles Calculations and Green's Function Transport Simulations \\
\underline{Supplemental Material}}
\author{Sridhar Sadasivam}
\affiliation{Department of Mechanical Engineering and Birck Nanotechnology Center, Purdue University, West Lafayette, IN 47907, USA}
\author{Ning Ye}
\affiliation{Department of Mechanical Engineering, University of Delaware, Newark, DE, 19716, USA}
\author{James Charles}
\affiliation{Network for Computational Nanotechnology and Department of Electrical and Computer Engineering, Purdue University, West Lafayette, Indiana 47907, USA}
\author{Kai Miao}
\affiliation{Network for Computational Nanotechnology and Department of Electrical and Computer Engineering, Purdue University, West Lafayette, Indiana 47907, USA}
\author{Joseph P. Feser}
\affiliation{Department of Mechanical Engineering, University of Delaware, Newark, DE, 19716, USA}
\author{Tillmann Kubis}
\affiliation{Network for Computational Nanotechnology (NCN), Purdue University, West Lafayette, Indiana 47907, USA}
\author{Timothy S. Fisher}
\email{tsfisher@purdue.edu}
\affiliation{Department of Mechanical Engineering and Birck Nanotechnology Center, Purdue University, West Lafayette, IN 47907, USA}
\maketitle
\section{Convergence of Cross-Interface Force Constants With Respect to Supercell Length}
\begin{figure}[!htb]
\centering
\subfloat[]{\includegraphics[height=25mm]{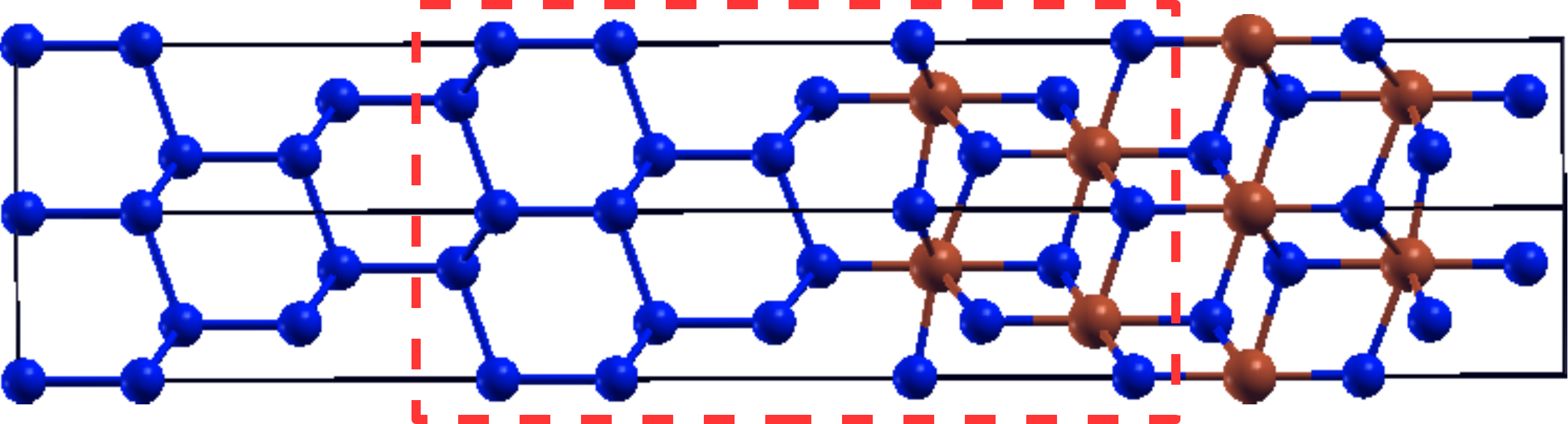}}\\
\subfloat[]{\includegraphics[height=25mm]{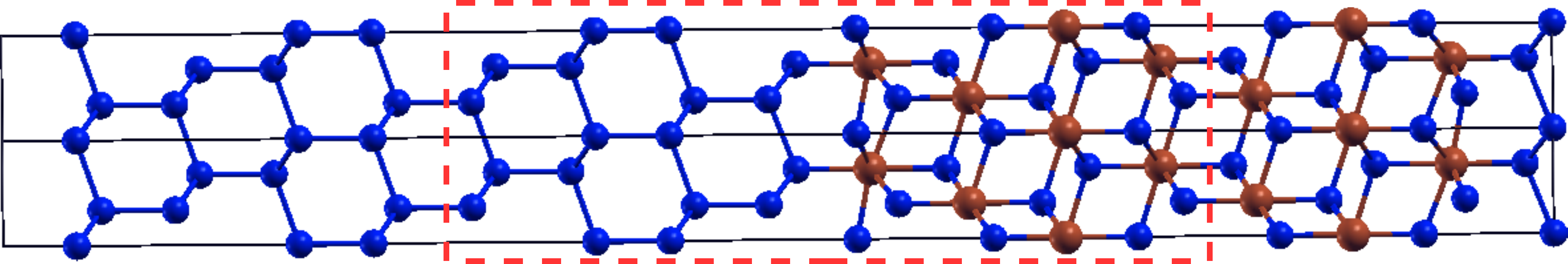}}
\caption{Supercells of the 8B interface configuration with different supercell lengths. The red dotted boxes indicate the region around the interface for which IFCs are extracted from the interface supercell calculation.}\label{interface_structure_CoSi2_appendix}
\end{figure}

\begin{figure}
\centering
\subfloat[]{\includegraphics[height=60mm]{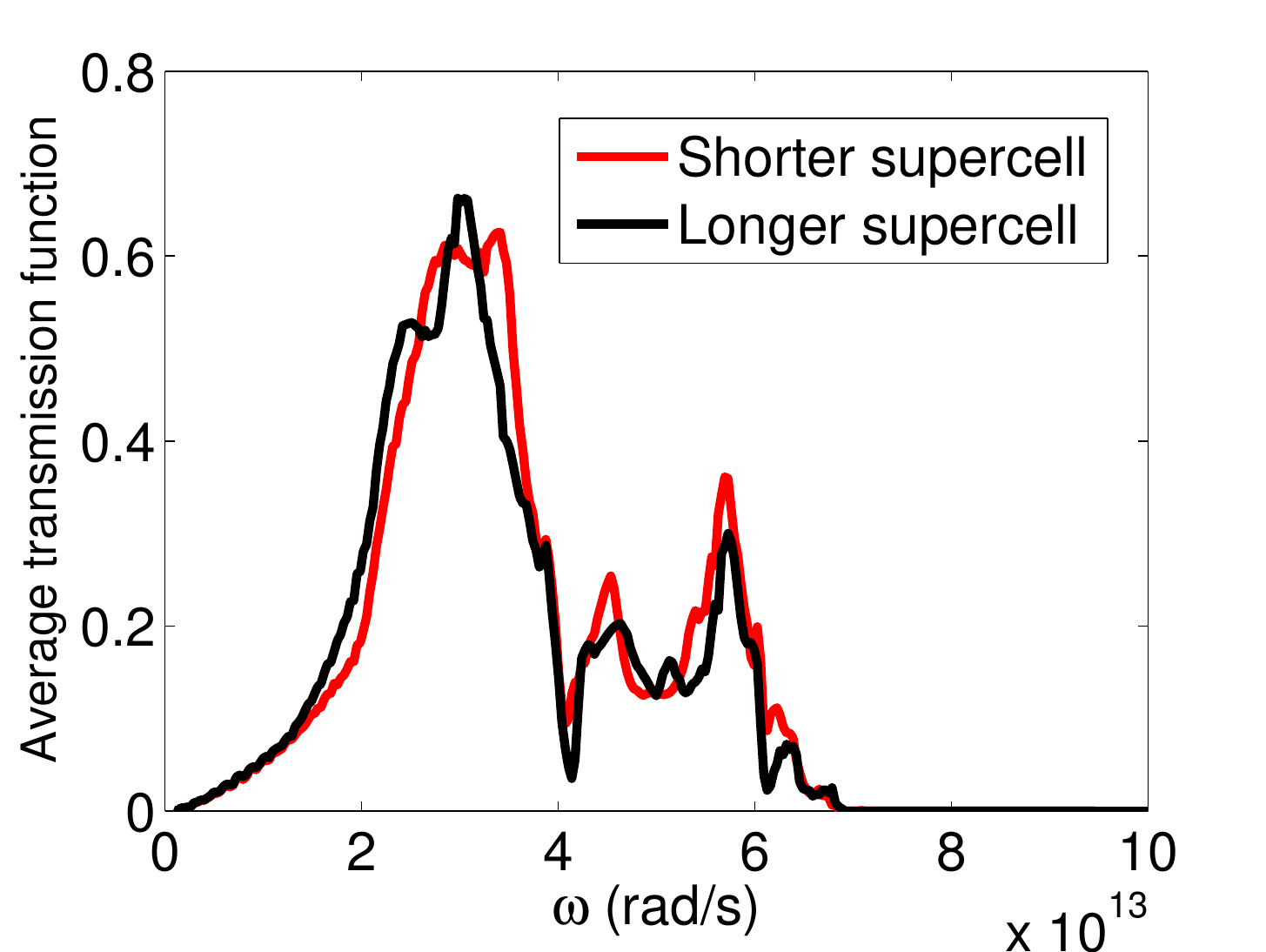}}\quad
\subfloat[]{\includegraphics[height=60mm]{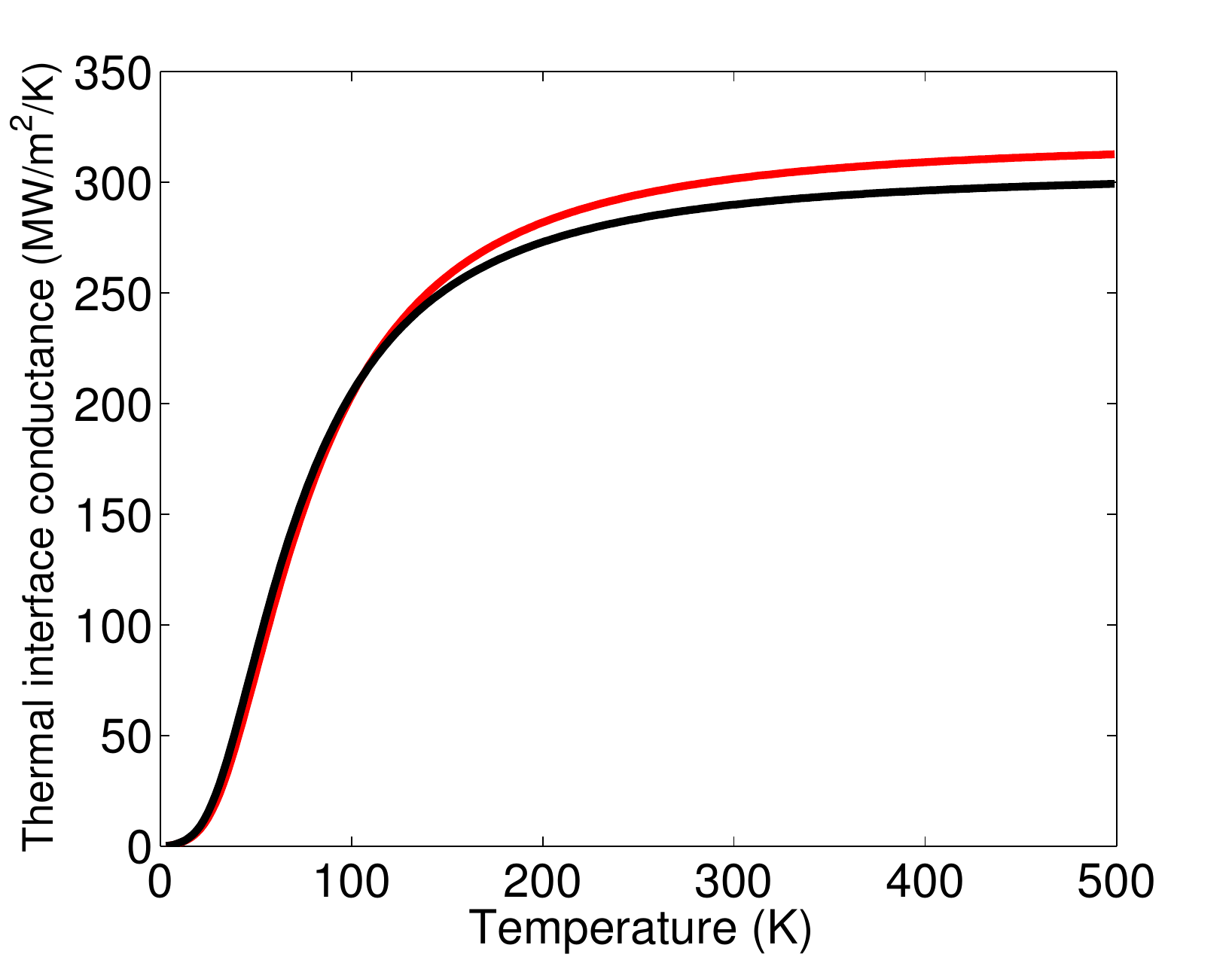}}
\caption{a) Phonon transmission function averaged over in-plane wavevectors for the 8B interface. Cross-interface force constants were obtained from DFPT calculations on supercells shown in Figure \ref{interface_structure_CoSi2_appendix}a,b. b) Thermal interface conductance of the 8B interface with cross-interface force constants obtained from DFPT calculations on supercells of different lengths.}\label{ballistic_results_appendix}
\end{figure}

Figure \ref{interface_structure_CoSi2_appendix} shows the various Si-CoSi$_2$ interface supercells of the 8B configuration considered for DFPT calculations. The red dotted box in Figures \ref{interface_structure_CoSi2_appendix}a,b show the length range of cross-interface interactions considered for the two supercells. The transport simulations assume that the force constants outside the red dotted box equal the bulk force constants of Si and CoSi$_2$. To test the convergence of interface conductance with respect to the spatial range of cross-interface interactions considered, Figure \ref{ballistic_results_appendix}b shows the transmission function averaged over in-plane wavevectors for two different supercell lengths. We observe that the transmission function and interface conductance (see Figure \ref{ballistic_results_appendix}b) are very similar for both supercells of the 8B interface shown in Figures \ref{interface_structure_CoSi2_appendix}a,b. This result proves that the IFCs between Si and CoSi$_2$ atoms outside the red dotted box are small enough to contribute insignificantly to the thermal interface conductance. 

\section{Solution Algorithm for Fourier Diffusion of Electrons Coupled with Phonons}
\label{diff_agf_algo_appendix}
\begin{figure}
\begin{center}
\includegraphics[height=65mm]{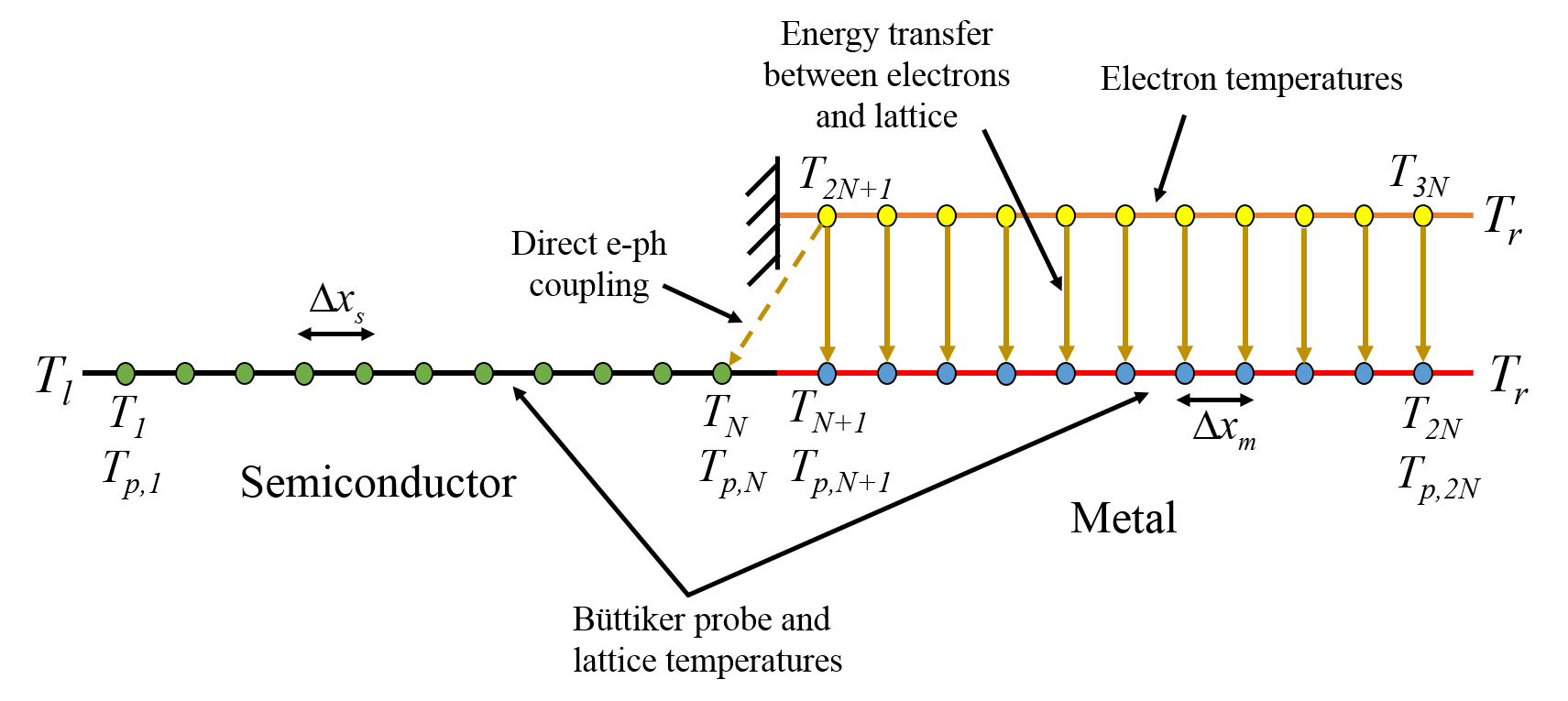}
\caption{Schematic of the various energy transfer processes between electrons in metal ($T_{2N+1} \dots T_{3N}$) and the lattice. Lattice temperatures are given by $T_{p,1} \dots T_{p,2N}$ and B\"{u}ttiker probe temperatures are denoted by $T_{1} \dots T_{2N}$. $T_l$ and $T_r$ denote the temperatures of the left and right contacts respectively. $\Delta x_s$, $\Delta x_m$ denote the spacing between adjacent grid points in the semiconductor and metal respectively.}\label{ttm_agf_schematic}
\end{center}
\end{figure}
The algorithm for coupled solution of Fourier diffusion equation for electrons along with the phonon B\"{u}ttiker probe temperatures is provided in this Appendix. The Newton-Raphson method is used to solve for the B\"{u}ttiker probe and electron temperatures simultaneously. Hence, the vector of unknown temperatures is of dimension $3N$ (see Figure \ref{ttm_agf_schematic}) where the first $N$ elements of the vector denote B\"{u}ttiker probe temperatures in the semiconductor, the next $N$ elements denote B\"{u}ttiker probe temperatures in the metal, and the last $N$ elements denote electron temperatures in the metal (we have assumed for simplicity that the number of B\"{u}ttiker probes or RGF blocks is the same in both metal and the semiconductor). 

The energy conservation requirement for any control volume $i$ ($2N+2\leq i \leq 3N-1$) in the electron grid is given by:
\begin{equation}
k_e\frac{T_{i+1}+T_{i-1}-2T_{i}}{\Delta x_m} = 2\pi\Delta x_m D(E_f)\int\limits_{0}^{\infty}{(\hbar\omega)^2\alpha^2F(\omega)[f_{BE}^o(T_{i})-f_{BE}^o(T_{p,i-N})]d\omega}
\label{energy_cons_electron}
\end{equation}
$T_{p}$ in the above equation denotes the local device temperature, and this could in general be different from the local B\"{u}ttiker probe temperature. The above equation applies for all control volumes in the electron grid except for the first ($i=2N+1$) and last ($i=3N$) control volumes where the adiabatic boundary condition (electrons are insulated at the metal-semiconductor interface) and the Dirichlet boundary condition of the right contact apply. $\Delta x_m$ denotes the spacing between grid points for the electron temperature in the metal and is same as the spacing between B\"{u}ttiker probes or the length of each `block' used in the RGF algorithm. The energy conservation equations for $i=2N+1$ and $i=3N$ are shown below:
\begin{equation}
\begin{split}
k_e\frac{T_{i+1}-T_{i}}{\Delta x_m} & = 2\pi\Delta x_m D(E_f)\int\limits_{0}^{\infty}{(\hbar\omega)^2\alpha^2F(\omega)[f_{BE}^o(T_{i})-f_{BE}^o(T_{p,i-N})]d\omega}\\
k_e\frac{T_r+T_{i-1}-2T_{i}}{\Delta x_m} & = 2\pi\Delta x_m D(E_f)\int\limits_{0}^{\infty}{(\hbar\omega)^2\alpha^2F(\omega)[f_{BE}^o(T_{i})-f_{BE}^o(T_{p,i-N})]d\omega}
\end{split}
\label{energy_cons_electron_first_last_pt}
\end{equation}
where $T_r$ denotes the temperature of the right (metal) contact. The energy current conservation equation for a B\"{u}ttiker probe $i$ in the metal side of the interface ($N+1\leq i \leq 2N$) is given by:
\begin{multline}
2\pi \Delta x_m D(E_f)\int\limits_{0}^{\infty}{(\hbar\omega)^2\alpha^2F(\omega)[f_{BE}^o(T_{i+N})-f_{BE}^o(T_{p,i})]d\omega} = \\ \sum\limits_{q_{||}}\int\limits_{0}^{\infty}{\frac{\hbar\omega}{2\pi}\Tr(\Sigma_i^{in}A-\Gamma_iG^n)d\omega}
\label{energy_cons_bp_metal}
\end{multline}
Energy current conservation equation for a B\"{u}ttiker probe $i$ in the semiconductor side of the interface ($1\leq i \leq N$) is given by:
\begin{equation}
\sum\limits_{q_{||}}\int\limits_{0}^{\infty}{\frac{\hbar\omega}{2\pi}\Tr(\Sigma_i^{in}A-\Gamma_iG^n)d\omega} = 0
\label{energy_cons_bp_semiconductor}
\end{equation}
The above equation assumes that electron-phonon energy transfer is zero for B\"{u}ttiker probes in the semiconductor. When direct coupling between metal electrons and the semiconductor lattice is considered, the above equation would need to be modified to include electron-phonon energy transfer for all the B\"{u}ttiker probes that are within the region of electron-phonon interaction in the semiconductor. 

Next, we provide expressions for elements of the Jacobian matrix $J$ that is needed in the Newton-Raphson method. For $1\leq i\leq N$, 
\begin{equation}
J_{i,j} = 
\begin{cases}
\sum\limits_{q_{||}}\int\limits_{0}^{\infty}{\frac{\hbar\omega}{2\pi}\Tr\left(\Gamma_iA_{ii}\frac{\partial f_{BE}^o}{\partial T}\bigg|_{T_j}\delta_{ij}-\Gamma_i\frac{\partial G^n_{ii}}{\partial T_j}\right)d\omega} \text{ if }j\leq2N \\ 
0 \text{ if }j>2N \\
\end{cases}
\label{Jac_ph_semiconductor}
\end{equation}
For $N+1\leq i \leq 2N$,
\begin{equation}
J_{i,j} = 
\begin{cases}
\sum\limits_{q_{||}}\int\limits_{0}^{\infty}{\frac{\hbar\omega}{2\pi}\Tr\left(\Gamma_iA_{ii}\frac{\partial f_{BE}^o}{\partial T}\bigg|_{T_j}\delta_{ij}-\Gamma_i\frac{\partial G^n_{ii}}{\partial T_j}\right)d\omega} + \\ 2\pi \Delta x_m D(E_f)\int\limits_{0}^{\infty}{(\hbar\omega)^2\alpha^2F(\omega)\frac{\partial f_{BE}^o}{\partial T}\bigg|_{T=T_{p,i}}\frac{\partial T_{p,i}}{\partial T_{j}}d\omega}\text{ if }j\leq2N \\ 
-2\pi \Delta x_m D(E_f)\int\limits_{0}^{\infty}{(\hbar\omega)^2\alpha^2F(\omega)\frac{\partial f_{BE}^o}{\partial T}\bigg|_{T=T_{i+N}}\delta_{i+N,j}} \text{ if }j>2N \\
\end{cases}
\label{Jac_ph_metal}
\end{equation}
For $2N+2 \leq i \leq 3N-1$, 
\begin{equation}
J_{i,j} = 
\begin{cases}
-2\frac{k_e}{\Delta x_m}- 2\pi \Delta x_m D(E_f)\int\limits_{0}^{\infty}{(\hbar\omega)^2\alpha^2F(\omega)\frac{\partial f_{BE}^o}{\partial T}\bigg|_{T=T_{i}}d\omega}\text{ if }j=i \\ 
\frac{k_e}{\Delta x_m}\text{ if }j=i+1 \\ 
\frac{k_e}{\Delta x_m}\text{ if }j=i-1 \\ 
2\pi \Delta x_m D(E_f)\int\limits_{0}^{\infty}{(\hbar\omega)^2\alpha^2F(\omega)\frac{\partial f_{BE}^o}{\partial T}\bigg|_{T=T_{p,i-N}}\frac{\partial T_{p,i-N}}{\partial T_{j}}d\omega} \text{ if }j\leq 2N \\
\end{cases}
\label{Jac_electron}
\end{equation}
Eqs.~(\ref{Jac_ph_metal}), (\ref{Jac_electron}) involve the term $\partial T_{p,i}/\partial T_{j}$ ($1\leq i,j \leq 2N$) which denotes the derivative of local device temperature with respect to the B\"{u}ttiker probe temperature:
\begin{equation}
\frac{\partial T_{p,i}}{\partial T_j} = \frac{\sum\limits_{q_{||}}\int\limits_{0}^{\infty}{\omega^2 \Tr\left[\frac{\partial G^n_{ii}(\omega;q_{||})}{\partial T_j}\right]d\omega}}{\sum\limits_{q_{||}}\int\limits_{0}^{\infty}{\omega^2 \Tr\left[A_{ii}(\omega;q_{||})\right]\frac{\partial f_{BE}^o}{\partial T}\bigg|_{T=T_{p,i}}d\omega}}
\end{equation}
The algorithm for iterative solution of the B\"{u}ttiker probe, lattice, and electron temperatures can be summarized as follows:
\begin{itemize}
\item Start with an initial guess for the B\"{u}ttiker probe and electron temperatures.
\item For each phonon frequency, compute $G^R_{ii}$, $G^n_{ii}$, and $\frac{\partial G^n_{ii}}{\partial T_j}$ using RGF.
\item Compute the local lattice temperature. 
\item Compute the net energy currents in each control volume of the electron grid, B\"{u}ttiker probes in the metal and semiconductor using Eqs.~(\ref{energy_cons_electron}), (\ref{energy_cons_electron_first_last_pt}), (\ref{energy_cons_bp_metal}) \& (\ref{energy_cons_bp_semiconductor}).
\item Compute the Jacobian matrix using Eqs.~(\ref{Jac_ph_semiconductor}), (\ref{Jac_ph_metal}) \& (\ref{Jac_electron}).
\item Update the temperature of B\"{u}ttiker probes and electrons using the Newton equation:
\begin{equation}
T_{new} = T_{old}-J^{-1}f
\end{equation}
\item If $\lVert T_{new}-T_{old} \rVert > \epsilon$, go back to Step 1. 
\end{itemize}
\section{Fitting of B\"{u}ttiker Probe Parameters to Bulk Thermal Conductivity}\label{thermal_cond_fit}
The energy-dependent scattering times used in the B\"{u}ttiker probe self-energy are chosen to ensure that the AGF-B\"{u}ttiker probe approach predicts the correct bulk thermal conductivity of Si and CoSi$_2$. We assume a quadratic dependence of the Umklapp scattering rate, i.e., $\tau^{-1}(\omega) = A\omega^2$ where the prefactor $A$ is fitted to reproduce the correct bulk thermal conductivity. A quadratic frequency dependence of scattering rate is assumed based on prior studies in the literature for Si \cite{singh2011effect,jeong2012thermal}. 

The procedure to extract thermal conductivity from AGF simulations involves the application of a temperature difference across a slab of the homoegeneous material, and computing the temperature profiles and energy currents for varying device lengths. For a specific length of the device, we extract the thermal conductivity from the heat flux $J$ and the slope of the temperature profile $dT/dx$ as $\kappa(L) = J/(dT/dx)$. $T$ denotes the local lattice temperature and not the B\"{u}ttiker probe temperature. The device lengths considered in the present work are much shorter than the maximum mean free path of bulk Si (mean free paths in bulk Si span four orders of magnitude \cite{minnich2011thermal}). Also, AGF simulations are computationally expensive to simulate large device lengths of the order of microns. 

In the present work, we adopt the linear extrapolation method that has been used to mitigate size effects in direct molecular dyanmics simulations where the inverse of thermal conductivity is plotted as a function of the inverse length, and a linear fit to the data is extrapolated to infinite length \cite{schelling2002comparison}. Although this method has been used in a number of prior studies \cite{ni2009thermal,pei2013tuning,hou2016modeling}, ref.~\onlinecite{sellan2010size} showed that a linear fit to data of $1/\kappa$ and $1/L$ is strictly appropriate only when the sample length is comparable to the maximum mean free path. In Si, the linear extrapolation technique was shown to under-predict thermal conductivity at temperatures of 500 and 1000 K. However, the maximum mean free path in bulk Si is of the order of microns and such large length scales are computationally intractable in AGF simulations. Also, no alternative methods to extract thermal conductivity (such as the Green Kubo method in molecular dynamics) exist to determine thermal conductivity in AGF simulations with inelastic scattering. Hence, we adopt the simple linear extrapolation method to fit the B\"{u}ttiker probe scattering parameters in the present work. Thermal conductivity from the extrapolation technique is also compared with that obtained from the inverse slope of a linear fit to the variation of thermal resistance of the device with increasing device length \cite{luisier2012atomistic}. 

\begin{figure}[htb]
\centering
\subfloat[]{\includegraphics[height=55mm]{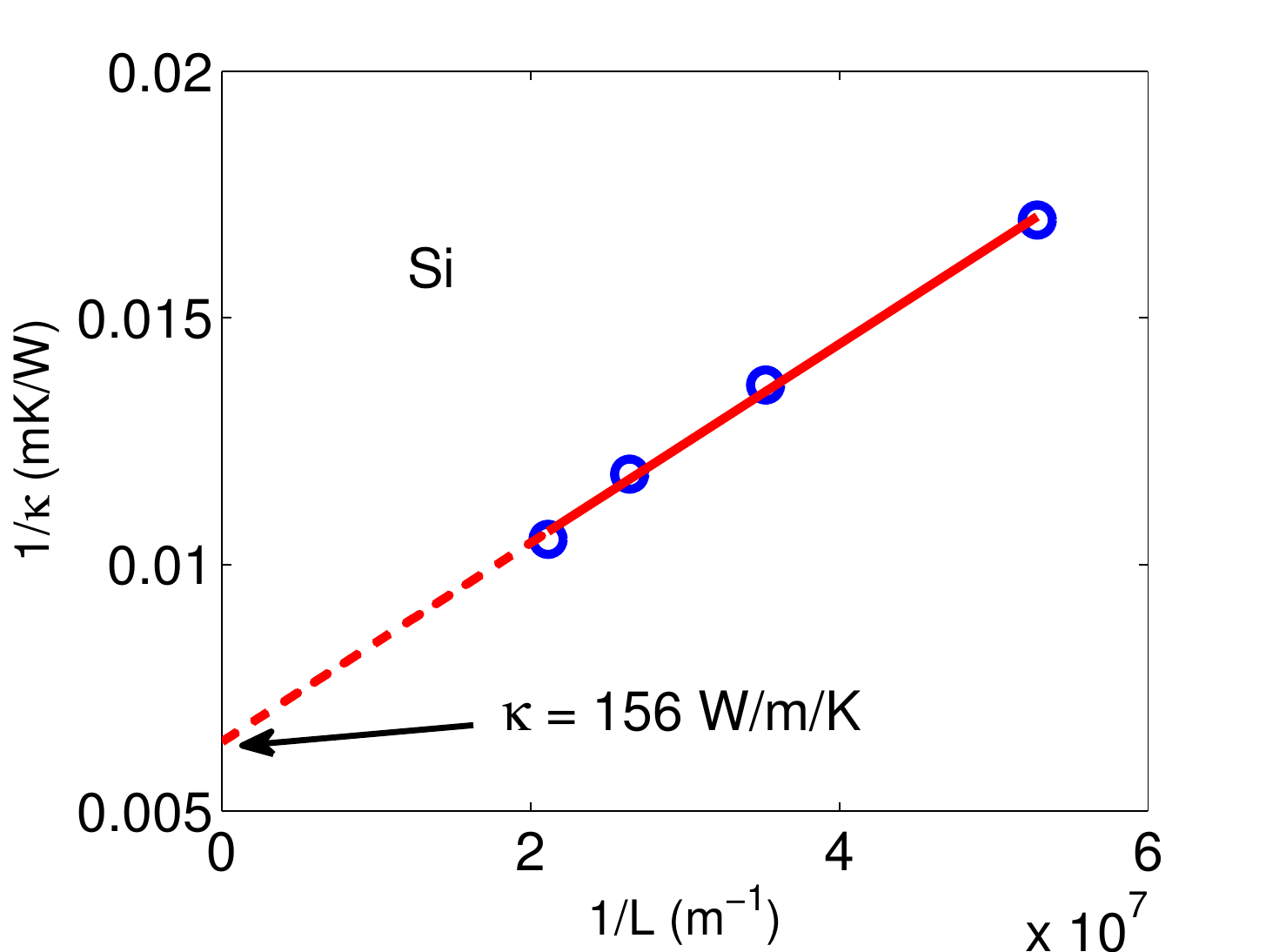}}
\subfloat[]{\includegraphics[height=55mm]{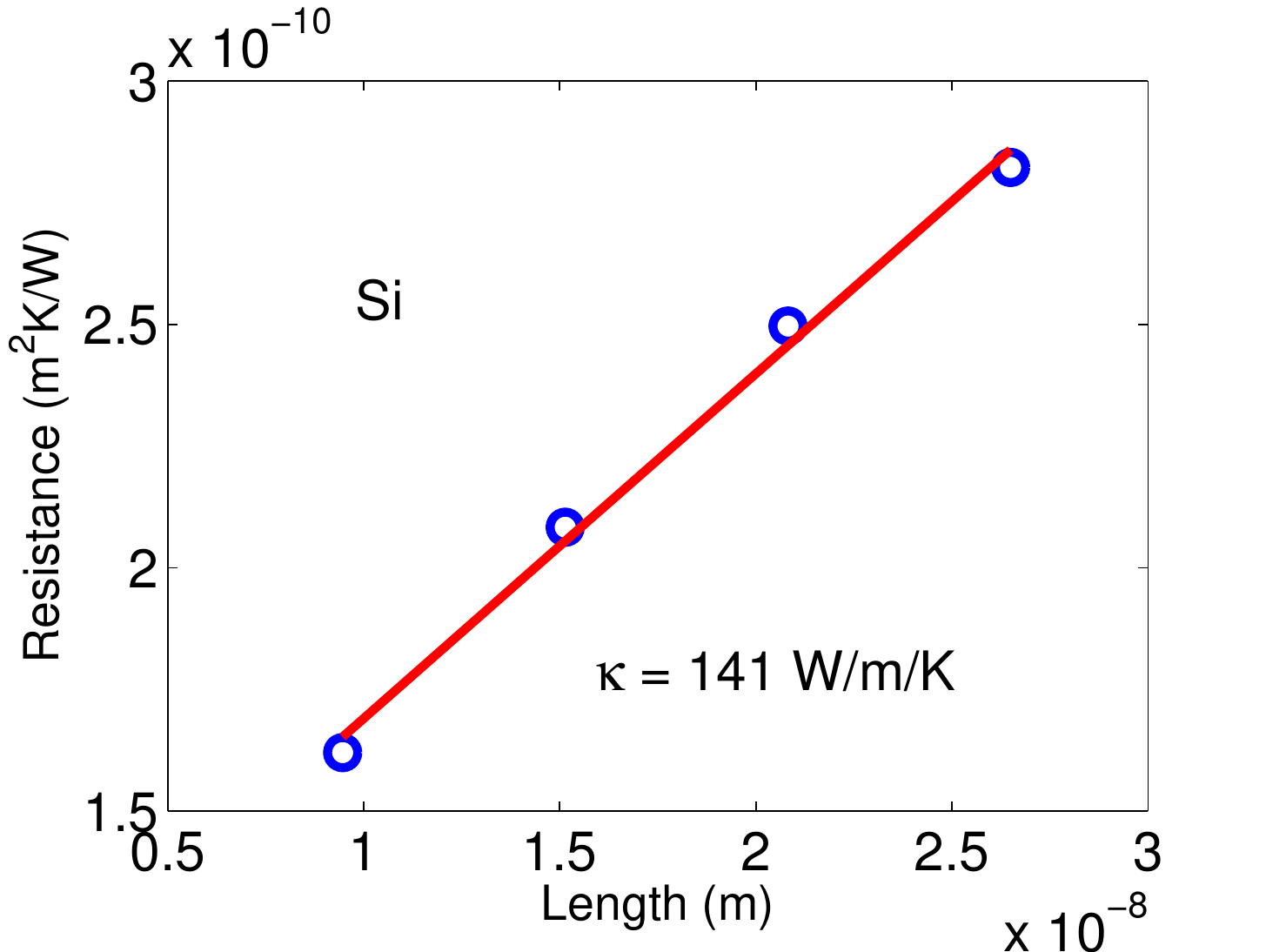}}\\
\subfloat[]{\includegraphics[height=55mm]{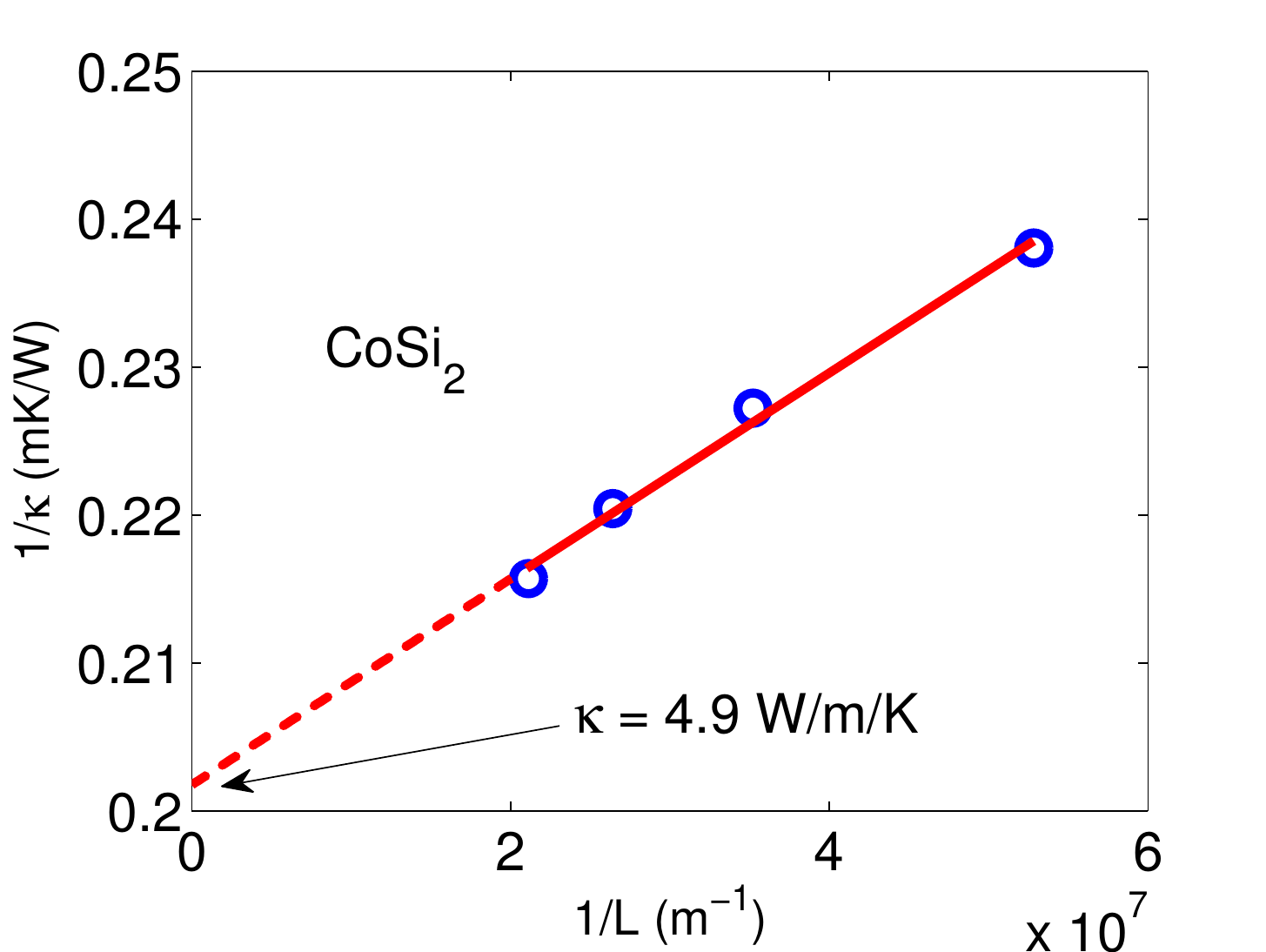}}
\subfloat[]{\includegraphics[height=55mm]{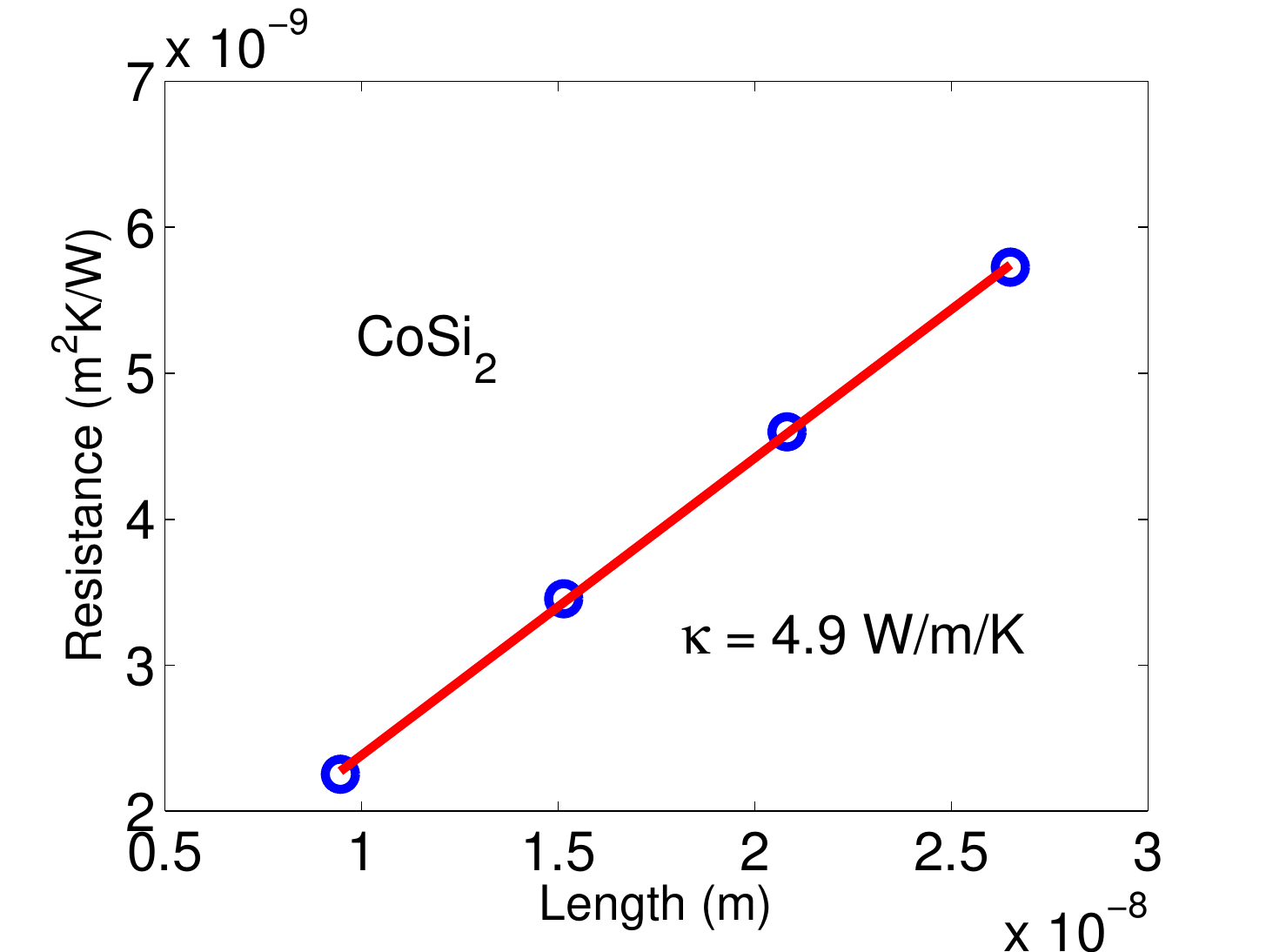}}
\caption{a,b) Thermal conductivity of Si obtained from linear extrapolation of the inverse length-dependent thermal conductivity and from linear fits to length-dependent device thermal resistance. c,d) Thermal conductivity of CoSi$_2$ obtained from linear extrapolation of the inverse length-dependent thermal conductivity and from linear fits to length-dependent device thermal resistance. Blue circles represent data extracted from the AGF simulations while the red lines are linear fits to the data.}\label{kappa_fit}
\end{figure}

Figure \ref{kappa_fit} shows the thermal conductivity at room temperature obtained from the linear extrapolation technique and using a linear fit to thermal resistance of the device region. For Si, the predictions from the two methods are within 10\% of each other while the two predictions show an excellent match for CoSi$_2$ which has a lower thermal conductivity with shorter mean free paths. The lattice contribution to thermal conductivity of CoSi$_2$ has been determined in prior literature by subtracting the electronic contribution to thermal conductivity (estimated using the Wiedemann-Franz law) from the total experimentally measured thermal conductivity. Ref.~\onlinecite{neshpor1968thermal} estimated the lattice thermal conductivity to be 4.8 W/m/K while ref.~\onlinecite{ditchek1984cobalt} estimated the lattice thermal conductivity to be negligible since the total thermal conductivity showed an excellent match with the electronic thermal conductivity from Wiedemann-Franz law. In the present work, we fit the scattering parameter for CoSi$_2$ to obtain a bulk thermal conductivity of 4.9 W/m/K at room temperature (see Figures \ref{kappa_fit}b,c). When electrons are included in the transport simulation, the thermal interface conductance is found to exhibit a weak dependence on the exact lattice thermal conductivity of CoSi$_2$. This is a direct consequence of electrons dominating the thermal conductivity of CoSi$_2$ ($\kappa_e = 46$ W/m/K) and even a ten-fold increase in the phonon-phonon scattering rate in CoSi$_2$ is found change the interface conductance by less than 10\%. Hence, the uncertainty in the exact lattice thermal conductivity of CoSi$_2$ does not have a significant influence on the predictions of thermal interface conductance. The scattering rate in Si is fitted to its thermal conductivity at a few different temperatures and the corresponding scattering parameters are shown in Table \ref{scat_rate_A}.

\begin{table}[ht]
\centering
\caption{Temperature-dependent B\"{u}ttiker probe scattering prefactor for Si and CoSi$_2$.}
\label{scat_rate_A}
\begin{tabular}{|l|l|l|l|l|l|}
\hline
Material & 100 K & 150 K                         & 200 K                          & 250 K                         & 300 K                         \\ \hline
Si       & $3\times 10^{-20}$ s$^{-1}$ & $10^{-19}$ s$^{-1}$  & $2.4 \times 10^{-19}$ s$^{-1}$ &  $6.6 \times 10^{-19}$ s$^{-1}$ & $1.1\times 10^{-18}$ s$^{-1}$ \\ \hline
CoSi$_2$ & $1.1\times 10^{-16}$ s$^{-1}$ & $1.1\times 10^{-16}$ s$^{-1}$ & $1.1\times 10^{-16}$ s$^{-1}$  & $1.1\times 10^{-16}$ s$^{-1}$ & $1.1\times10^{-16}$ s$^{-1}$ \\ \hline
\end{tabular}
\end{table}

\section{Dependence of the Eliashberg Function on Smearing and k-Space Grid}
\begin{figure}[htb]
\centering
\subfloat[]{\includegraphics[height=60mm]{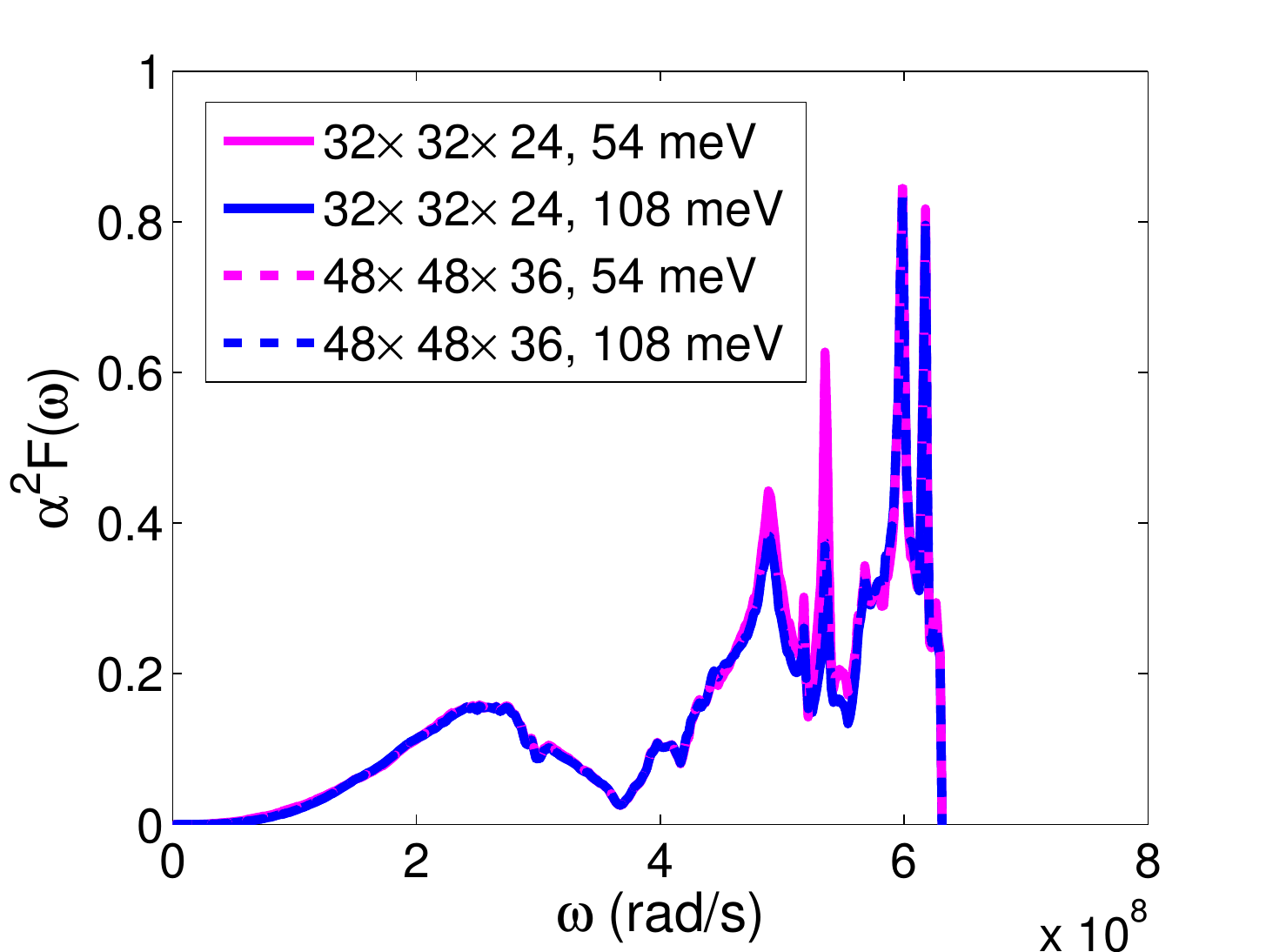}}\\
\subfloat[]{\includegraphics[height=55mm]{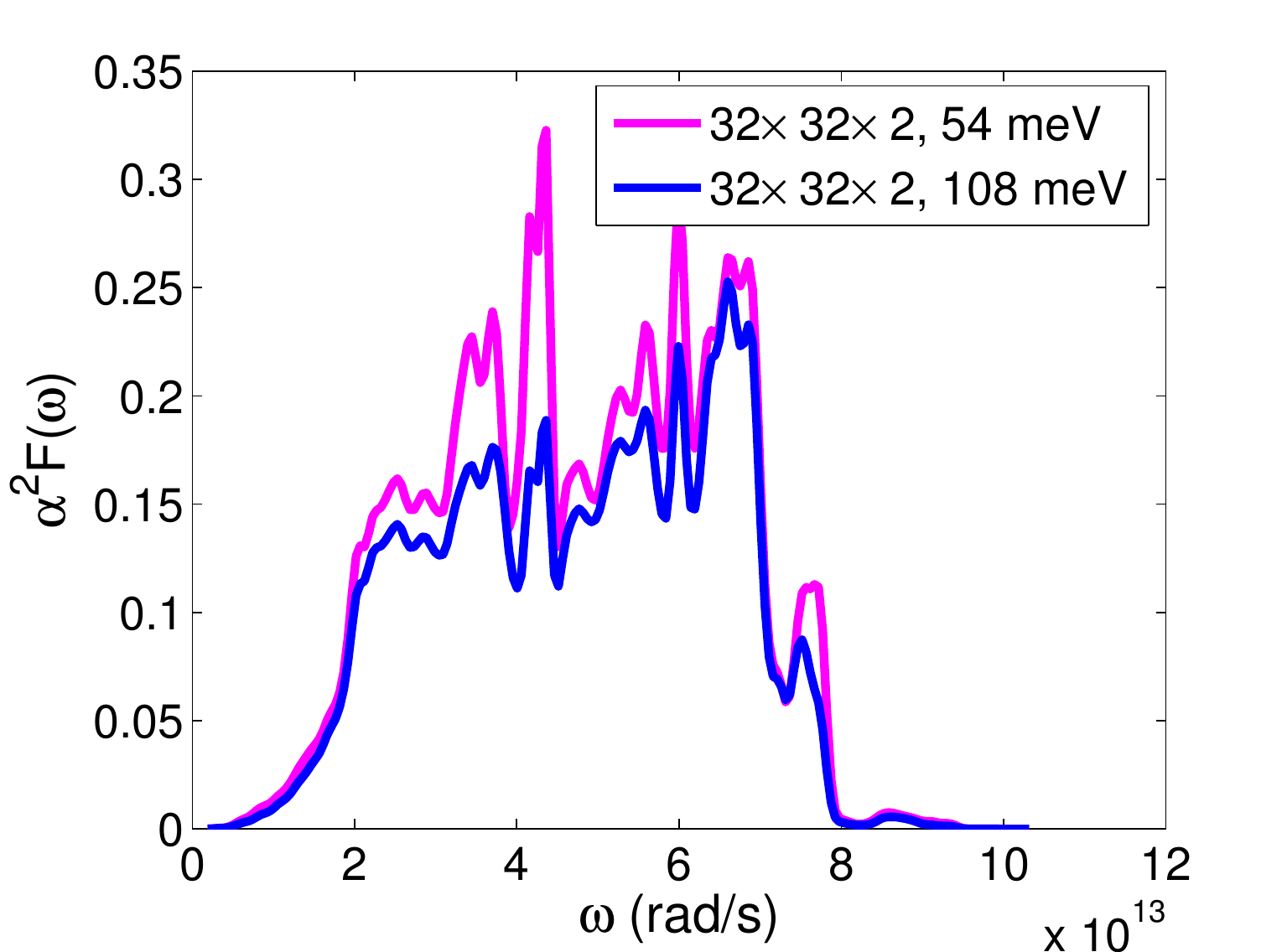}}
\subfloat[]{\includegraphics[height=55mm]{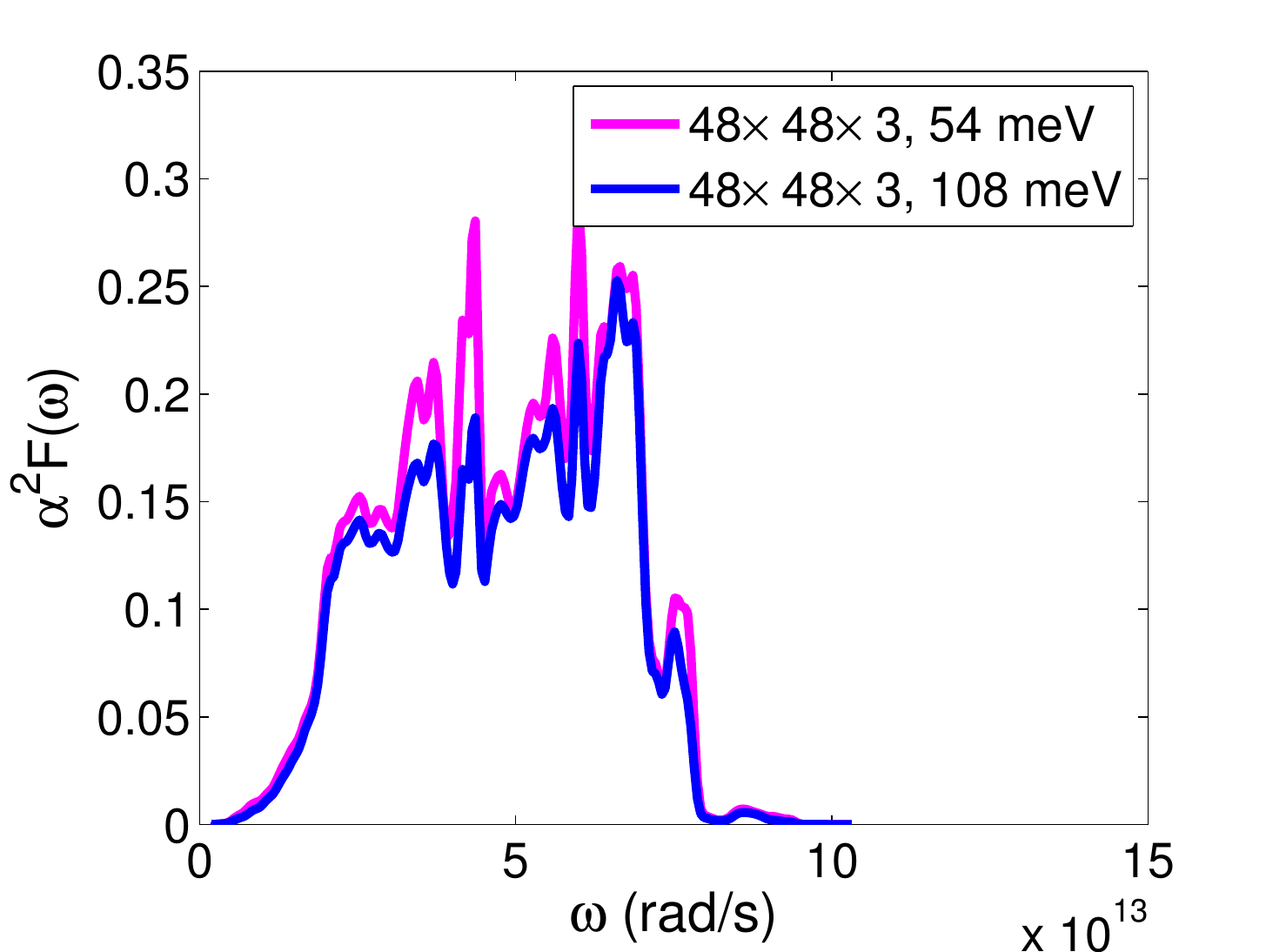}}
\caption{Eliashberg function of bulk strained CoSi$_2$ (a) and the interface supercell (b,c). The legends correspond to different k-point grids (for linear interpolation of electron-phonon matrix elements) and smearing used in the calculation of phonon linewidths.}\label{a2F_broadening}
\end{figure}

The computation of phonon linewidth and the associated Eliashberg function involves the integration of a double-delta function around the Fermi surface. Convergence of the phonon linewidth requires a fine k-point grid and interpolation techniques such as the Wannier-Fourier interpolation of electron-phonon matrix elements have been proposed in the literature \cite{giustino2007electron}. However, Wannierization of the complicated electronic bandstructure associated with a metal-semiconductor interface supercell is computationally expensive and disentanglement of the bands around Fermi energy is expected to be difficult. Hence, we use a simple linear interpolation of electron-phonon matrix elements in k-space while Fourier interpolation is used to interpolate the phonon linewidths in q-space using the approach outlined in ref.~\onlinecite{wierzbowska2005origins}. Self-consistent calculations for bulk CoSi$_2$ and the Si-CoSi$_2$ interface supercell have been performed on a fine k-point grid of $16\times 16\times 12$ and $16 \times 16 \times 1$ respectively and the electron-phonon matrix elements computed on this k-point grid (for a given phonon wavevector) are interpolated to finer k-point grids before the calculation of phonon linewidths. To check the numerical convergence of phonon linewidths, we plot the Eliashberg function of bulk CoSi$_2$ and the interface supercell (see Figure \ref{a2F_broadening}) for different k-point grids used in the interpolation of electron-phonon matrix elements and the smearing values used in the Gaussian approximation of delta functions. Figure \ref{a2F_broadening} shows that the Eliashberg function and the associated electron-phonon coupling coefficient are well converged with respect to the k-point grid and the smearing value except for small differences in the peaks of the Eliashberg function for different smearing parameters which are expected to reduce with further refinement of k-point grids. 
\section{Interface Conductance from Huberman \& Overhauser Model}
In this section, we predict the contribution to interface conductance from direct electron-phonon coupling using the analytical model developed by Huberman \& Overhauser \cite{huberman1994electronic}. Their analytical model considers the joint phonon modes as evanescent waves which decay with a characteristic length that equals the bulk mean free path of the materials forming the interface. The bulk phonon mean free path of Si is about 40 nm while the phonon mean free path of CoSi$_2$ is about 1 nm, estimated using the kinetic theory formula for thermal conductivity $\kappa = (1/3)C_vv_g\Lambda$, where the phonon thermal conductivity of CoSi$_2$ is 4.9 W/m/K, heat capacity $C_v=2.75 \times 10^6$ J/m$^3$/K, phonon group velocity $v_g = 5134$ m/s (averaged over the TA and LA phonon group velocities). Hence, the primary assumption of the Huberman \& Overhauser model, i.e., the joint phonon mode energy density is concentrated on the semiconductor side of the interface is expected to hold for the Si-CoSi$_2$ interface if the length scale of joint phonon modes equals the bulk phonon mean free path. 

\begin{figure}
\centering
\includegraphics[height=60mm]{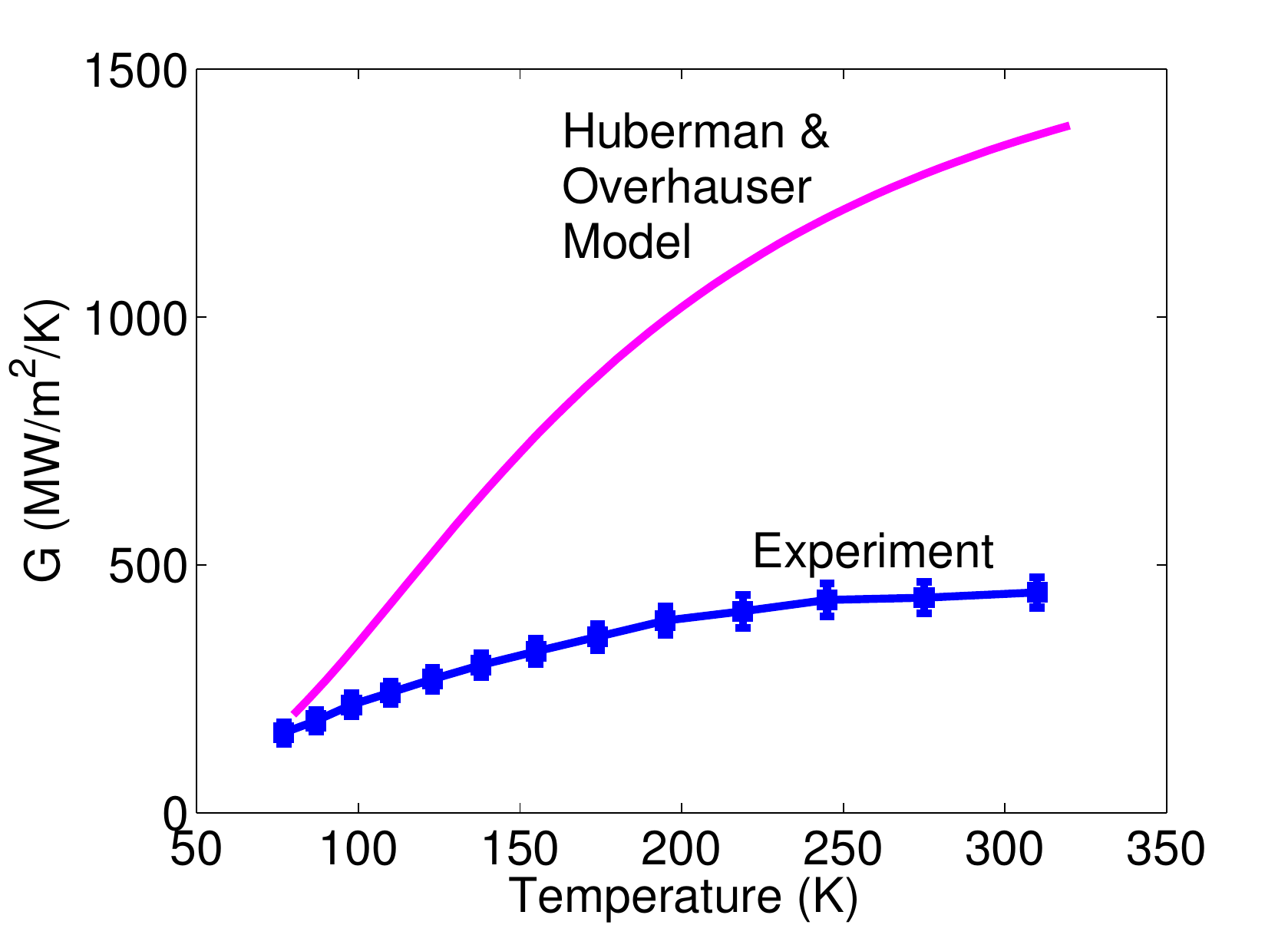}
\caption{Comparison of interface conductance obtained from the analytical model of Huberman \& Overhauser with experimental data.}\label{G_huberman}
\end{figure}

The final result for the interface conductance from cross-interface electron-phonon coupling in ref.~\cite{huberman1994electronic} is given by:
\begin{equation}
\label{huberman_model}
G_{ep,i} = INk_Bv_{g,l}f(\theta_{D,l}/T)
\end{equation}
where $I$ is a function that depends on the phonon group velocities, mass density, and elastic properties of the materials forming the interface. $N$ denotes the unit-cell density in CoSi$_2$ and $v_{g,l}$ is the longitudinal phonon group velocity in CoSi$_2$. $\theta_{D,l}$ is the Debye temperature of longitudinal phonons in CoSi$_2$ and the function $f(y)$ is defined as follows:
\begin{equation}
f(y) = \frac{3}{y^3}\int\limits_{0}^{y}\frac{e^x}{(e^x-1)^2}x^4dx
\end{equation}
The interface conductance obtained from application of Eq.~(\ref{huberman_model}) for the Si-CoSi$_2$ interface is shown in Figure \ref{G_huberman} along with the experimental data. The simple analytical model of Huberman \& Overhauser is found to significantly over-estimate the contribution of direct electron-phonon coupling to the interface conductance. The assumption that the length scale of joint phonon modes equals the bulk mean free path is likely the main contributor to the large error in interface conductance. From the simulations performed in this paper, a length scale of 1.9 nm is extracted by fitting the simuation predictions to experimental data (see main text). Such a length scale is significantly smaller than the bulk phonon mean free path in Si. Also, the use of a deformation potential model for coupling between electrons and interfacial phonon modes may not be quantitatively accurate as such models are typically applied for electron-phonon scattering in bulk materials. We conclude that although the Huberman \& Overhauser model is useful to understand the underlying physics of interfacial phonon modes and its coupling with electrons, the results from the model may not be quantitatively accurate due to several simplifying assumptions. 
\section{Electron-Phonon Coupling in Bulk Silicon}
\begin{figure}
\centering
\includegraphics[height=60mm]{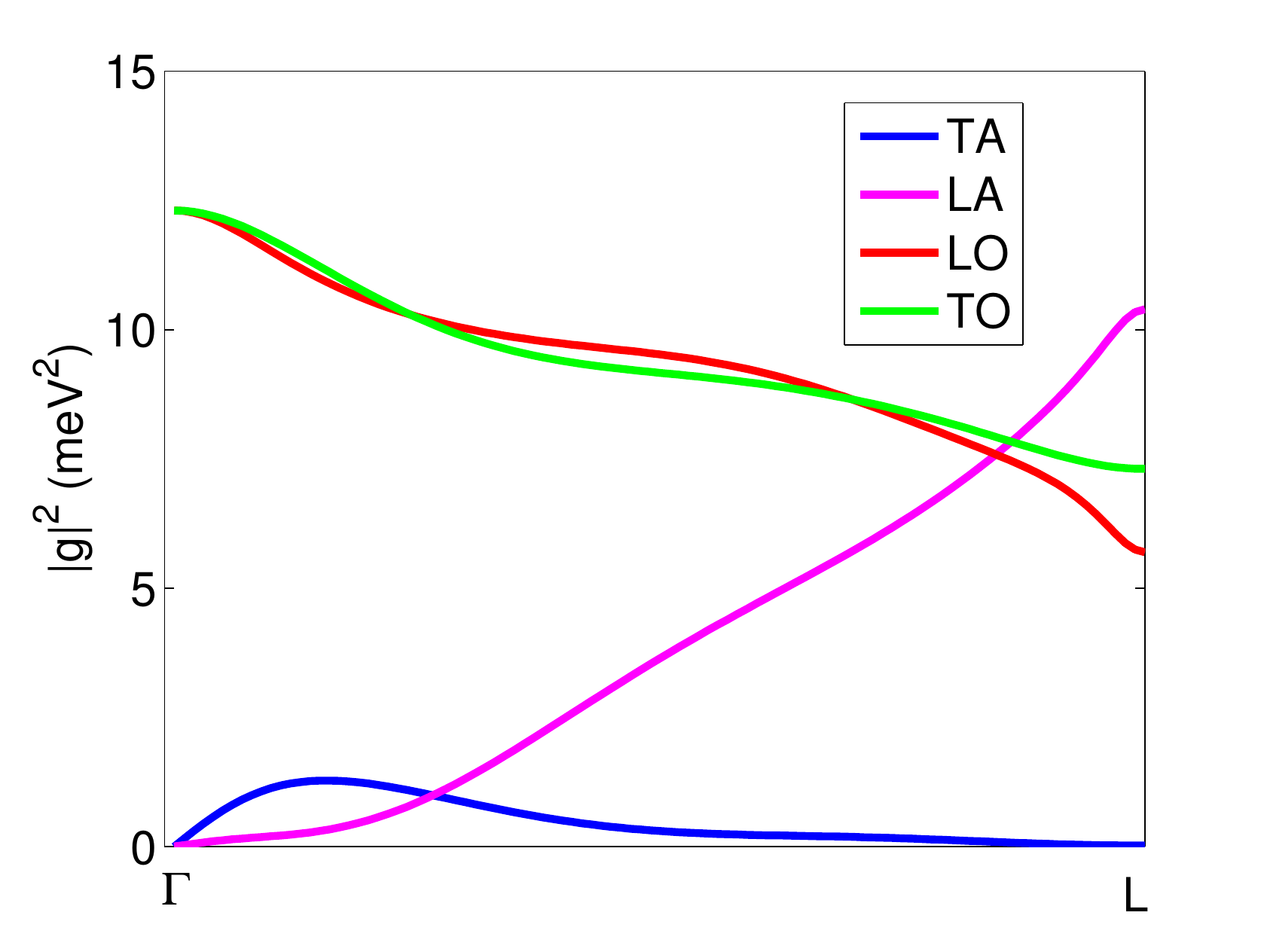}
\caption{Square of the absolute magnitude of electron-phonon matrix elements for phonons along the $\Gamma$-L direction and electrons at the valence band maximum of Si.}\label{eph_matrix_plot}
\end{figure}

In this section, we compare the electron-phonon scattering matrix elements for acoustic and optical phonon modes of Si. The scattering matrix elements are computed using the first-principles linear response code Quantum Espresso \cite{Giannozzi_JPhysics_2009} along with the Electron-Phonon Wannier (EPW) \cite{Ponce2016116} code for Wannier interpolation of electron-phonon matrix elements. Figure \ref{eph_matrix_plot} shows the square of the absolute magnitude of the matrix elements as a function of phonon wavevector along the $\Gamma$-L direction for an electron state at the valence band maximum of Si (averaged over degenerate states at the valence band maximum). We observe that the scattering matrix elements for optical phonons are larger than that for acoustic phonons near the $\Gamma$ point and the matrix elements are of commensurate magnitude near the edge of the Brillouin zone. This result is in contrast with electron-phonon coupling for interfacial Si modes where acoustic delocalized phonon modes dominate coupling with metal electrons. 
\bibliography{references}